\documentclass[onecolumn,superscriptaddress,nofootinbib,prd,preprintnumbers,amsmath,amssymb]{revtex4}

\usepackage[pdftex]{graphicx}
\usepackage{dcolumn}
\usepackage{bm}
\usepackage{graphics}
\usepackage{rotating}
\usepackage{amssymb} 
\usepackage{comment}
\usepackage{multirow}
\usepackage{footnote}
\usepackage{comment}

\newcommand{\nc}{\newcommand}  

\def\Acknowledgements{\bigskip  \bigskip \begin{center} \begin{large}
             \bf ACKNOWLEDGEMENTS \end{large}\end{center}}

\def\beq{\begin{equation}}
\def\eeq#1{\label{#1}\end{equation}}
\def\eeqn{\end{equation}}

\newenvironment{Eqnarray}%
   {\arraycolsep 0.14em\begin{eqnarray}}{\end{eqnarray}}
\def\beqa{\begin{Eqnarray}}
\def\eeqa#1{\label{#1}\end{Eqnarray}}
\def\eeqan{\end{Eqnarray}}

\nc{\ra}{\rightarrow}  
\nc{\slsh}{\slash\hspace*{-0.22cm}}
\def\Re{{\cal R \mskip-4mu \lower.1ex \hbox{\it e}\,}}
\def\Im{{\cal I \mskip-5mu \lower.1ex \hbox{\it m}\,}}

\nc{\vev}[1]{ \left\langle {#1} \right\rangle }
\nc{\bra}[1]{ \langle {#1} | }
\nc{\ket}[1]{ | {#1} \rangle }
\nc{\fb}{\,{\rm fb}^{-1}}
\nc{\ev}{{\rm eV}}
\nc{\kev}{{\rm keV}}
\nc{\Mev}{{\rm MeV}}
\nc{\gev}{{\rm GeV}}
\nc{\tev}{{\rm TeV}}
\nc{\mev}{{\rm MeV}}

\def\del{\partial}
\def\Dslash{\not{\hbox{\kern-4pt $D$}}}
\def\dslash{\not{\hbox{\kern-2pt $\del$}}}
\def\pslash{\not{\hbox{\kern-2pt $p$}}}
\def\ETmiss{ \not{\hbox{\kern-4pt $E$}}_T }

\def\msb{{\bar{\ssstyle M \kern -1pt S}}}

\newcommand{\TeV}{\rm{ TeV}}
\newcommand{\GeV}{\rm{ GeV}}

\newcommand{\iab}{\rm{ ab}^{-1}}
\newcommand{\ab}{\rm{ ab}}

\newcommand{\nb}{\rm{ nb}}
\newcommand{\OO}{\mathcal{O}}

\begin{document}

\title{Methods and Results for Standard Model Event Generation at \\ $\sqrt{s}$ = 14 TeV, 33 TeV and 100 TeV Proton Colliders: \\A Snowmass Whitepaper}

\affiliation{Boston University, Boston, MA, USA}
\affiliation{SLAC National Accelerator Laboratory, Menlo Park, USA}
\affiliation{Fermi National Accelerator Laboratory, Batavia, USA}
\affiliation{Panjab University, Chandigarh, India}
\affiliation{Stanford Institute for Theoretical Physics, Stanford University, Stanford, USA}
\affiliation{University of Nebraska, Lincoln, USA }
\affiliation{Brown University, Providence, USA}
\affiliation{University of California, San Diego, USA}
\affiliation{Purdue University Calumet, Hammond, USA}

\author{Aram Avetisyan}\affiliation{Boston University, Boston, MA, USA}
\author{John M. Campbell}\affiliation{Fermi National Accelerator Laboratory, Batavia, USA}
\author{Timothy Cohen}\affiliation{SLAC National Accelerator Laboratory, Menlo Park, USA}
\author{Nitish Dhingra}\affiliation{Panjab University, Chandigarh, India}
\author{James Hirschauer}\affiliation{Fermi National Accelerator Laboratory, Batavia, USA}
\author{Kiel Howe}\affiliation{Stanford Institute for Theoretical Physics, Stanford University, Stanford, USA}
\author{Sudhir Malik}\affiliation{University of Nebraska, Lincoln, USA }
\author{Meenakshi Narain}\affiliation{Brown University, Providence, USA}
\author{Sanjay Padhi}\affiliation{University of California, San Diego, USA}
\author{Michael E. Peskin}\affiliation{SLAC National Accelerator Laboratory, Menlo Park, USA}
\author{John Stupak III}\affiliation{Purdue University Calumet, Hammond, USA}
\author{Jay G. Wacker}\affiliation{SLAC National Accelerator Laboratory, Menlo Park, USA}



\begin{abstract}
\vspace{50 pt}
\begin{center}
{\bf Abstract}
\end{center}
This document describes the novel techniques used to simulate the common Snowmass 2013 Energy Frontier Standard Model backgrounds for future hadron colliders.  The purpose of many Energy Frontier studies is to explore the reach of high luminosity data sets at a variety of high energy colliders.  The generation of high statistics samples which accurately model large integrated luminosities for multiple center-of-mass energies and pile-up environments is not possible using an unweighted event generation strategy --- an approach which relies on event weighting was necessary.  Even with these improvements in efficiency, extensive computing resources were required.  This document describes the specific approach to event generation using {\tt Madgraph}5 to produce parton-level processes, followed by parton showering and hadronization with {\tt Pythia}6, and pile-up and detector simulation with {\tt Delphes}3.  
The majority of Standard Model processes for $pp$ interactions at $\sqrt{s}=$ 14, 33, and 100 TeV with 0, 50, and 140 additional pile-up interactions are publicly available.
\end{abstract}

\maketitle

\newpage


\section{Introduction}
\label{sec:sim-intro}

The Large Hadrnon Collider (LHC) had an extraordinary first run, culminating in the discovery of a Higgs boson \cite{ATLAS-Higgs, CMS-Higgs}.  Yet, so far none of the many searches for physics beyond the Standard Model (SM) have reported any hints of other new particles.  In the absence of additional discoveries, we remain without answers to many of the lines of inquiry which motivated building the LHC in the first place: What (if any) mechanism stabilizes the hierarchy between the weak scale and the Planck scale?  What are the particles that constitute dark matter?  Is the Higgs elementary or composite? Is supersymmetry realized at any scale?  Do the gauge couplings unify?  Attempting to answer these and other questions, both known and unknown, is the driving purpose of the Energy Frontier.  Our flagship experiment will be the upcoming 14 TeV LHC, hopefully followed by a 33 TeV and/or a 100 TeV proton collider.

Even in light of null results from the LHC with 7/8 TeV $pp$ collisions, there are still many exciting possibilities for future discoveries of beyond the SM physics.  Take the example of supersymmetry.  In this context, the naive expectations derived from the requirement of a minimally tuned electroweak sector have not yet born any fruit.  A qualitative argument can be made that the scalar superpartners are heavier than  previously anticipated in order to accommodate a 125 GeV Higgs boson; the new states could lie just out of reach of the 8 TeV LHC.   Furthermore, the fact that it is possible to obscure new signals in models with, e.g.~compressed spectra or $R$-parity violation, demonstrates that a ``leave no stone unturned" attitude is required.  It is clear that we must ensure that we have constrained all possible scenarios for new physics at and above the TeV scale.  There is a compelling case to be made for continuing the mission of the Energy Frontier by pushing collider experiments into the highest possible energy and luminosity regimes.

A central goal of ``The Path Beyond the Standard Model" Snowmass working group is to assess the discovery potential of future accelerators, and detectors.  Quantitative assessments of the reach in mass and cross section for these machines requires generating a Monte Carlo SM background sample which has sufficiently high statistics  that it can accurately model the tails of distributions.  Specifically, proton-proton colliders with center of mass energies at 14, 33, and~100~TeV~are being evaluated with total integrated luminosity of $3\,\ab^{-1}$ at each machine.  To complicate matters further, it is critical to evaluate how the impact of these different colliders will depend on the pile-up environment.  

Specifically, the Snowmass background samples must encompass the following future accelerator scenarios:
\begin{itemize}
\item LHC Phase I: $p\,p$ collisions at $\sqrt{s} =14 \,\TeV$.   The final integrated luminosity will be 300 ${\rm fb}^{-1}$, with an average of about 50 pile-up interactions per bunch crossing.
\item HL-LHC or LHC Phase II: high luminosity $p\,p$ collisions at $\sqrt{s} = 14 \,\TeV$.  The final integrated luminosity will be 3 ${\rm ab}^{-1}$ with an average of 140 pile-up interactions per bunch crossing.
\item HE-LHC: High luminosity $p\,p$ collisions at $\sqrt{s} = 33 \,\TeV$.  The final integrated luminosity will be 3 ${\rm ab}^{-1}$ with an average of 140 pile-up interactions per bunch crossing.
\item VLHC: $p\,p$ collisions at $\sqrt{s} = 100 \,\TeV$.  The planned integrated luminosity is
 1 ${\rm ab}^{-1}$ with  an average of 40 pile-up interactions per bunch crossing.
\end{itemize}

To understand the challenges associated with generating \emph{inclusive} SM Monte Carlo for a project with this scope, consider $t\,\overline{t}$ production at $\sqrt{s}=$ 14 TeV.  The top pair production cross section at NLO is $\sigma_{14}(p\,p\rightarrow t\,\overline{t}) \simeq 1 \,\text{nb}$ \cite{MCFM}.  If we want a factor of 10 more Monte Carlo events than expected events in $3\,\iab$, this requires generating $\OO(10^{10})$ events per pileup setting.  This is clearly not feasible both from a computational and, given that each event is $\OO(1\,\text{kb})$, a data storage/distribution point of view.

To solve this problem requires developing techniques which rely on \emph{weighted} event generation.  These techniques will allow us to produce a $t \overline{t}$ sample that is adequate for a $3\,\ab^{-1}$ study with a factor of $\sim 2000$ fewer events than in the unweighted case.  This is possible using built in features of existing tools (sometimes with small modifications). This document describes the specific procedure used for the Snowmass SM backgrounds in detail.  Our goal is that a user of these samples understands in detail how they were generated.  We also hope that 
these samples will serve as a benchmark for future projects which require Monte Carlo generation on this scale.  

The rest of this note is organized as follows.  Section \ref{sec:generator} describes the explicit background processes which were generated and their organization into several exclusive categories. Section \ref{sec:Weighting} describes the weighting procedure developed for Snowmass which relies on features of both \texttt{Madgraph} and \texttt{BRIDGE}. Section \ref{sec:studies} presents studies of the kinematic features of the backgrounds including the dependence on pile-up conditions. The main result of this work is a set of fully reconstructed weighted event files for the dominant inclusive SM backgrounds at 14, 33, and 100 TeV $p\,p$ collisions in different pile-up conditions and with adequate statistics for high luminosity studies and described in section~\ref{SMsamples}.  

A detailed description of the \texttt{Delphes} simulation of the ``Snowmass'' detector response and object reconstruction,  
can be found in Ref.~\cite{SnowmassPerformanceWP}. In Ref~\cite{Snowmass-OSG}, a detailed description of the software developed for event generation and
processing using the  Open Science Grid infrastructure is provided.

\section{Organization of Standard Model Background Generation}
\label{sec:generator}
The background event generation required for Snowmass studies searching for physics beyond the SM presents a tremendous challenge.  There are a huge number of processes which must be produced at multiple center-of-mass energies, and run through a fast detector simulator at multiple pile-up settings.  To minimize book-keeping and ensure coverage of all final states, multiple background processes with similar cross sections and final states have been grouped together in single event files in a simple way. 

The Snowmass backgrounds have been organized in terms of five particle ``containers":
\begin{eqnarray}
J &=& \{ g, u, \bar{u}, d, \bar{d}, s,\bar{s}, c,\bar{c}, b,\bar{b}\}\\
L &=& \{ e^+, e^-, \mu^+, \mu^-, \tau^+, \tau^-, \nu_e, \nu_\mu, \nu_\tau\,\}\\
V &=& \{W^+, W^-, Z^0, \gamma\}\\
T &=& \{t, \bar{t}\}\\
H &=& \{h^0\}
\end{eqnarray}
which, define the background processes at generator level. 
In order to determine which backgrounds are most important for a given study, it is also useful to organize processes by the strength of the couplings involved.  Specifically, three separate coupling constants are used for this purpose:
\begin{eqnarray}\label{eq:couplings}
\alpha_s, \,\alpha_w,\, \alpha_h
\end{eqnarray}
The first, $\alpha_s$, is the strong coupling constant and its order will typically be omitted.  
The second is the weak coupling constant and represents all couplings of the $W^\pm$ and $Z^0$ bosons along with the top Yukawa coupling to the Higgs boson since it is of similar strength.  The coupling $\alpha_h$ represents both the Higgs couplings through the $g\,g\,h^0$ dimension-5 effective operator and the bottom Yukawa coupling.   
Higgs production is included with the corresponding multi-vector boson processes when the cross sections are comparable.

This organization is presented in Table~\ref{tab:bkgstatus}, where the names of the Snowmass backgrounds datasets are given along with a brief description of the main processes included. The generator-level final states and order in the couplings for each category are also given, which completely specify the included processes. Note that the final states are given at generator-level with on-shell heavy resonances treated as stable particles. In all diagrams internal on-shell $V,\,H,$ and $T$ resonances are excluded, and thus all of the background categories are orthogonal.  For all background categories, QCD radiation of generator-level  jets is allowed up to a total of four final state partons (e.g. $TT + nJ$ means $TT + 0J/1J/2J$), with appropriate matching to the parton shower as described later. Background categories are also denoted as ``dominant" or ``subdominant" depending on the size of their production cross sections. 

\clearpage
\begin{savenotes}
\begin{center}
\begin{table}[htp]
\renewcommand{\arraystretch}{1.4}
\setlength{\tabcolsep}{7pt}
\begin{tabular}{c|c|c|c}
\hline
\hline
Dataset Name & Main Processes & Final States & Order  \\
\hline
\hline
\multicolumn{4}{c}{Dominant Backgrounds}\\
\hline
{B-4p, Bj-4p\footnote{For technical reasons, the $V+0J$ contribution (B-4P) has been generated separately from the $V+$(1-3)$J$ contribution (BJ-4p). These two samples are orthogonal and should both be included to obtain the full $V+$(0-3)$J$ contribution.} }  & vector boson + jets& $V+ n J$ &$\OO(\alpha_s^{n} \alpha_w)$\\
\hline
BB-4p & divector + jets & $VV+n J$ & $\OO(\alpha_s^{n} \alpha_w^2)$\\
 \hline
TT-4p & top pair + jets & $TT+nJ$ & $\OO(\alpha_s^{2+n})$\\
 \hline
TB-4p & top pair off-shell $T^*\to Wj$ + jets & $TV+nJ$ &  $\OO(\alpha_s^{n+1}\alpha_w)$\\
\hline
TJ-4p & single top (s and t-channel) + jets &  $T+nJ$ &  $\OO(\alpha_s^{n-1} \alpha_w^2)$\\
 \hline
LL-4p & off-shell $V^* \to LL$ + jets &  $LL+nJ$  [$m_{ll} > 20$ GeV] & $\OO(\alpha_s^{n} \alpha_w^2)$\\
 \hline
 \multicolumn{4}{c}{Subdominant Backgrounds}\\
 \hline 
TTB-4p & top pair + boson &  $(TTV+nJ), (TTH+nJ)$  & $\OO(\alpha_s^{2+n}\alpha_w)$\\
\hline
BLL-4p & off-shell divector $V^* \to LL$ + jets  & $VLL+nJ$ [$m_{ll} > 20$ GeV] &$\OO(\alpha_s^{n} \alpha_w^3)$\\
\hline
BBB-4p & tri-vector + jets, Higgs associated + jets & $(VVV+nJ), (VH + nj) $&$\OO(\alpha_s^{n}\alpha_w^3)$\\
\hline
H-4p & gluon fusion + jets &  $H + n J$ &$\OO(\alpha_s^{n} \alpha_h)$\\
 \hline
BJJ-vbf-4p & vector boson fusion + jets  &  $(V+nJ),(H+nJ)\;\; [n \geq 2]$ & $\OO(\alpha_s^{n-2} \alpha_w^3)$\\
\hline
\hline
\end{tabular}
\caption{A list of the Standard Model background categories generated for Snowmass. The main processes contributing to each category are described. The generator-level final states and order in couplings for each category is also given, which completely specifies all of the included processes. All of these events are available online; see the Snowmass twiki \cite{SnowmassSimulationsTwiki}.
\label{tab:bkgstatus} 
}
\end{table}
\end{center}
\end{savenotes}

\section{Weighting Procedure}
\label{sec:Weighting}

At the generator level, the SM backgrounds were simulated using \texttt{Madgraph 1.5.10}~\cite{Madgraph}. On-shell heavy resonances $(t,\,\overline{t},\,W^\pm,\,Z^0,\,h^0)$ were treated as stable at the generator level and decayed using \texttt{BRIDGE}~\cite{bridge}. Five-flavor MLM matching was performed using the \texttt{Madgraph Pythia}~\cite{Madgraph} interface in the shower-kt scheme, which uses \texttt{PYTHIA6}~\cite{Pythia6} for showering and hadronization.  The matching parameter QCUT=XQCUT was set as in Table~\ref{tab:xqcut}, and a photon-jet $\Delta R$ cut of DRAJ=0.3 was used. A Higgs mass of $m_h = 125{\rm~GeV}$ was chosen.

\begin{table}[t]
\renewcommand{\arraystretch}{1.4}
\setlength{\tabcolsep}{7pt}
\begin{center}
\begin{tabular}{l|l} 
\hline
\hline
Dataset Names &  QCUT  \\ \hline
B-4p, BJ-4p, BJJ-vbf-4p, BB-4p, BBB-4p, LL-4p, LLB-4p, H-4p  &  40 GeV \\ 
TJ-4p, TB-4p & 60 GeV \\
TT-4p, TTB-4p & 80 GeV\\
\hline
\hline
\end{tabular}
\caption{Values of QCUT=XQCUT for matching procedure for each background category. The same values were used at 14, 33, and 100 TeV.}
\label{tab:xqcut}
\end{center}
\end{table}

Both weighted and unweighted background datasets were generated, with the weighted background sets providing statistics to cover integrated luminosity in the ab$^{-1}$ range. In the weighted events dataset, each background category is split into bins of the variable $S^*_{T}$, which we define as the scalar sum of the $p_T$ of all generator level particles, as shown in Tables~\ref{tab:HTbinXsec14} and \ref{tab:HTbinXsec33}.  Note that these cross sections are the matched cross section for each $S^*_{T}$ bin -- $K$-factors are included as a part of the weight for each event as described below.  The entire generation process through reconstruction is carried out separately in each bin. Each bin is orthogonal and includes individual event weights $w_i$ (stored in the \texttt{Event.Weight} leaf) such that for each bin $\alpha$ containing $N_\alpha$ events
\begin{equation}
\frac{1}{N_\alpha} \sum_{i}^{N_\alpha} w_i =  \sigma_{{\rm NLO},\alpha},
\end{equation}
where $\sigma_{{\rm NLO},\alpha}$ is the NLO cross section for the $S^*_{T}$ bin. The individual weights $w_i$ contain NLO $K$-factors and branching fractions as described in the following.

\subsection{Weighted Event Generation}

The bins of $S^*_{T}$ are chosen for each background category so that roughly one decade of cross section falls in each bin. \texttt{Madgraph5 v1.5.10} was modified to implement a generator level cut on $S^*_{T}$, and we will 
refer to cuts on this variable as \texttt{htmin}, \texttt{htmax}.

In detail, the process for each parameter point is:
\begin{enumerate}
\item \emph{Compute the approximate differential cross section with respect to $S^*_{T}$}. We run \texttt{Madgraph} in ``survey" mode (using the command \texttt{bin/madevent survey}) while incrementing the \texttt{htmin} cut to determine the cross section as a function of this cut, $\sigma_i \equiv \sigma(S^*_{T} > \texttt{htmin}_i)$ with
\begin{equation*}
\texttt{htmin}_{i=0\dots n} = \{0 \mbox{ GeV},\,100\mbox{ GeV},\,200\mbox{ GeV},\dots\}
\end{equation*}
We find that subsequent steps of 100 GeV provide an accurate enough characterization of the cross section for our purposes. We increase the cut until $\sigma_i < 1/\mathcal{L}$ where $\mathcal{L}$ is the luminosity for which good statistics are desired. The differential cross section is calculated from the differences:
\begin{equation*} 
\text{d}\sigma_i = \sigma_{i+1}-\sigma_{i}.
\end{equation*}

\item \emph{Determine bins of $S^*_{T}$ for event generation.} We define ${\rm bin}_{\alpha=1\dots m}$ by $\texttt{htmin}_\alpha\le S^*_{T} < \texttt{htmax}_\alpha$. We choose bin edges based on a ``weight fraction" $x$ with $0<x\leq1$ as follows:

\begin{enumerate}
\item The lower edge of the first bin is $\texttt{htmin}_1 = 0{\rm~GeV}$.
\item The upper edge of the first bin $\texttt{htmax}_1$ is chosen to be the smallest value such that $\sigma_1 \ge x\times \sigma_{\rm tot}$.
\item The remaining upper bin edges $\texttt{htmax}_{\alpha=2\dots m}$ are chosen similarly with each bin as small as possible such that  
\begin{equation}
\sigma_\alpha \equiv  \sigma(\texttt{htmax}_\alpha > S^*_{T} > \texttt{htmin}_\alpha) > x \times\sigma(S^*_{T} > \texttt{htmin}_\alpha),
\end{equation}
with $\texttt{htmin}_{\alpha}=\texttt{htmax}_{\alpha-1}$, where $\sigma({\rm bin}_k)$ is the sum over $\text{d}\sigma_i$ for the range associated with bin$_k$.
\item The final bin is inclusive and determined by $\sigma({\rm bin}_m) \times \mathcal{L} < N/10$, where $N$ is the total number of events to be generated in the final bin.
\end{enumerate}

Note that $x = 0.9$ is used for the Snowmass backgrounds.

\item \emph{Generate weighted events.} We generate $N\simeq5\times10^6$ generator-level events in each of the $m$ bins.  For each bin separately, the events are showered, decayed, and matched in $\texttt{Pythia}$ and reconstructed in $\texttt{Delphes}$. After matching, each bin has $n_k \le N$ events and an associated matched cross section  $\sigma_{{\rm LO-matched}}$.  In Sec.~\ref{sec:weightnorm} below we discuss how this cross section is incorporated into the individual event weights to give the correct relative normalization of the bins.

\item \emph{Output format}. The final output for each background channel is a set of \texttt{Delphes 3} format 
ROOT~\cite{rootA} files corresponding to the bins of $S^*_{T}$. 

\end{enumerate}

The values of the bin edges and the matched cross section in each bin for the background categories can be seen in Tables~\ref{tab:HTbinXsec14},~\ref{tab:HTbinXsec33}, and~\ref{tab:HTbinXsec100}. In Sec. \ref{sec:studies} we discuss the efficacy of this procedure and validate it against an unweighted event sample.

\begin{table}[h!]
\renewcommand{\arraystretch}{1.3}
\begin{center}
{\small
\begin{tabular}{ l  l | c | c | c | c | c | c | c l }
 \hline\hline
       $S^*_T$ bin: &$\alpha=$       & 1     &  2 & 3 & 4  & 5  & 6  & 7  &   \\\hline 
       \multirow{3}{*}{B-4p} & $S^*_T$ (GeV) & {$0-1$}     &  &  &  &  &  &  &   \\ 
                  & $\sigma_{\alpha}$ (pb) & $200944.68$ &  &  &  &  &  &  &   \\ 
                         & $N_{\alpha}\, (\times 10^6)$ & $19, 8.9, 8.7$ &  &  &  &  &  &  &   \\ [0.5ex] \hline           
       \multirow{3}{*}{BB-4p} &$S^*_T$ (TeV) & {$0-0.3$}   & {$0.3-0.7$} & {$0.7-1.3$} & {$1.3-2.1$}  & {$2.1-10^5$}  &  &  &   \\ 
       & $\sigma_{\alpha}$ (pb)  & $249.98$    & $35.23$     & $4.14$       & $0.42$         & $0.047$          &  &  &   \\ 
              & $N_{\alpha}\, (\times 10^6)$  &    $17, 13, 13$    & $14, 10, 10$     & $12, 6.3, 6.4$       & $11, 7.8, 7.8$         & $4.6, 3.3, 3.3$          &  &  &   \\ [0.5ex] \hline
       \multirow{3}{*}{BBB-4p}  & $S^*_T$ (TeV)  & {$0-0.6$}   & {$0.6-1.3$} & {$1.3-10^5$}  &   &   &  &  &   \\
       & $\sigma_{\alpha}$ (pb)  & $2.57$      & $0.15$       & $0.013$          &   &   &  &  &   \\ 
       & $N_{\alpha}\, (\times 10^6)$  & $36, 28, 14$      & $12, 7.2, 7.2$       & $18, 23, 23$          &   &   &  &  &   \\ [0.5ex] \hline       
       \multirow{3}{*}{BJ-4p} & $S^*_T$ (TeV)    & {$0-0.3$}   & {$0.3-0.6$} & {$0.6-1.1$} & {$1.1-1.8$}  & {$1.8-2.7$}  & {$2.7-3.7$} & {$3.7-10^5$} &   \\
       & $\sigma_{\alpha}$ (pb)  & $34409.92$  & $2642.85$   & $294.12$     &  $25.95$       & $2.42$         & $0.23$        & $0.028$ &   \\ 
       & $N_{\alpha}\, (\times 10^6)$  & $63, 12, 12$  & $45, 8.3, 8.1$   & $40, 7.3, 7.2$     &  $40, 7.4, 7.1$    & $36, 7.2, 6.5$  & $36, 6.7, 6.5$ & $37, 6.8, 6.6$ &   \\ [0.5ex] \hline       
       \multirow{3}{*}{BJJ-vbf-4p} & $S^*_T$ (TeV) & {$0-0.7$}   & {$0.7-1.4$} & {$1.4-2.3$} & {$2.3-3.4$} & {$3.4-10^5$}  &  &  &   \\
       & $\sigma_{\alpha}$ (pb)  & $86.46$     & $4.35$       & $0.32$        & $0.030$       & $0.003$ & & &   \\ 
       & $N_{\alpha}\, (\times 10^6)$  & $17, 8.9, 9$      & $13, 6.5, 9$       & $11, 5.5, 5.4$        & $4.2, 2, 2$       & $1.3, 1.1, 1.1$ & & &   \\ [0.5ex] \hline       
       \multirow{3}{*}{H-4p}    & $S^*_T$ (TeV)      & {$0-0.3$}   & {$0.3-0.8$}  & {$0.8-1.5$} & {$1.5-10^5$}  & & & &   \\
       & $\sigma_{\alpha}$ (pb)  & $21.56$     & $1.11$       & $0.092$      & $0.0101$         & & & &  \\ 
       & $N_{\alpha}\, (\times 10^6)$  & $14, 6.8, 6.8$  & $13, 4.2, 4.2$       & $6.5, 3.5, 3.5$      & $    6.4, 3.2, 3.2$         & & & &  \\ [0.5ex] \hline       
       \multirow{3}{*}{LL-4p}   & $S^*_T$ (TeV)        & {$0-0.1$} & {$0.1-0.2$}  & {$0.2-0.5$} & {$0.5-0.9$}  & {$0.9-1.4$}  & {$1.4-10^5$} & &   \\
       & $\sigma_{\alpha}$ (pb)  & $1341.37$ & $156.29$     & $42.40$     & $2.84$       & $0.21$        & $0.029$ & &   \\ 
       & $N_{\alpha}\, (\times 10^6)$  & $21, 10, 10$   & $17, 8.8, 8.5$     & $15, 7.5, 7.4$    & $14, 7, 6.9$  & $14, 7.3, 7.2$  & $14, 6.9, 6.9$ & &   \\ [0.5ex] \hline       
       \multirow{3}{*}{LLB-4p}  & $S^*_T$ (TeV)        & {$0-0.4$} & {$0.4-0.9$}  & {$0.9-10^5$}  &  &  &  &  &   \\
       & $\sigma_{\alpha}$ (pb)  & $2.97$    & $0.23$       & $0.021$         &  &  &  &  &   \\ 
       & $N_{\alpha}\, (\times 10^6)$  & $17, 8.4, 8.6$    & $17, 8.2, 8.3$     & $15, 7.5, 7.7$         &  &  &  &  &   \\ [0.5ex] \hline
       \multirow{3}{*}{TB-4p}    & $S^*_T$ (TeV)       & {$0-0.5$} & {$0.5-0.9$}  & {$0.9-1.5$}  & {$1.5-2.2$}  & {$2.2-10^5$}  &  &  &   \\
       & $\sigma_{\alpha}$ (pb)  & $63.89$   & $7.12$       & $0.98$        & $0.084$        & $0.0095$         &  &  &   \\ 
       & $N_{\alpha}\, (\times 10^6)$  & $15, 6.5, 7.4$   & $13, 25, 6.5$       & $30, 6.1, 6.2$        & $12, 6.2, 6.2$        & $12, 6.2, 6,1$         &  &  &   \\ [0.5ex] \hline       
       \multirow{3}{*}{TJ-4p}   & $S^*_T$ (TeV)        & {$0-0.5$} & {$0.5-1.0$} & {$1.0-1.6$} & {$1.6-2.4$} & {$2.4-10^5$}  &  &  &   \\
       & $\sigma_{\alpha}$ (pb)  & $109.74$  & $5.99$       & $0.38$        & $0.035$       & $0.0031$ &  &  &   \\ 
       & $N_{\alpha}\, (\times 10^6)$  & $9.7, 4.4, 3.2$  & $33, 26, 24$       & $12, 5.9, 5.8$        & $10, 5.1, 5.1$       & $9.4, 4.7, 4.6$ &  &  &   \\ [0.5ex] \hline       
       \multirow{3}{*}{TT-4p}   & $S^*_T$ (TeV)        & {$0-0.6$} & {$0.6-1.1$} & {$1.1-1.7$} & {$1.7-2.5$} & {$2.5-10^5$}  &  &  &   \\
       & $\sigma_{\alpha}$ (pb)  & $530.89$  & $42.55$      & $4.48$        & $0.53$        & $0.054$          & & &   \\ 
       & $N_{\alpha}\, (\times 10^6)$  & $24, 21, 22$  & $31, 25, 26$      & $32, 26, 24$        & $32, 23, 13$        & $30, 25, 25$          & & &   \\ [0.5ex] \hline
        \multirow{3}{*}{TTB-4p}   & $S^*_T$ (TeV)        & {$0-0.9$} & {$0.9-1.6$} & {$1.6-2.5$} & {$2.5-10^5$} &  &  &  &   \\
       & $\sigma_{\alpha}$ (pb) & $2.67$ & $0.25$    & $0.024$     & $0.0021$    &          &  &  &   \\ 
       & $N_{\alpha}\, (\times 10^6)$  & $6.1, 6.9, 6.6$  & $6.2, 6.7, 7$      & $6.2, 6.7, 6.9$        & $6.2, 7.1, 6.8$        &           & & &   \\ [0.5ex] \hline

 \hline
 \hline

\end{tabular}
\caption{Matched cross section $\sigma_{\alpha}$ (in pb) and events $N_{\alpha}$ (in millions) for various SM processes,
 in bins of $S^*_T$ (in TeV), generated at $\sqrt{s}$ = 14 \TeV. The three values of $N_{\alpha}$ listed for each
 $S^*_T$ bin correspond to an average of 0, 50, and 140 additional pileup events respectively.
\label{tab:HTbinXsec14}}
}
\end{center}
\end{table}

\begin{table}[h!]
\renewcommand{\arraystretch}{1.3}
\begin{center}
{\small
\begin{tabular}{ l l | c | c | c | c | c | c | cl }
 \hline\hline
       $S^*_T$ bin: &$\alpha=$       & 1     &  2 & 3 & 4  & 5  & 6  & 7 &   \\\hline 
       \multirow{3}{*}{B-4p}   & $S^*_T$ (GeV)  & {$0-1$} &  &  &  &  &  &  &   \\ 
       & $\sigma_{\alpha}$ (pb)  & $537755.099$ &  &  &  &  &  &  &   \\
       & $N_{\alpha}\, (\times 10^6)$  & $1.3, 1.5, 1.6$ &  &  &  &  &  &  &   \\ [0.5ex] \hline       
       \multirow{3}{*}{BB-4p}   & $S^*_T$ (TeV)    & {$0-0.4$}   & {$0.4-1.0$} & {$1.0-2.0$} & {$2.0-3.4$}  & {$3.4-10^5$}  &  &  &   \\ 
       & $\sigma_{\alpha}$ (pb)  & $776.0040$  & $106.85$     & $10.17$       & $0.86$         & $0.095$          &  &  &   \\ 
       & $N_{\alpha}\, (\times 10^6)$  & $3.1, 2.4, 2.4$  & $2.7, 2.1, 2$     & $4.6, 3.5, 3.5$       & $13, 13, 13$         & $2.8, 2.7, 2.8$          &  &  &   \\ [0.5ex] \hline       
       \multirow{3}{*}{BBB-4p}   & $S^*_T$ (TeV)   & {$0-0.8$} & {$0.8-2.0$} & {$2.0-3.6$} & {$3.6-10^5$}  &   &  &  &   \\
       & $\sigma_{\alpha}$ (pb)  & $8.68$    & $0.45$       & $0.027$       & $0.0025$         &   &   &   &   \\ 
       & $N_{\alpha}\, (\times 10^6)$  & $5.8, 4.3, 4.2$    & $5.2, 3.8, 3.9$       & $11, 11, 11$       & $1.8, 1.8, 1.8$         &   &   &   &   \\ [0.5ex] \hline       
       \multirow{3}{*}{BJ-4p}   & $S^*_T$ (TeV)    & {$0-0.4$}    & {$0.4-1.0$} & {$1.0-1.8$} & {$1.8-3.0$} & {$3.0-4.6$} & {$4.6-6.6$} & {$6.6-10^5$} &   \\
       & $\sigma_{\alpha}$ (pb)  & $129824.79$  & $6598.37$    & $325.56$      & $32.45$       & $3.17$        & $0.33$        & $0.04045$       &   \\ 
       & $N_{\alpha}\, (\times 10^6)$  & $6.1, 4.5, 4.4$  & $4.2, 3.2, 3.1$    & $3.4, 2.6, 2.6$  & $3.1, 2.3, 2.3$   & $2.8, 2.1, 2.1$  & $2.7, 2, 2$   & $2.2, 2, 2$       &   \\ [0.5ex] \hline       
       \multirow{3}{*}{BJJ-vbf-4p}   & $S^*_T$ (TeV) & {$0-0.8$} & {$0.8-1.6$} & {$1.6-3.0$} & {$3.0-4.8$} & {$4.8-10^5$} &  &  &   \\
       & $\sigma_{\alpha}$ (pb)    & $302.56$  & $16.41$      & $1.74$        & $0.14$         & $0.016$         &  &  &   \\ 
       & $N_{\alpha}\, (\times 10^6)$  & $6.2, 4.6, 4.5$  & $4.7, 3.4, 3.5$    & $3, 2.4, 2.4$   & $0.83, 0.8, 0.8$    & $0.25, 0.25, 0.25$ &  &  &   \\ [0.5ex] \hline       
       \multirow{3}{*}{H-4p}   & $S^*_T$ (TeV)     & {$0-0.4$} & {$0.4-1.2$} & {$1.2-2.4$} & {$2.4-4.2$}  & {$4.2-10^5$} & & &   \\
       & $\sigma_{\alpha}$ (pb)  & $96.32$   & $5.83$       & $0.43$        & $0.043$        & $0.0046$        & & &  \\
       & $N_{\alpha}\, (\times 10^6)$  & $4.9, 3.7, 3.6$   & $3.1, 2.3, 2.2$  & $ 2.5, 1.9, 1.9$  & $2.2, 1.7, 1.7$    & $2.1, 1.6, 1.6$ & & &  \\ [0.5ex] \hline       
       \multirow{3}{*}{LL-4p}   & $S^*_T$ (TeV)    & {$0-0.2$}  & {$0.2-0.6$} & {$0.6-1.2$} & {$1.2-1.8$}  & {$1.8-10^5$}  &  & &   \\
       & $\sigma_{\alpha}$ (pb)  & $3060.081$ & $139.42$    & $6.46$       & $0.24$         & $0.069$          &  &  &   \\
       & $N_{\alpha}\, (\times 10^6)$  & $3.1, 2.3, 2.2$  & $5.6, 4.3, 4.1$    & $5.2, 3.9, 3.9$       & $5, 3.8, 3.9$  & $4.6, 3.5, 3.4$ &  &  &   \\ [0.5ex] \hline       
       \multirow{3}{*}{LLB-4p}   & $S^*_T$ (TeV)  & {$0-0.8$} & {$0.8-2.0$}  & {$2.0-10^5$}  &  &  &  &  &   \\
       & $\sigma_{\alpha}$ (pb) & $8.19$    & $0.16$        & $0.0072$         &  &  &  &  &   \\ 
       & $N_{\alpha}\, (\times 10^6)$ & $6.6, 5.1, 4.8$   & $5.6, 4.2, 4.1$        & $5.7, 5.5, 5.5$     &  &  &  &  &   \\ [0.5ex] \hline       
       \multirow{3}{*}{TB-4p}   & $S^*_T$ (TeV)   & {$0-0.6$} & {$0.6-1.2$}  & {$1.2-2.0$}  & {$2.0-3.6$} & {$3.6-10^5$}  &  &  &   \\
       & $\sigma_{\alpha}$ (pb) & $432.36$  & $53.98$       & $5.61$         & $0.63$        & $0.059$          &  &  &   \\ 
       & $N_{\alpha}\, (\times 10^6)$ & $1.1, 1.1, 1.2$   & $1.4, 1.3, 1.2$       & $0.97, 0.92, 0.99$  & $4, 4.1, 3.6$ & $3.8, 3.6, 3.7$ &  &  &   \\ [0.5ex] \hline       
       \multirow{3}{*}{TJ-4p}   & $S^*_T$ (TeV)  & {$0-0.6$}    & {$0.6-1.2$} & {$1.2-2.2$} & {$2.2-3.6$} & {$3.6-10^5$}  &  &  &   \\
       & $\sigma_{\alpha}$ (pb)  & $493.505$  & $23.52$      & $1.97$        & $0.13$        & $0.011$          &  &  &   \\ 
       & $N_{\alpha}\, (\times 10^6)$  & $3.2, 2.9, 3.5$  & $4.8, 3.6, 3.6$   & $3.9, 2.9, 2.8$   & $3.2, 2.5, 2.3$   & $2.6, 2, 2$       &  &  &   \\ [0.5ex] \hline
       \multirow{3}{*}{TT-4p}   & $S^*_T$ (TeV)    & {$0-0.6$} & {$0.6-1.2$}  & {$1.2-2.0$} & {$2.0-3.2$}  & {$3.2-4.8$}  & {$4.8-10^5$}  &  &   \\
       & $\sigma_{\alpha}$ (pb)  & $3438.71$ & $505.82$      & $61.82$       & $7.66$         & $0.73$         & $0.071$          &  &   \\ 
       & $N_{\alpha}\, (\times 10^6)$  & $5.1, 4.3, 4.9$  & $4.3, 4.1, 4.1$  & $16, 17, 17$   & $17, 18, 18$    & $8.7, 17, 18$  & $16, 18, 18$          &  &   \\ [0.5ex] \hline       
        \multirow{3}{*}{TTB-4p}   & $S^*_T$ (TeV) & {$0-1.2$} & {$1.2-2.2$} & {$2.2-3.6$} & {$3.6-10^5$} &  &  &  &   \\
       & $\sigma_{\alpha}$ (pb)                 & $23.91$     & $1.58$ & $0.16$           & $0.017$              &  &  &  &   \\ 
       & $N_{\alpha}\, (\times 10^6)$             & $57, 54, 55$       & $58, 60, 57$  & $1.8, 1.9, 1.6$  & $3.8, 3.6, 3.8$               &  &  &  &   \\ [0.5ex] \hline

 \hline
 \hline
\end{tabular}
\caption{Matched cross section $\sigma_{\alpha}$ (in pb) and events $N_{\alpha}$ (in millions) for various SM processes,
 in bins of $S^*_T$ (in TeV), generated at $\sqrt{s}$ = 33 \TeV. The three values of $N_{\alpha}$ listed for each
 $S^*_T$ bin correspond to an average of 0, 50, and 140 additional pileup events respectively.
\label{tab:HTbinXsec33}}
}
\end{center}
\end{table}

\begin{table}[h!]
\renewcommand{\arraystretch}{1.3}
\begin{center}
{\small
\begin{tabular}{ l  l | c | c | c | c | c | c   l }
 \hline\hline
       $S^*_T$ bin: $\alpha=$&       & 1     &  2 & 3 & 4  & 5  & 6   &   \\\hline 
       \multirow{3}{*}{B-4p}   & $S^*_T$ (TeV)  & {$0-1$} &  &  &  &  &    &   \\
       & $\sigma_{\alpha}$ (pb)  & $1501690.0$ &  &  &  &  &    & \\
       &$N_{\alpha}\, (\times 10^6)$ & $0.73, 0.7$ &  &  &  &  &    &   \\ 	[0.5ex] \hline
       \multirow{3}{*}{BB-4p} & $S^*_T$ (TeV)  & {$0-0.5$} & {$0.5-1.5$} & {$1.5-3.0$} & {$3.0-5.5$} & {$5.5-9.0$} & {$9.0-10^5$} &     \\ 
       & $\sigma_{\alpha}$ (pb)  & $2867.87$ & $405.20$ & $22.84$ & $2.22$ & $0.20$ & $0.024$ &  \\	
       &$N_{\alpha}\, (\times 10^6)$ & $6.7, 6.9$  & $5.9, 5.3$  & $4.2, 4.0$ & $1.1, 1.2$ & $0.21, 0.2$ &  &     \\ [0.5ex] \hline
       \multirow{3}{*}{BBB-4p}   & $S^*_T$ (TeV)   & {$0-1.0$} & {$1.0-3.0$} & {$3.0-6.0$} & {$6.0-10^5$}  &   &  &     \\
       & $\sigma_{\alpha}$ (pb)  & $34.45$ & $1.86$ & $0.09$ & $0.0073$ & & &  \\ 
       &$N_{\alpha}\, (\times 10^6)$ & $6.0, 6.1$  & $0.69, 0.67$  & $0.085, 0.081$  & $0.056, 0.057$ &   &   &      \\ [0.5ex] \hline
       \multirow{3}{*}{Bj-4p}   & $S^*_T$ (TeV)    & {$0-0.5$}    & {$0.5-1.5$} & {$1.5-3.0$} & {$3.0-5.5$} & {$5.5-9.0$} & {$9.0-10^5$} &     \\
       & $\sigma_{\alpha}$ (pb)  & $485362.0$ & $20395.9$ & $635.88$ & $47.62$ &$3.48$ & $0.36$ &  \\
       &$N_{\alpha}\, (\times 10^6)$ & $25, 24$  & $16, 15$ & $11, 12$  & $9.4, 9.7$ & $7.6, 7.5$  & $2.1, 2.0$  &    \\ [0.5ex] \hline
       \multirow{3}{*}{Bjj-vbf-4p}   & $S^*_T$ (TeV) & {$0-1.0$} & {$1.0-2.5$} & {$2.5-5.0$} & {$5.0-8.5$} & {$8.5-10^5$} &  &     \\
       & $\sigma_{\alpha}$ (pb)  & $1088.85$ & $52.70$ & $2.84$ & $0.204$ &$0.023$ & &  \\
       &$N_{\alpha}\, (\times 10^6)$ & $6.2, 6.8$  & $2.9, 2.8$  & $0.084, 0.09$   & $0.022, 0.022$    & $0.0086, 0.0082$ &  &     \\ [0.5ex] \hline
       \multirow{3}{*}{H-4p}   & $S^*_T$ (TeV)     & {$0-0.5$} & {$0.5-1.5$} & {$1.5-3.5$} & {$3.5-7.0$}  & {$7.0-10^5$} & &    \\
       & $\sigma_{\alpha}$ (pb)  & $476.14$ & $33.81$ & $3.80$ & $0.38$ &$0.040$ & &  \\
       &$N_{\alpha}\, (\times 10^6)$ & $5.0, 5.2$   & $1.8, 1.7$  &  $1.6, 1.6$  & $1.8, 1.9$    & $1.6, 1.6$ & &   \\ [0.5ex] \hline
       \multirow{3}{*}{LL-4p}   & $S^*_T$ (TeV)    & {$0-0.4$}  & {$0.4-1.2$} & {$1.2-2.0$} & {$2.0-3.2$}  & {$3.2-10^5$}  &  &    \\
       & $\sigma_{\alpha}$ (pb)  & $7914.07$ & $118.01$ & $1.53$ & $0.22$ & $0.05$ & &  \\
       &$N_{\alpha}\, (\times 10^6)$ & $5.3, 5.6$  & $3.6, 3.7$    & $3.7, 3.9$    & $2.0, 2.0$  & $0.66, 0.74$ &  &     \\ [0.5ex] \hline
       \multirow{3}{*}{LLB-4p}   & $S^*_T$ (TeV)  & {$0.8-2.0$} & {$2.0-4.0$} & {$4.0-10^5$}  &  &  &  &     \\
       & $\sigma_{\alpha}$ (pb)  & $0.81$ & $0.053$ & $0.0053$ & & & &  \\
       & $N_{\alpha}\, (\times 10^6)$ & $4.2, 4.1$      & $0.71, 0.67$   & $0.11,0.099$ &  &  &  &   \\ [0.5ex] \hline
       \multirow{3}{*}{tB-4p}   & $S^*_T$ (TeV)   & {$0-1.0$} & {$1.0-2.0$}  & {$2.0-3.5$}  & {$3.5-6.0$} & {$6.0-9.0$}  & {$9.0-10^5$} &     \\    
       & $\sigma_{\alpha}$ (pb)  & $3399.65$ & $165.25$ & $15.57$ & $1.59$ & $0.11$ & $0.013$ &  \\
       & $N_{\alpha} \, (\times 10^6)$ & $5.6, 5.6$   & $5.2, 5.5$  & $4.7, 4.9$  & $4.1, 4.2$ & $3.9, 3.5$ & $1.0,1.1$  &     \\ [0.5ex] \hline
       \multirow{3}{*}{tj-4p}   & $S^*_T$ (TeV)  & {$0-1.0$}    & {$1.0-2.0$} & {$2.0-4.0$} & {$4.0-7.0$} & {$7.0-10^5$}  &  &     \\
       & $\sigma_{\alpha}$ (pb)  & $2323.9$ & $34.71$ & $2.89$ & $0.163$ & $0.013$ & &  \\
       &$N_{\alpha}\, (\times 10^6)$ & $6.9, 7.3$  & $4.1, 4.3$   & $3.0, 3.1$   & $1.5, 1.5$   & $0.18, 0.17$  &  &     \\ [0.5ex] \hline
       \multirow{3}{*}{tt-4p}   & $S^*_T$ (TeV)    & {$0-1.0$} & {$1.0-2.0$}  & {$2.0-3.5$} & {$3.5-5.5$}  & {$5.5-8.5$}  & {$8.5-10^5$}  &     \\
       & $\sigma_{\alpha}$ (pb)  & $29141.30$ & $1777.28$ & $185.22$ & $18.92$ & $2.39$ & $0.277 $ &  \\
       &$N_{\alpha}\, (\times 10^6)$ & $7.6, 7.7$  & $7.2, 7.2$  & $7.3, 7.1$   & $7.1, 7.0$    & $13, 13$  & $20, 20$ &     \\ [0.5ex] \hline
        \multirow{3}{*}{ttB-4p}   & $S^*_T$ (TeV) & {$0-1.5$} & {$1.5-3.0$} & {$3.0-5.5$} & {$5.5-9.0$} & {$9.0-10^5$} &  &     \\
       & $\sigma_{\alpha}$ (pb)  & $206.01$ & $12.58$ & $1.18$ & $0.092$ & $0.009$ & &  \\
       &$N_{\alpha}\, (\times 10^6)$      & $7, 7$   & $7.4, 7.9$  & $6.9, 7.9$  & $7.7, 7.9$ & $2.1, 2.3$ &  &     \\ [0.5ex] \hline
 \hline
 \hline
\end{tabular}
\caption{Matched cross section $\sigma_{\alpha}$ (in pb) and events $N_{\alpha}$ (in millions) for various SM processes,
 in bins of $S^*_T$ (in TeV), generated at $\sqrt{s}$ = 100 \TeV. The two values of $N_{\alpha}$ listed for each
 $S^*_T$ bin correspond to an average of 40, and 140 additional pileup events respectively.
\label{tab:HTbinXsec100}}
}
\end{center}
\end{table}

\subsection{$K$-Factors}
Given computational constraints and the scope of the event generation required for the Snowmass process it was only feasible to utilize leading order cross sections (with parton shower matrix element matching) for event generation.  As is well known, performing calculations at next-to-leading order can give sizable corrections to the overall rate, while shapes of various kinematic distributions tend to be more stable when going from leading order to next-to-leading order.  Therefore, the standard procedure of including a constant $K$-factors which rescales the overall cross section for a given process has been implemented for the  processes outlined in Table \ref{tab:bkgstatus} above.

Specifically, the procedure used for the Snowmass background is as follows.  For a given process, an inclusive cross section is computed using both {\tt Madgraph} \cite{Madgraph} and {\tt MCFM} \cite{MCFM} , where the generator level cuts were chosen to be as similar as possible between the two programs.  Note that we used the default PDF choices for the two programs: {\tt Madgraph} uses the CTEQ6l1 PDF while {\tt MCFM} uses the CTEQ6m PDF.  The default scale choices were taken in both programs.

For example, the inclusive $t\,\overline{t}$ production cross section at $\sqrt{s}=14\,\TeV$ is
\begin{eqnarray}
\sigma_{14}(p\,p\rightarrow t \,\overline{t}+X)_{\tt{MG}} = 0.60\,\text{nb} \qquad \sigma_{14}(p\,p \rightarrow t \,\overline{t}+X)_{\tt{MCFM}} = 0.75\,\nb
\end{eqnarray}
This yields a $K$-factor
\begin{eqnarray}
K_{14}(t\bar{t}) = \frac{\sigma_{14}(p\,p\rightarrow t\bar{t} +X)_{\tt{MCFM}}}{ \sigma_{14}(pp\rightarrow t \bar{t}+X)_{{\tt MG}}} =1.24.
\end{eqnarray}
For each process, the $K$ factor determined in this way is applied uniformly to the event generation in all of the $\alpha=1\dots n$ bins of $S^*_T$ to obtain the estimated NLO cross section from the LO-matched cross section:
\begin{equation}
\sigma_{{\rm NLO},\alpha} \equiv K \times \sigma_{{\rm LO-matched},\alpha}
\end{equation}

To define the $K$-factors in Table \ref{tab:sigmaMCFM}, phase space cuts had to be specified at the generator level in order to allow a meaningful comparison between the NLO calculations of {\tt MCFM} and the matched cross sections from {\tt MadGraph}.  Some of these cuts were taken to be $\sqrt{s}$ dependent, which leads to some non-trivial scaling behavior for the $K$-factors as a function of center-of-mass energy.  We also note that for $\sqrt{s} = 33 \text{ TeV}$ and 100 TeV some $K$-factors are less than one.  This is related to the choice of QCUT specified in Table \ref{tab:xqcut}.  The matched {\tt MadGraph} cross section is a function of this parameter which implies that the $K$-factors will also have non-trivial dependence on this variable.  For consistency between colliders, we allow $K$-factors to be less than one to account for the particular choices made in this study as our goal is to attempt to faithfully reproduce the NLO cross sections from {\tt MCFM}.  If one wishes to implement cross sections which are computed to higher order, it is possible to use the generator level information contained in the {\tt Delphes} output along with the information provided in Table \ref{tab:sigmaMCFM} to apply $K$-factors by hand.

\begin{center}
\begin{table}[h!]
\renewcommand{\arraystretch}{1.4}
\setlength{\tabcolsep}{7pt}
\begin{tabular}{c|c|c|c}
\hline
\hline
Process & $\sqrt{s} = 14 \text{ TeV}$ & $\sqrt{s} = 33 \text{ TeV}$ & $\sqrt{s} = 100 \text{ TeV}$ \\
 \hline
$t\,\overline{t}$ & 1.24 & 1.10 & 0.96 \\
$W^+\,j$              & 1.17 & 0.85 & 0.74 \\
$W^-\,j$               & 1.20 & 0.89  & 0.75 \\
$Z^0\,j$               & 1.17 & 0.87 & 0.76 \\
$\gamma\,j$      & 1.54 & 1.04 & 0.89 \\
$W^+\,W^-$      & 1.25 & 1.08 & 1.0 \\
$W^+\,Z^0$       & 1.24 & 1.06 & 0.95 \\
$W^-\,Z^0$        & 1.26 & 1.09 & 0.97 \\
$Z^0\,Z^0$         & 1.37 & 1.29 & 1.21 \\
$W^+\,\gamma$          & 1.22 & 0.80 & 0.67 \\
$W^-\,\gamma$           & 1.33 & 0.83 & 0.67 \\
$Z^0\,\gamma$           & 1.24 & 0.95 & 0.76 \\
$\gamma\,\gamma$ & 1.34 & 1.08 & 0.98 \\
$t\,W^-$                        & 1.0 & 0.77 & 0.78 \\
$\overline{t}\,W^+$    & 1.0 & 0.77 & 0.78 \\
$t\,\overline{b}$           & 1.76 & 1.72 & 1.94 \\
$\overline{t}\,b$           & 1.88 & 1.73 & 1.78 \\
$\ell^+\,\ell^-$             & 1.20 & 1.16 & 1.20 \\
\hline
\hline
\end{tabular}
\caption{The $K$-factors for all dominant processes computed using {\tt MCFM} and {\tt MadGraph}.  For sub-dominant processes, the $K$-factors are taken to be unity.}
\label{tab:sigmaMCFM}
\end{table}
\end{center}

\subsection{Decay branching ratios}

We use a  simple procedure to enhance statistics in rare decay modes. For simplicity consider an event with a single unstable heavy particle $X$.  Instead of decaying this heavy particle according to its branching ratios, for a given event we select with equal probability from $N^X_{\rm modes}$ possible decay modes.  This introduces an equal ``effective branching ratio" for each final state of $1/N^X_{\rm modes}$.  In order to re-weight the event by the actual branching ratio (and to cancel the ``effective branching ratio") we need to multiply the original event weight by
\begin{equation}
N^X_{\rm modes} \times {{\rm BR}^X_k}
\end{equation} 
where ${\rm BR}^X_k$ is the branching ratio for $X$ to decay into the selected mode $k$.   

Since some events involve many unstable particles $X_j$, we need to perform this procedure recursively until all particles are decayed.  Then the overall weight of an event $i$ needs to be multiplied by
\begin{equation}
f_i = \prod_{j} N^{X_j}_{\rm modes} \times {{\rm BR}^{X_j}_{d_j}}
\end{equation}
where $X_j$ are the heavy particles present in the event, and $d_j$ is the selected decay mode for the $j$th particle.

 The actual decays are performed at the LHE level using a modified version of BRIDGE~\cite{bridge} to repeatedly decay all heavy particles until no more are present in the event. The heavy particles and decay modes are listed in Table \ref{Tab: BRs}.

\begin{table}
\renewcommand{\arraystretch}{1.3}
\setlength{\tabcolsep}{7pt}
\begin{center}
\begin{tabular}{c||c|c|c}
\hline
\hline
Particle & $N_{\rm modes}$ & Decay Modes & BR \\
\hline\hline
\scalebox{1.5}{$t$} & 1 & $b\,W^+$ & 1.0 \\
\hline
& & $d\,\overline{u}$ & 0.33\\
 & & $s\,\overline{c}$ & 0.33 \\
\scalebox{1.5}{$W^\pm$} & 5 & $e^-\,\overline{\nu}_e$ & 0.11 \\
 & & $\mu^-\,\overline{\nu}_\mu$ & 0.11 \\ 
  & & $\tau^-\,\overline{\nu}_\tau$ & 0.11 \\
  \hline
 & & $u\,\overline{u}$ & 0.12 \\
 & & $d\,\overline{d}$ & 0.15 \\
 & & $c\,\overline{c}$ & 0.12 \\
 & & $s\,\overline{s}$ & 0.15 \\
 & & $b\,\overline{b}$ & 0.15 \\
\scalebox{1.5}{$Z^0$} & 11  & $e^+\,e^-$ & 0.034 \\
 & & $\mu^+\,\mu^-$ & 0.034 \\
 & & $\tau^+\,\tau^-$ & 0.034 \\
 & & $\nu_e \,\overline{\nu}_e$ & 0.068 \\
 & & $\nu_\mu \,\overline{\nu}_\mu$ & 0.069 \\
 & & $\nu_\tau \,\overline{\nu}_\tau$ & 0.068 \\
\hline
 &  & $\tau^+\,\tau^-$ & 0.066 \\
& & $ b\,\overline{b} $ & 0.60 \\
& & $ g\,g $ & 0.088 \\
& & $ \gamma\,\gamma $ & $2.4\times 10^{-3}$\\
& & $ \gamma\,Z $ & $1.6\times 10^{-3}$ \\
& & $ W^-\,u\,\overline{d} $ & 0.036\\
& & $ W^-\,c\,\overline{s} $ & 0.037\\
& & $ W^-\,e^+\,\overline{\nu}_e $ & 0.012\\
& & $ W^-\,\mu^+\,\overline{\nu}_\mu $ & 0.012\\
& & $ W^-\,\tau^+\,\overline{\nu}_\tau $ & 0.012\\
& & $ W^+\,\overline{u}\,d $ & 0.036\\
& & $ W^+\,\overline{c}\,s $ & 0.037\\
\scalebox{1.5}{$h^0$} & 26 & $ W^+\,e^-\,\overline{\nu}_e $ & 0.012\\
& & $ W^+\,\mu^-\,\overline{\nu}_\mu $ & 0.012\\
& & $ W^+\,\tau^-\,\overline{\nu}_\tau $ & 0.012\\
& & $ Z\,u\,\overline{u} $ & $3.3\times 10^{-3}$\\
& & $ Z\,d\,\overline{d} $ & $4.3\times 10^{-3}$\\
& & $ Z\,c\,\overline{c} $ & $3.3\times 10^{-3}$\\
& & $ Z\,s\,\overline{s} $ & $4.3\times 10^{-3}$\\
& & $ Z\,b\,\overline{b} $ & $3.4\times 10^{-3}$\\
& & $ Z\,e^+\,e^- $ & $9.6\times 10^{-4}$ \\
& & $ Z\,\mu^+\, \mu^- $ & $9.7\times 10^{-4}$ \\
& & $ Z\,\tau^+\, \tau^- $ & $9.3\times 10^{-4}$ \\
& & $ Z\,\nu_e\, \overline{\nu}_e$ & $1.9\times 10^{-3}$ \\
& & $ Z\,\nu_\mu\, \overline{\nu}_\mu$ & $1.9\times 10^{-3}$ \\
& & $ Z\,\nu_\tau \, \overline{\nu}_\tau$ & $1.9\times 10^{-3}$ \\
\hline
\hline
\end{tabular}
\caption{Table of branching ratios as computed by BRIDGE.}
\label{Tab: BRs}
\end{center}
\end{table}

To get a sense of the impact this procedure has on the number of events which must be generated, we present the improvements for $t\,\overline{t}$ and $W^\pm\,Z^0$ in Table~\ref{tab:BRexamples}.  We see that flattening the branching ratios saves a factor of 2.25 for dileptonic $t\,\overline{t}$ decays which implies we can generate roughly 2.5 million events per $S^*_T$ bin instead of 5 million.   For trileptons coming from $W^\pm\,Z^0$ production, the impact is even more dramatic, saving a factor of almost 5.

\begin{table}
\renewcommand{\arraystretch}{1.4}
\setlength{\tabcolsep}{7pt}
\begin{center}
\begin{tabular}{c|c|c|c}
\hline
\hline
$t\,\overline{t} \rightarrow$ & weight & in sample & change \\
\hline
hadronic & 44\% & 25\% & 0.56 \\
semi-leptonic & 44\% & 50\% & 1.13 \\
di-leptonic & 11\% & 25\% & 2.25 \\
 \hline
 \hline
 $W^\pm\,Z^0 \rightarrow$ & weight & in sample & change \\
\hline
$1\,\ell$ & 30\% & 44\% & 1.4 \\
$2\,\ell$ & 6.7\% & 11\% & 1.6 \\
$3\,\ell$ & 3.3\%& 16\% & 4.8 \\
\hline
\hline
\end{tabular}
\caption{Examples which demonstrate the impact of flattening the branching ratios.}
\label{tab:BRexamples}
\end{center}
\end{table}

\subsection{Overall weight normalization}
\label{sec:weightnorm}

Before running \texttt{Pythia}, each individual event has a weight stored in the final output combining the $K$-factor for the production process $K_i$ and the branching ratio for the decay mode of the event $f_i$ :
\begin{equation}
w^*_i = K_{i} \times f_{i}
\end{equation}
where $K_{i}$ and $f_{i}$ are defined above.

After \texttt{Pythia} is used to shower the events (including the matching procedure) in a given $S^*_T$ bin $\alpha$, it provides an adjusted cross section $\sigma_{{\rm LO-matched},\alpha}$.  The final weight provided in each event $i$ is given by
\begin{equation}
w_i = \sigma_{{\rm LO-matched},\alpha} w^*_i,
\end{equation}
where $\sigma_{{\rm LO-matched},\alpha}$ is the madgraph MLM matched cross section for the $S^*_T$ bin $\alpha$ as given in Tables \ref{tab:HTbinXsec14} and \ref{tab:HTbinXsec33}.   

In order to use these events in an analysis, one must keep track of the number of events $N_\alpha$ used for a given $S^*_T$ bin.  
Then each event $i$ contributes to a histogram with a weight
\begin{equation}
\frac{w_i}{N_\alpha}.
\end{equation}

With this normalization, for a given final state
\begin{equation}
\sum_{i}^{N_\alpha} \frac{w_i}{N_\alpha} =  \sigma_{{\rm NLO},\alpha} \times {\rm BR}.
\end{equation}

\section{SM background Samples}
\label{SMsamples}

The weighted output events from \texttt{Madgraph-Pythia} are processed further by  v3.0.9 of the 
\texttt{Delphes} \cite{Delphes} framework for
fast simulation using the ``Snowmass'' detector and object reconstruction \cite{SnowmassPerformanceWP}. 
During event processing by \texttt{Delphes}, datasets corresponding to 
 proton-proton collisions at $\sqrt{s}$ =  14, 33 and 100 TeV, are produced with three different additional pile-up 
conditions: no pile-up interactions, an average of 50, and an average of 140 additional pile-up  interactions per bunch crossing.
The \texttt{Delphes} \texttt{ROOT}-format files are available along with more detailed information on the 
Snowmass twiki \cite{SnowmassSimulationsTwiki} 
and at the locations given in Table~\ref{tab:storage}.
The generated sample sizes for each $S^*_T$ bin number $\alpha$, $N_{\alpha}$, 
are listed in Tables~\ref{tab:HTbinXsec14} and~\ref{tab:HTbinXsec33}.

\begin{table}[h]
\renewcommand{\arraystretch}{1.4}
\setlength{\tabcolsep}{7pt}
\begin{center}
\begin{tabular}{l|l} \hline\hline
\multicolumn{2}{l}{Events at $\sqrt{s}$=14 and 33  \TeV} \\ \hline
PU $\left<\mu\right>$=0 & \url{ http://red-gridftp11.unl.edu/Snowmass/HTBinned/Delphes-3.0.9.1/NoPileUp} \\
PU $\left<\mu\right>$=50 & \url{http://red-gridftp11.unl.edu/Snowmass/HTBinned/Delphes-3.0.9.1/50PileUp} \\
PU $\left<\mu\right>$=140 & \url{http://red-gridftp11.unl.edu/Snowmass/HTBinned/Delphes-3.0.9.1/140PileUp} \\  \hline
PU $\left<\mu\right>$=0 &  \url{http://red-gridftp11.unl.edu/Snowmass/HTBinned/Delphes-3.0.9.2/NoPileUp} \\
PU $\left<\mu\right>$=50 &  \url{http://red-gridftp11.unl.edu/Snowmass/HTBinned/Delphes-3.0.9.2/50PileUp}\\
PU $\left<\mu\right>$=140 &  \url{http://red-gridftp11.unl.edu/Snowmass/HTBinned/Delphes-3.0.9.2/140PileUp}\\ \hline
\hline
\end{tabular}
\caption{Storage location of generated event samples (in \texttt{Delphes} output \texttt{ROOT} format).}
\label{tab:storage}
\end{center}
\end{table}

\section{Distributions of Kinematic Variables for $t\overline tV/H + nJ$ events }
\label{sec:studies}

\begin{figure}[h!]
\centering
\includegraphics[width=0.3\textwidth]{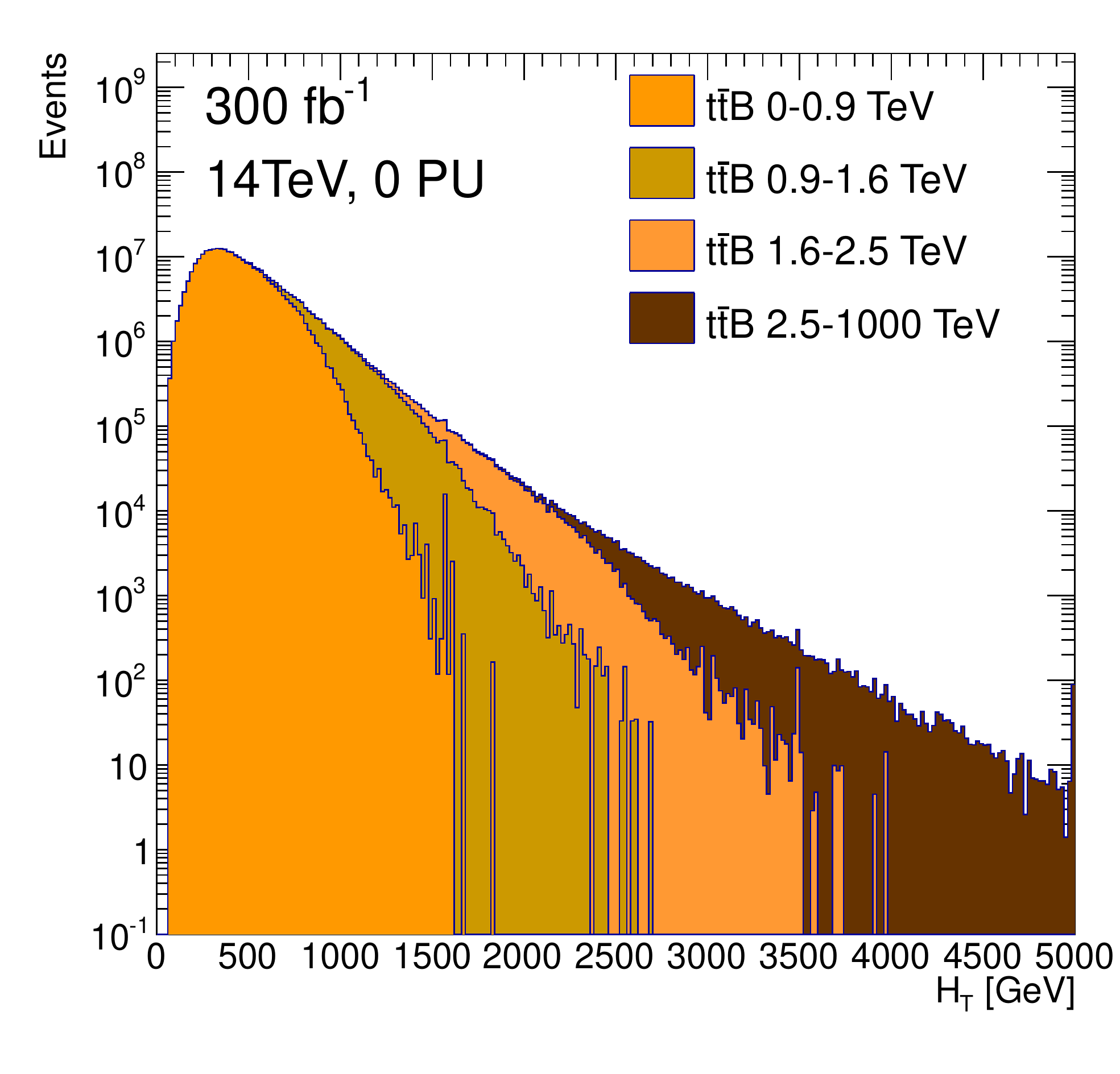}
\includegraphics[width=0.3\textwidth]{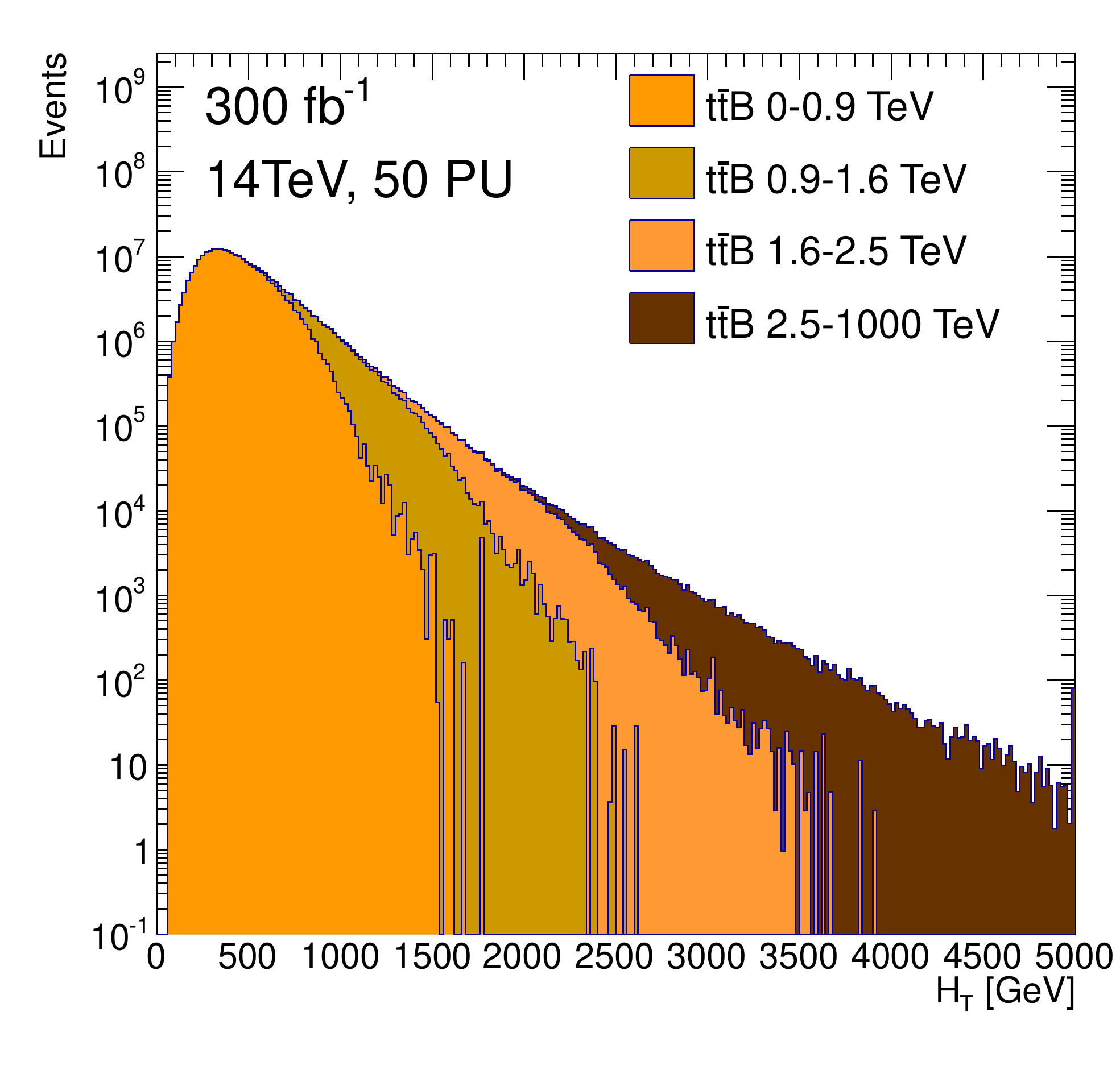}
\includegraphics[width=0.3\textwidth]{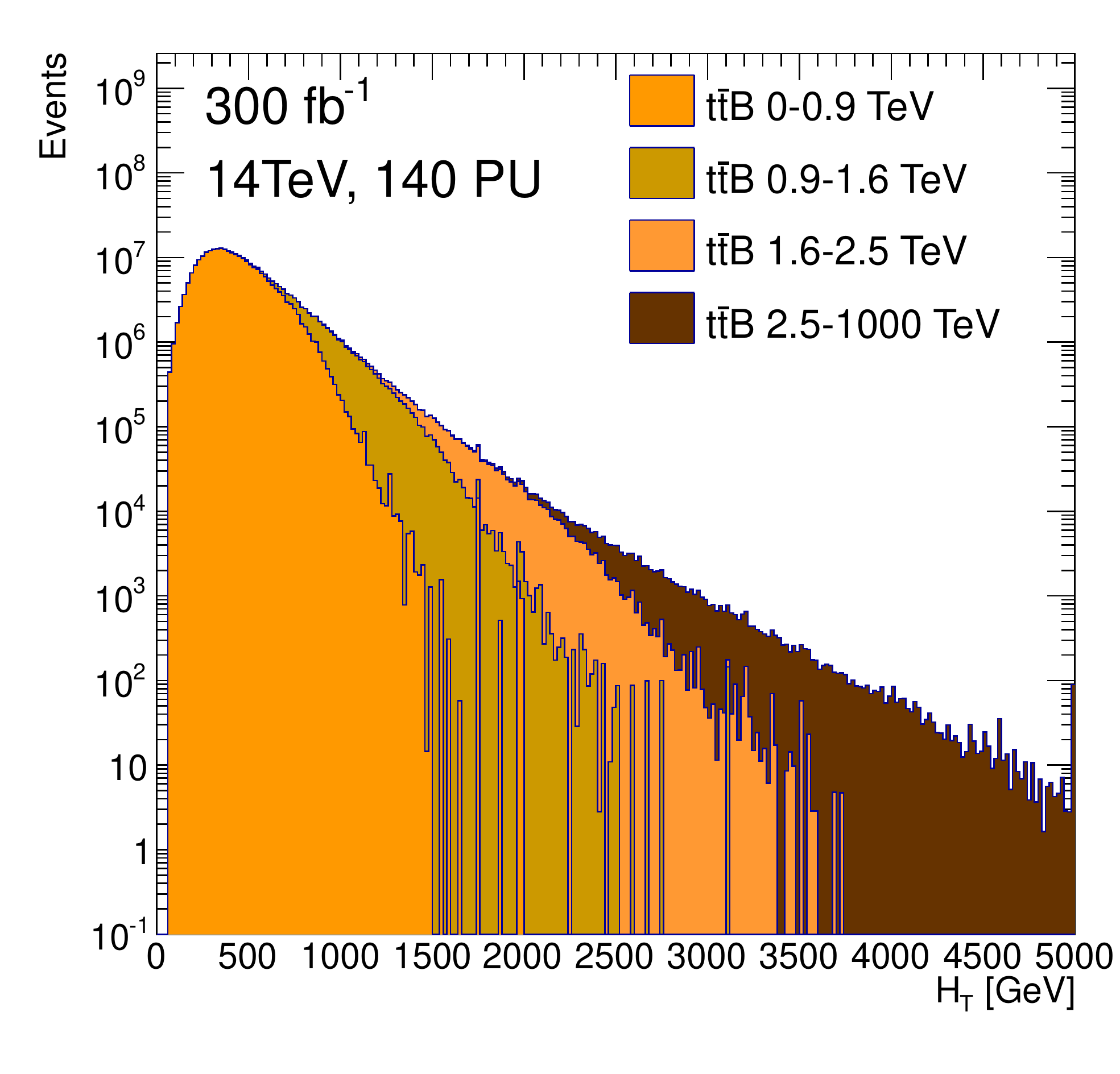}\\
\includegraphics[width=0.3\textwidth]{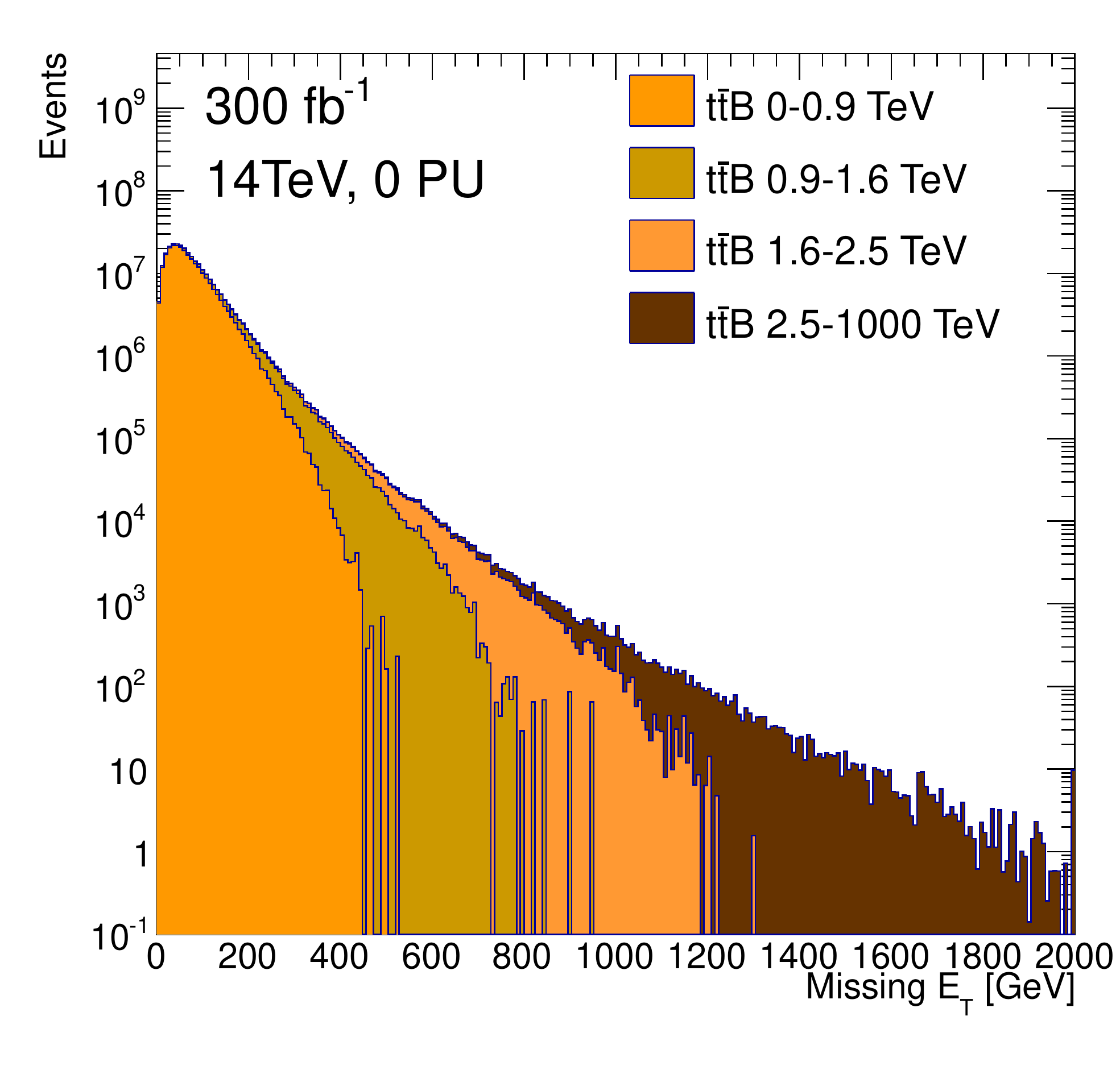}
\includegraphics[width=0.3\textwidth]{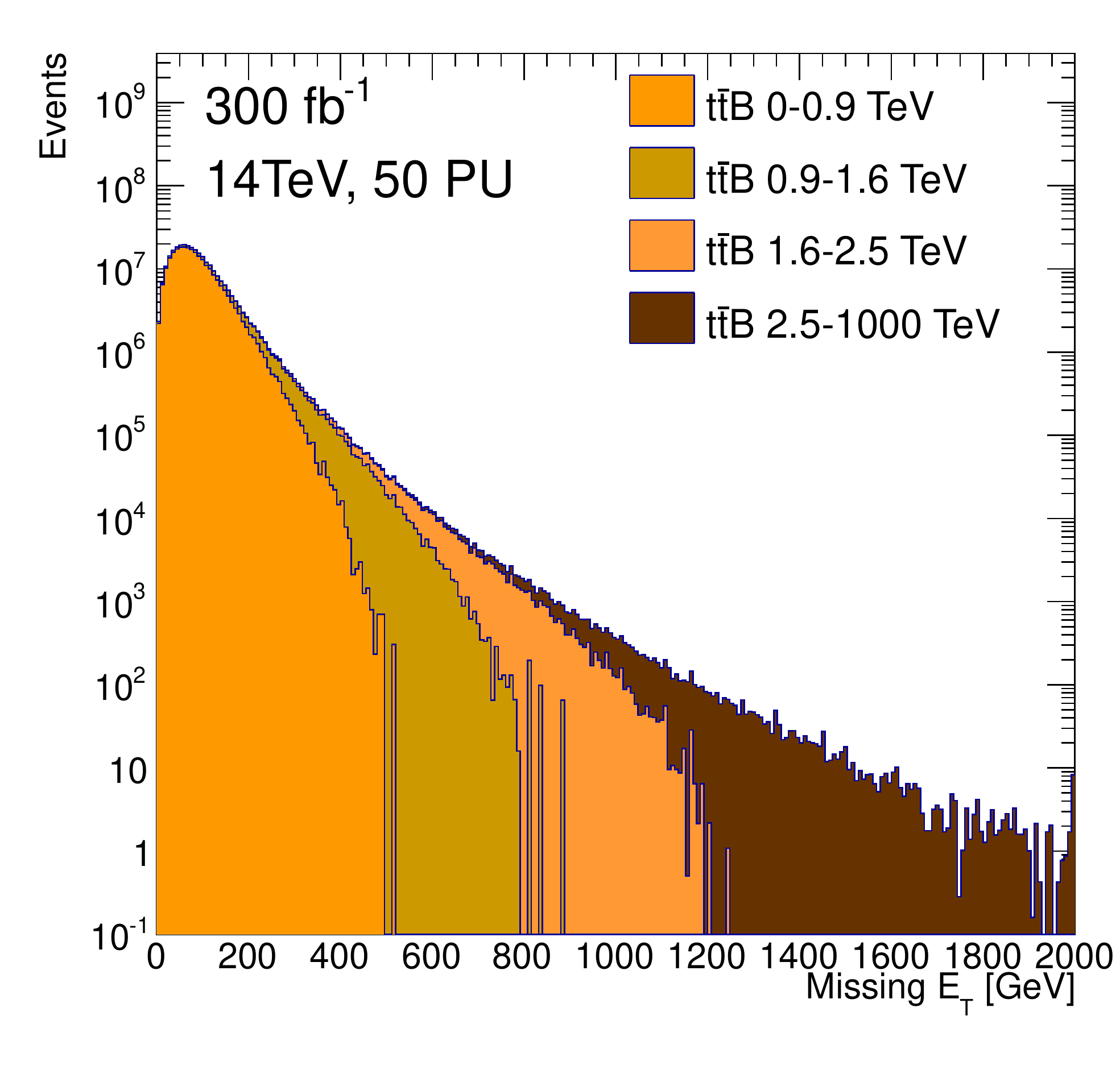}
\includegraphics[width=0.3\textwidth]{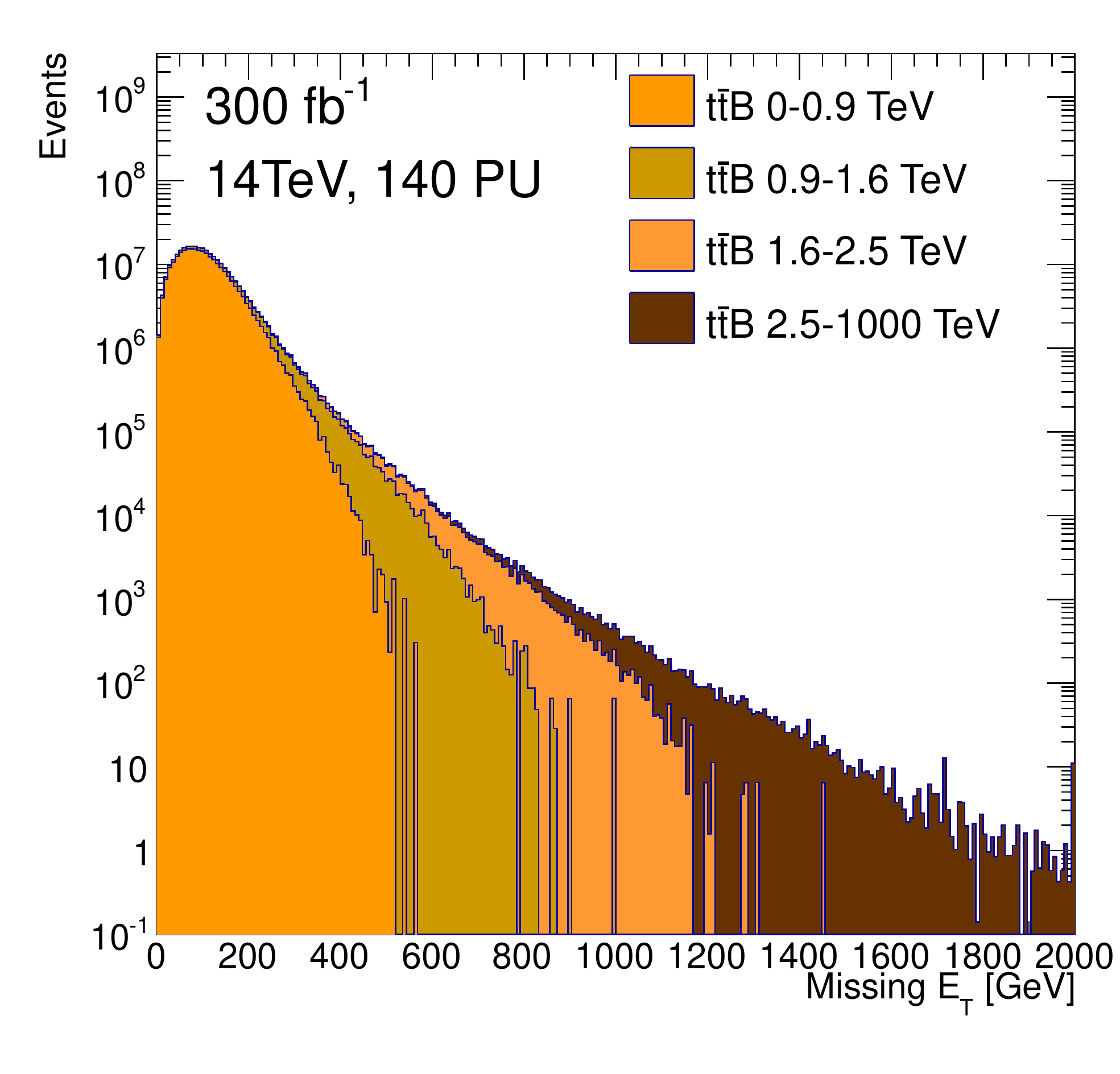}
\caption{Distribution of $H_T$ (top row), and $\ETmiss$ (bottom row)  
in $t\overline tV+nJ$ events. The three different columns represent conditions with 
an average of zero, 50 and 140 additional pile-up interactions.
\label{fig:binned01}}
\end{figure}

\begin{figure}[h!]
\centering
\includegraphics[width=0.3\textwidth]{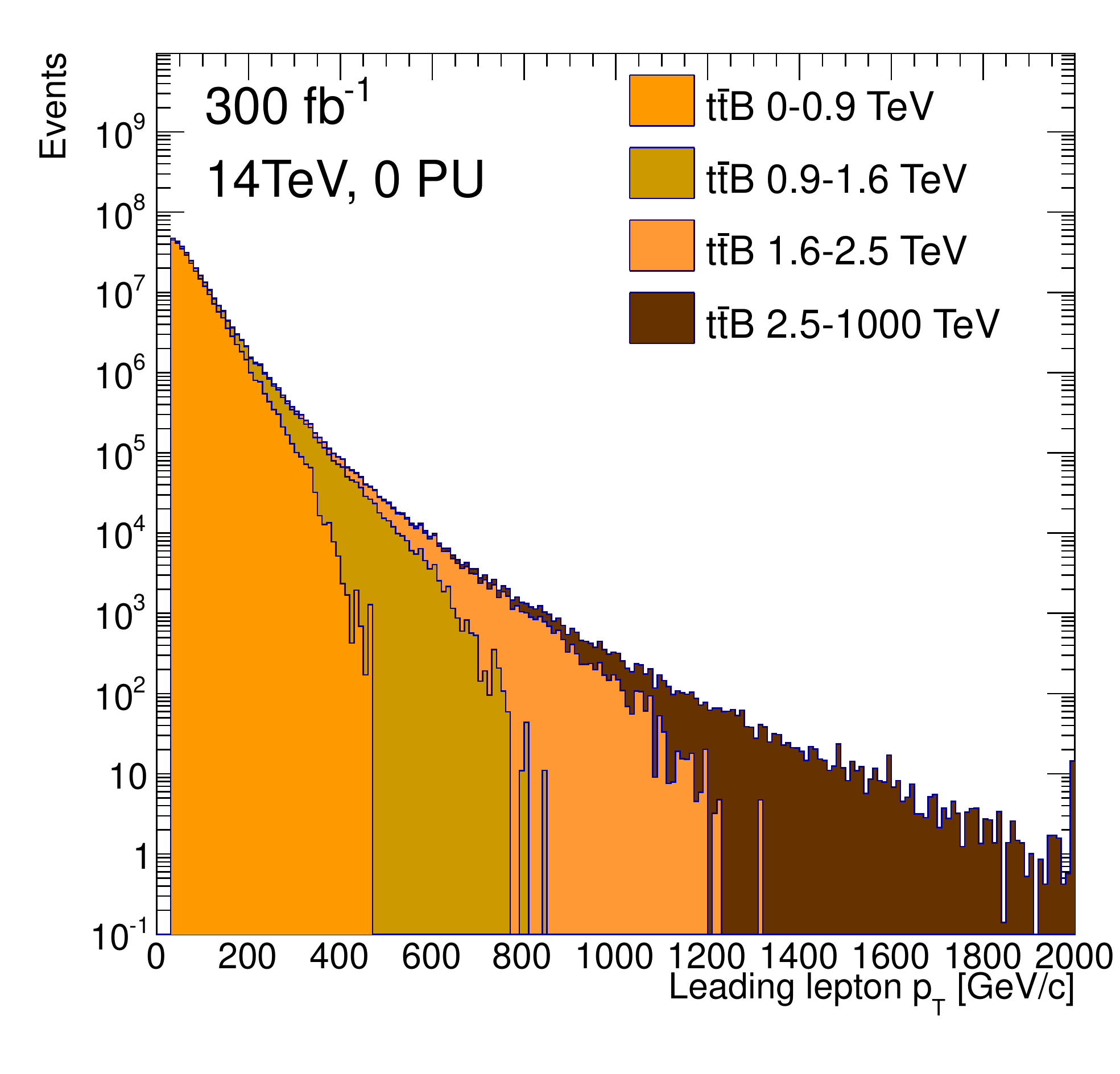}
\includegraphics[width=0.3\textwidth]{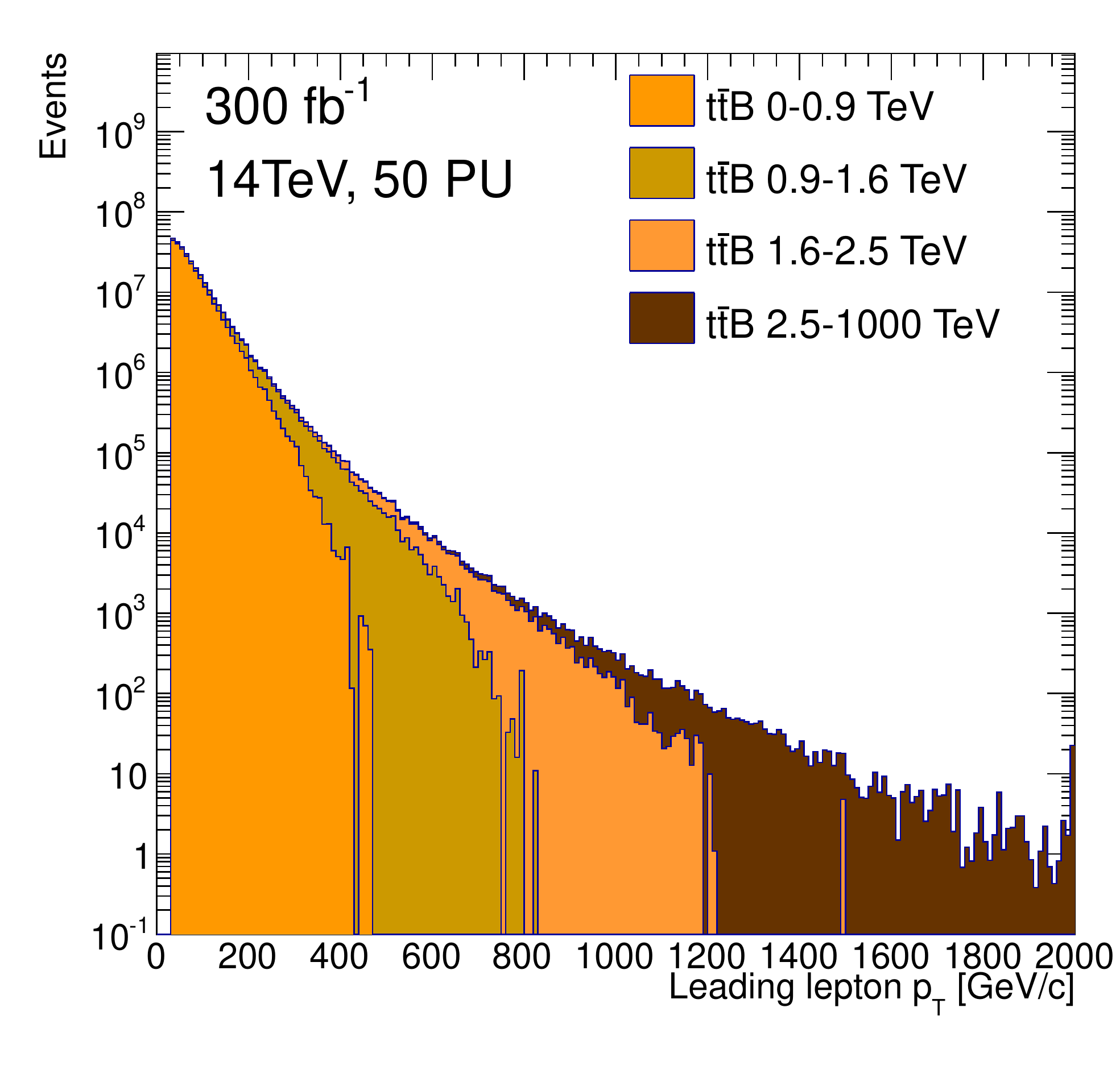}
\includegraphics[width=0.3\textwidth]{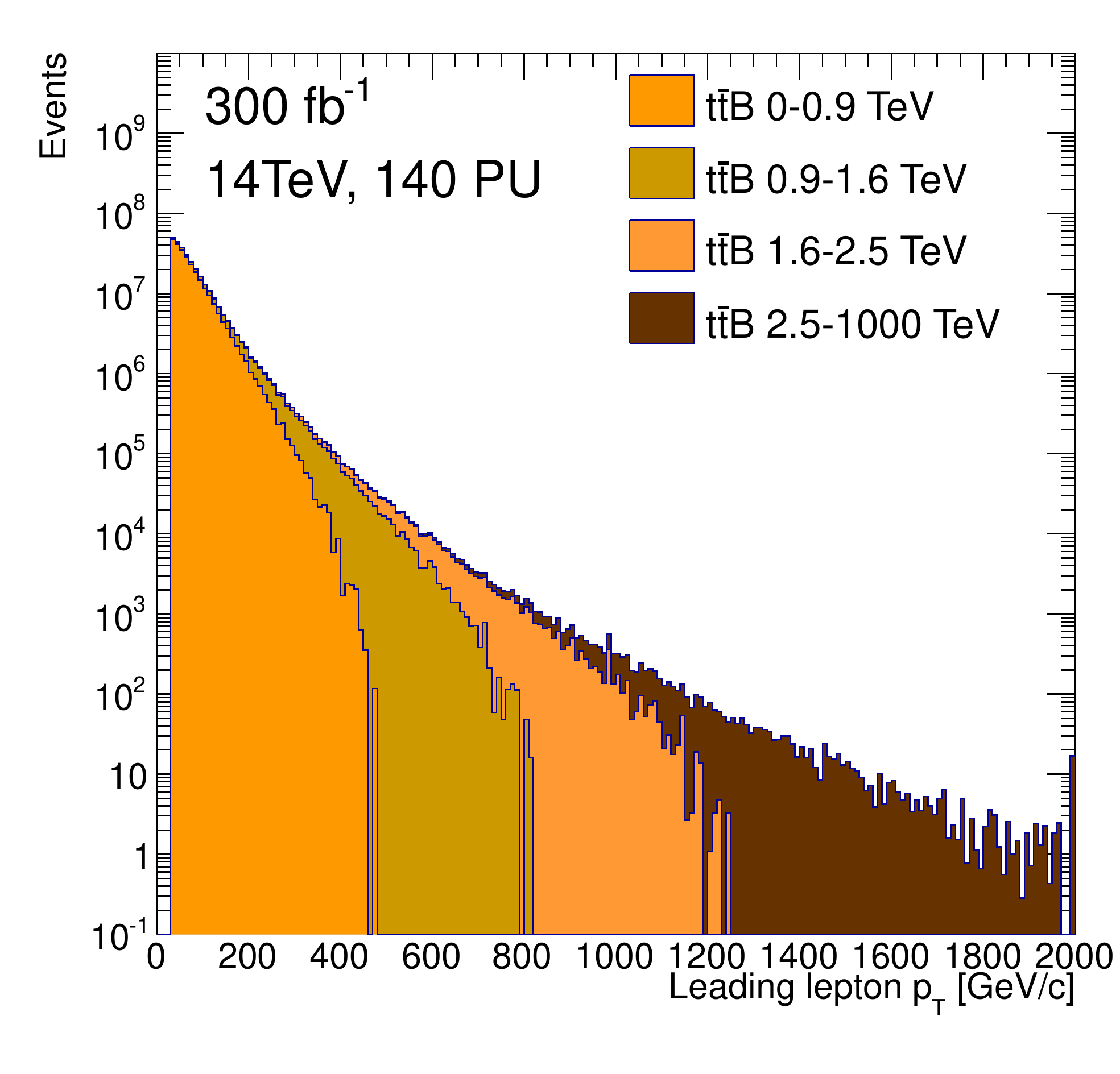}\\
\includegraphics[width=0.3\textwidth]{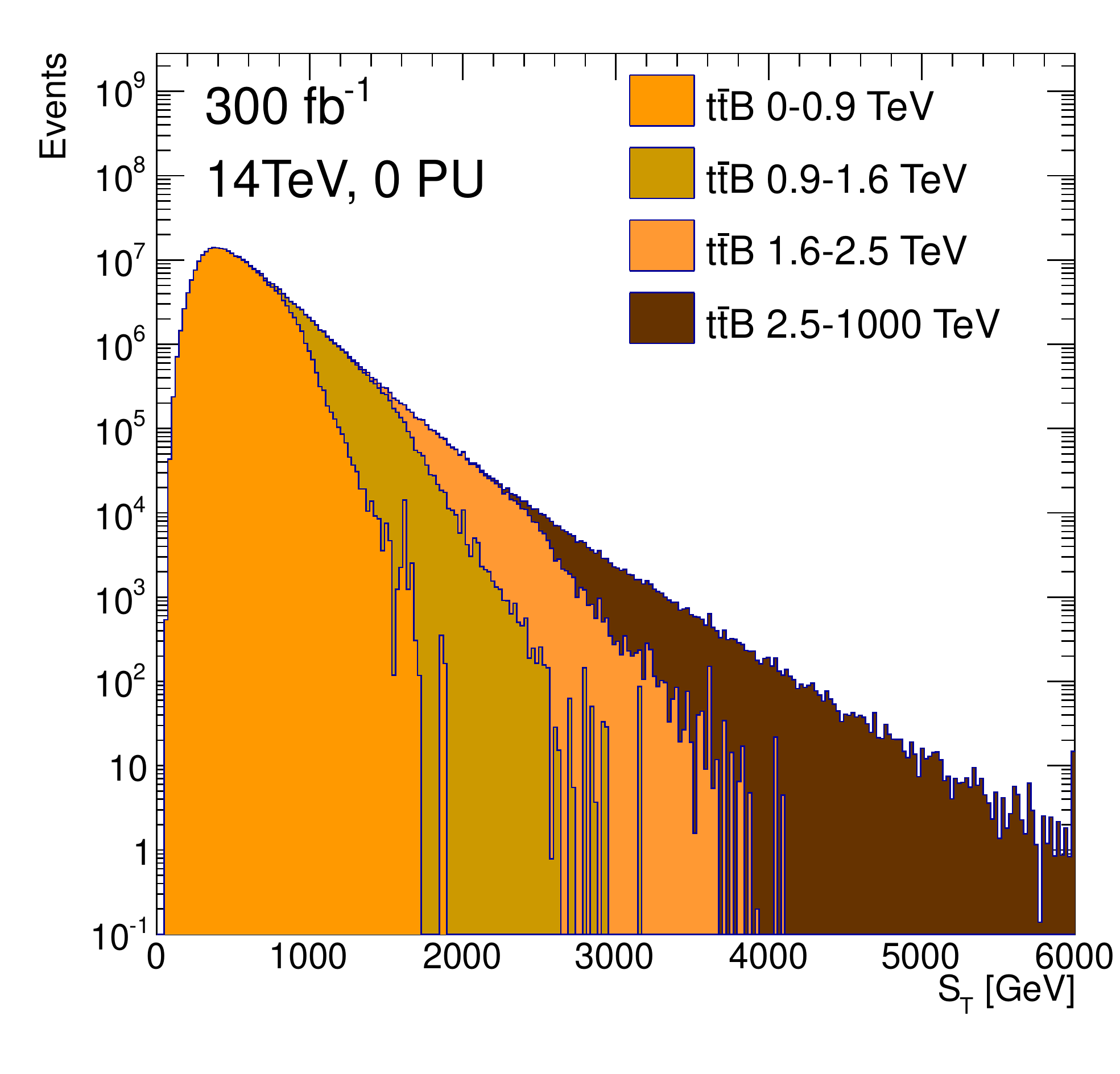}
\includegraphics[width=0.3\textwidth]{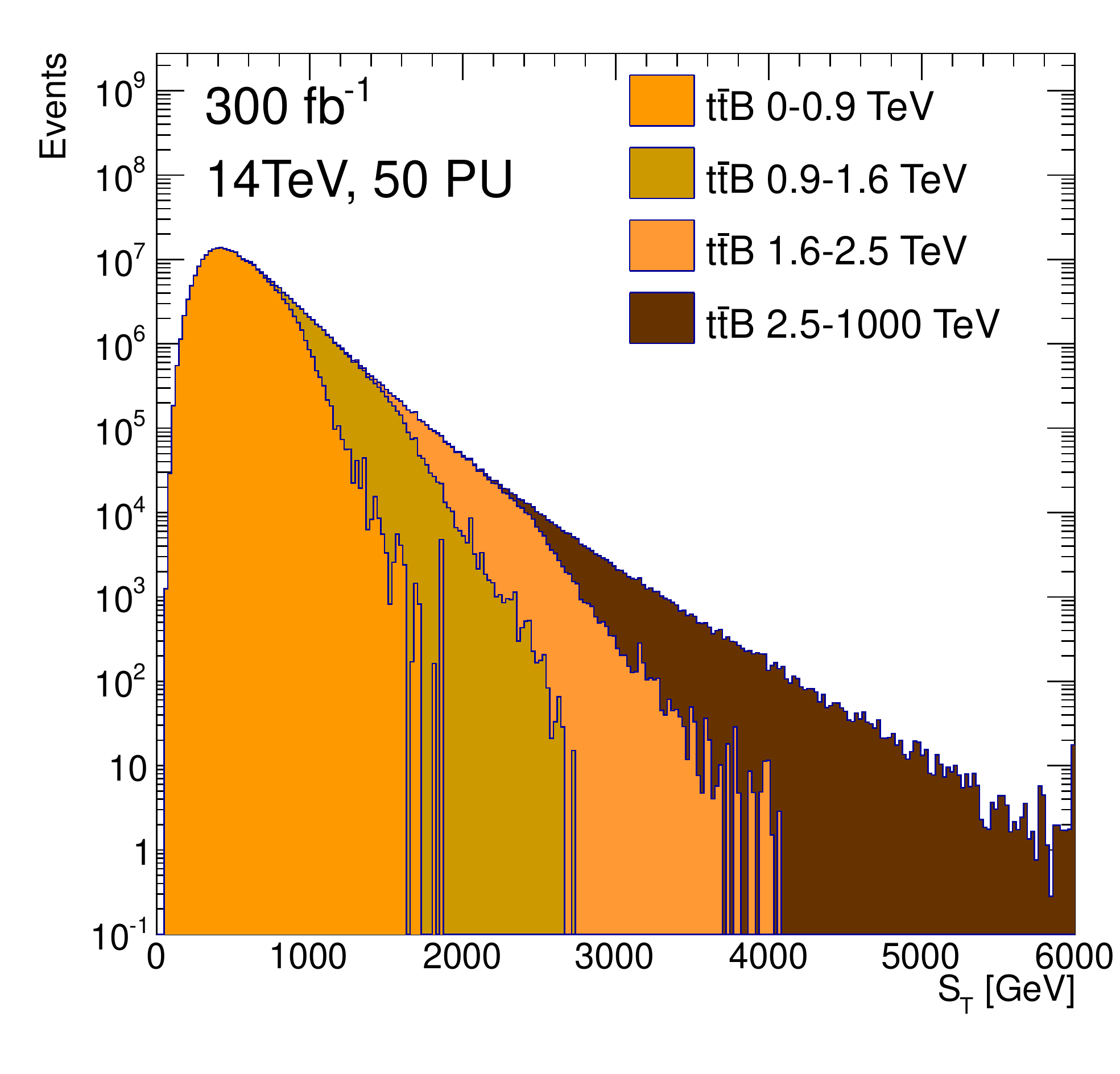}
\includegraphics[width=0.3\textwidth]{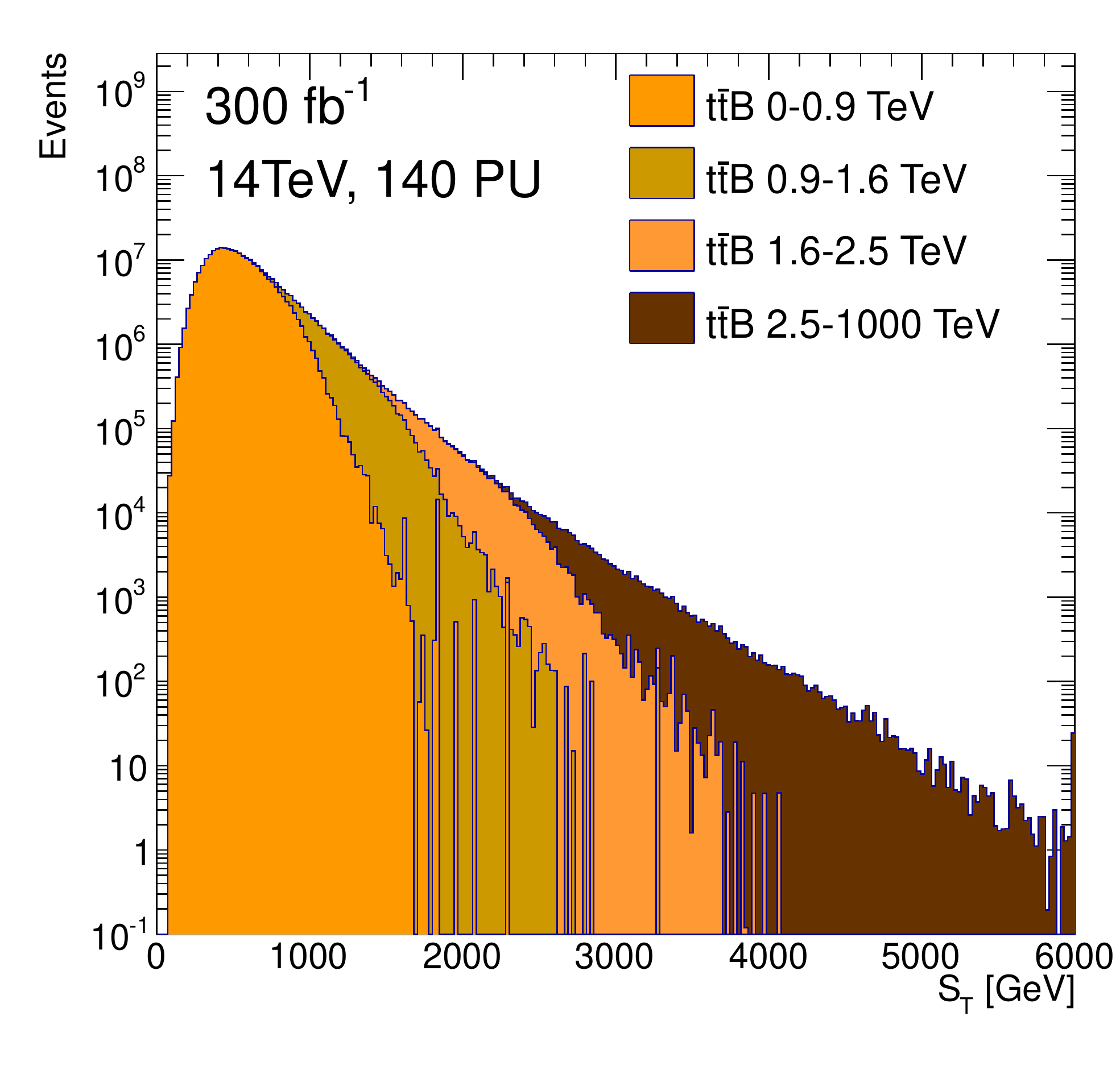}
\caption{Distribution of lepton $p_T$ (top row), and $S_T$ (bottom row)  
in $t\overline tV+nJ$ events. The three different columns represent conditions with 
an average of zero, 50 and 140 additional pile-up interactions.
\label{fig:binned02}}
\end{figure}

\begin{figure}[h!]
\centering
\includegraphics[width=0.3\textwidth]{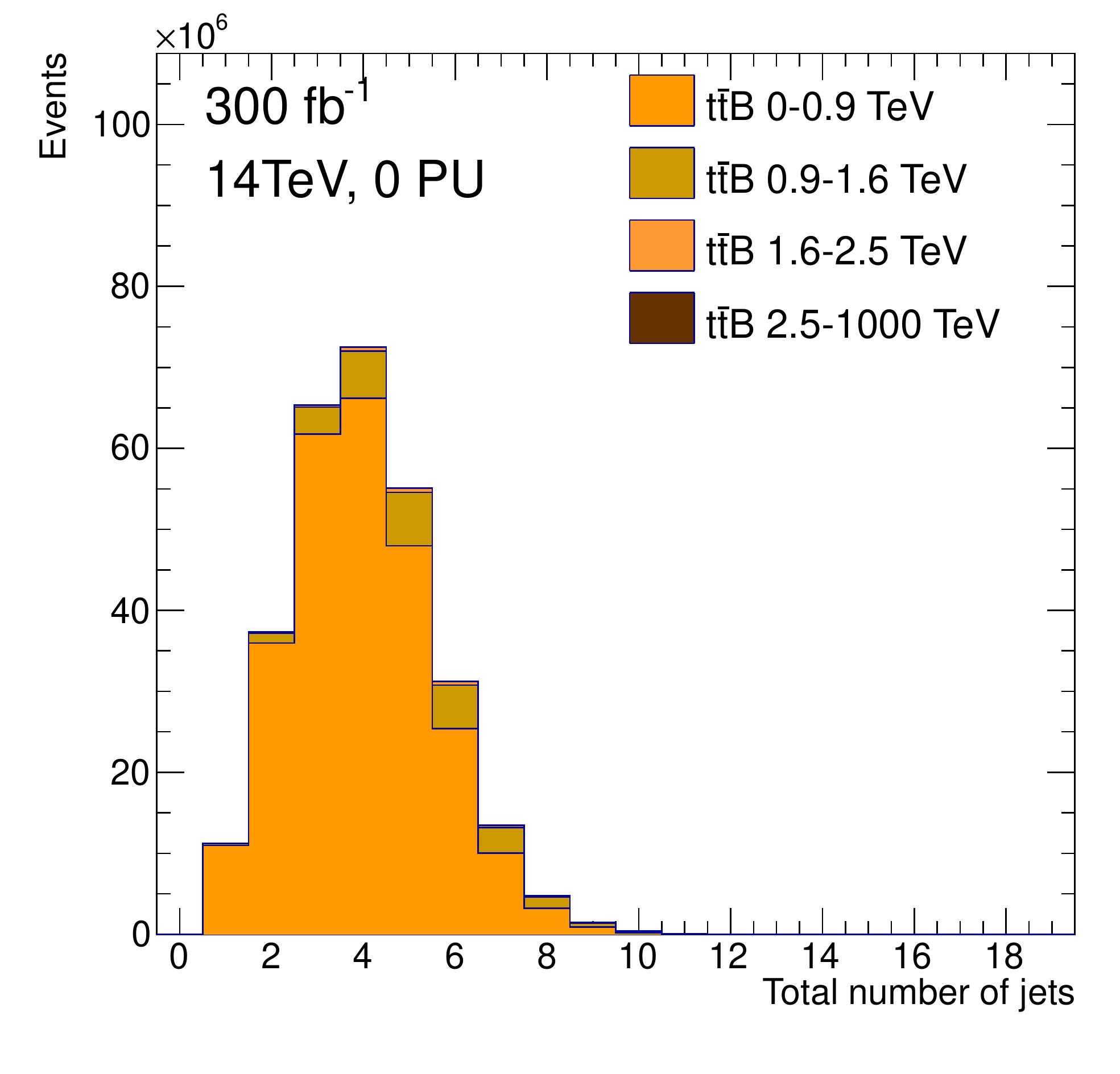}
\includegraphics[width=0.3\textwidth]{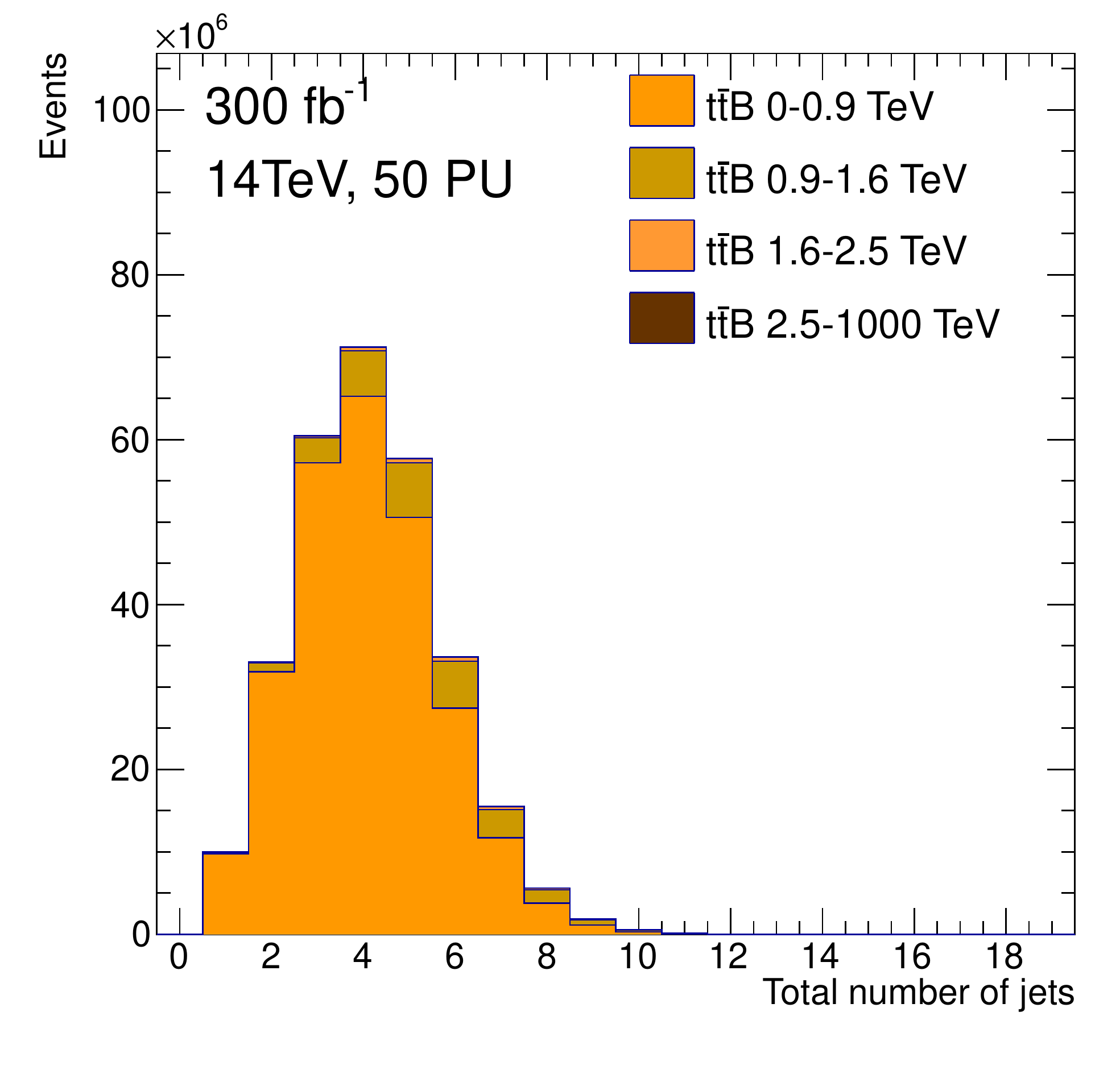}
\includegraphics[width=0.3\textwidth]{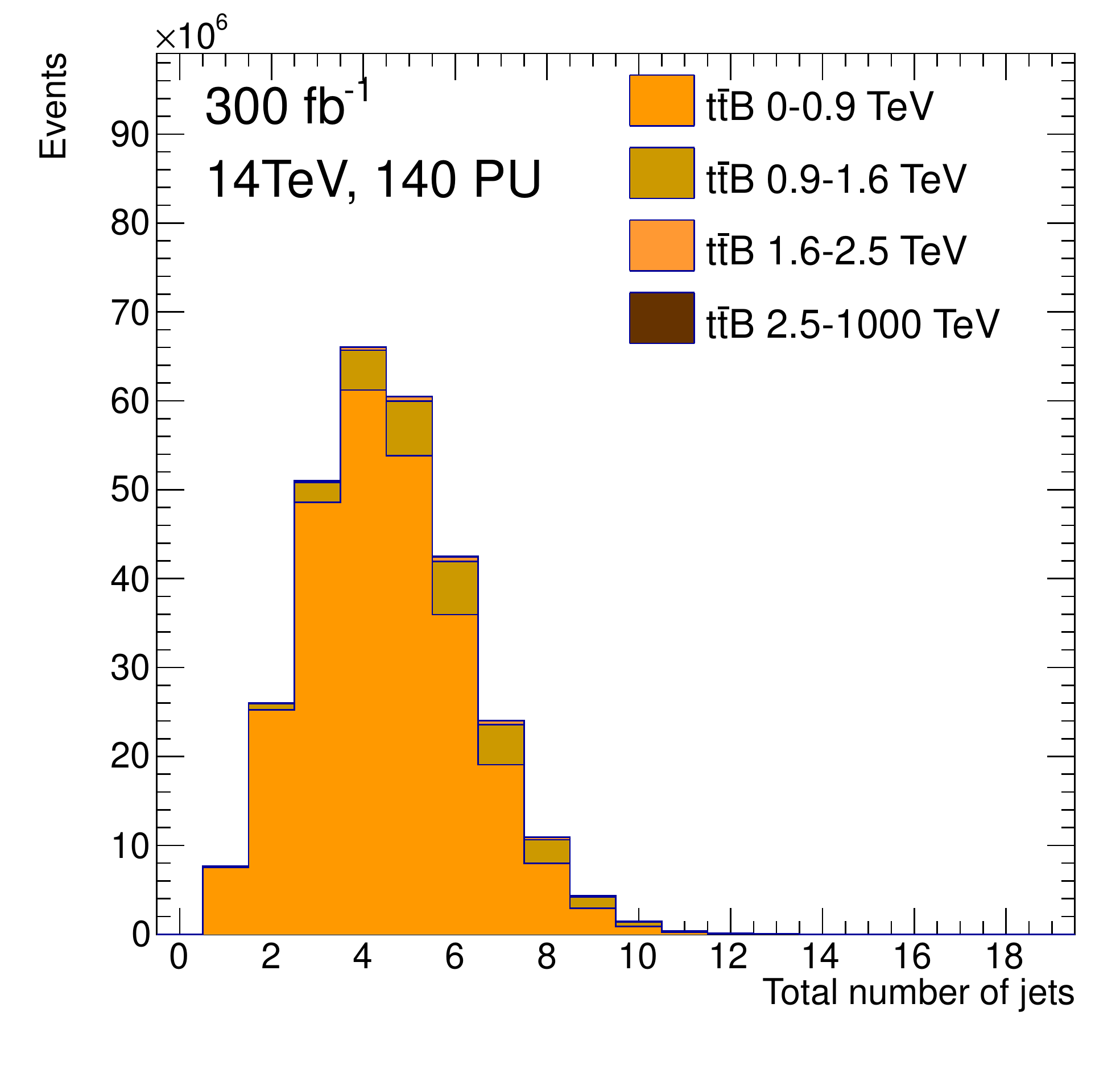}\\
\includegraphics[width=0.3\textwidth]{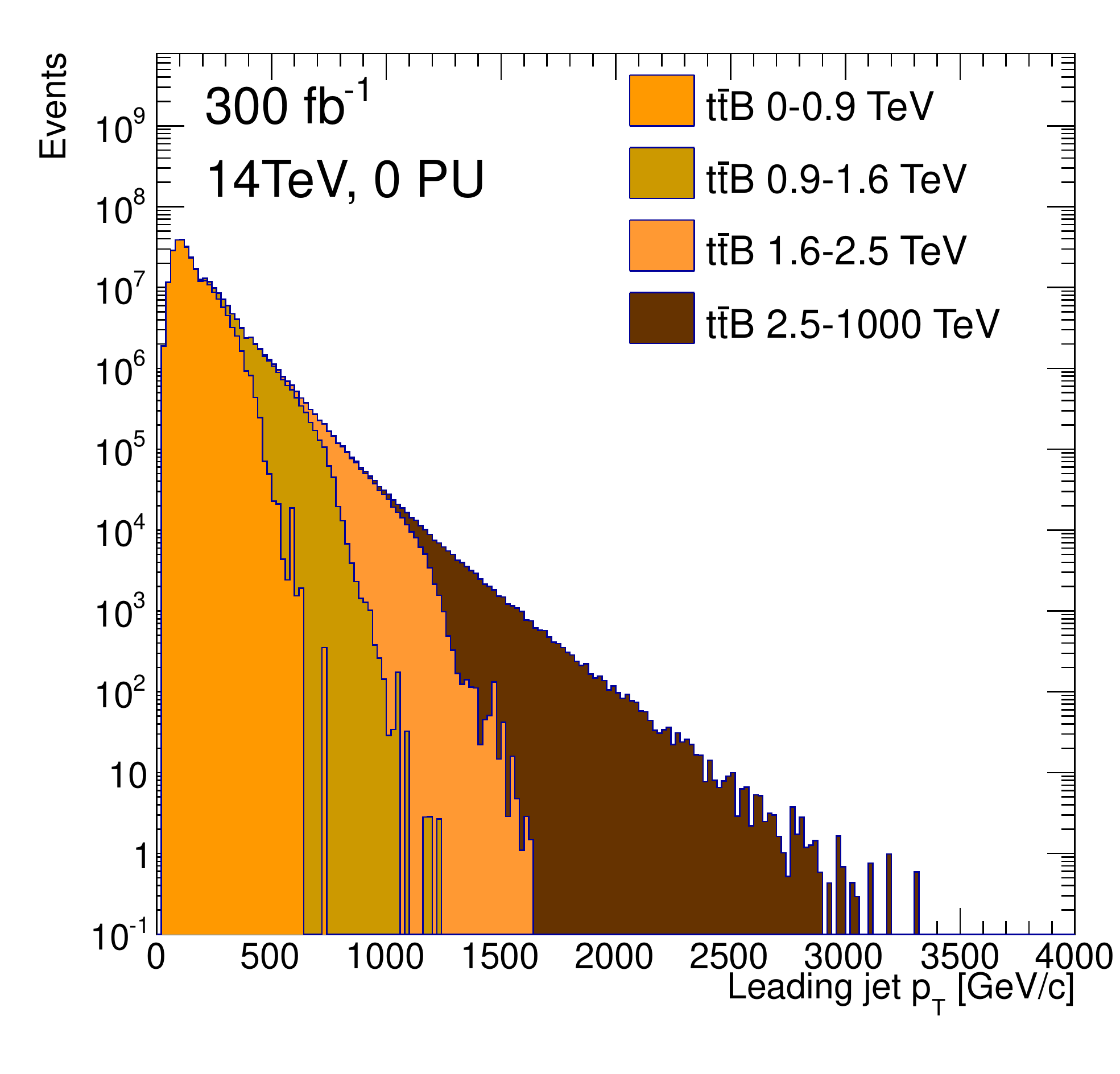}
\includegraphics[width=0.3\textwidth]{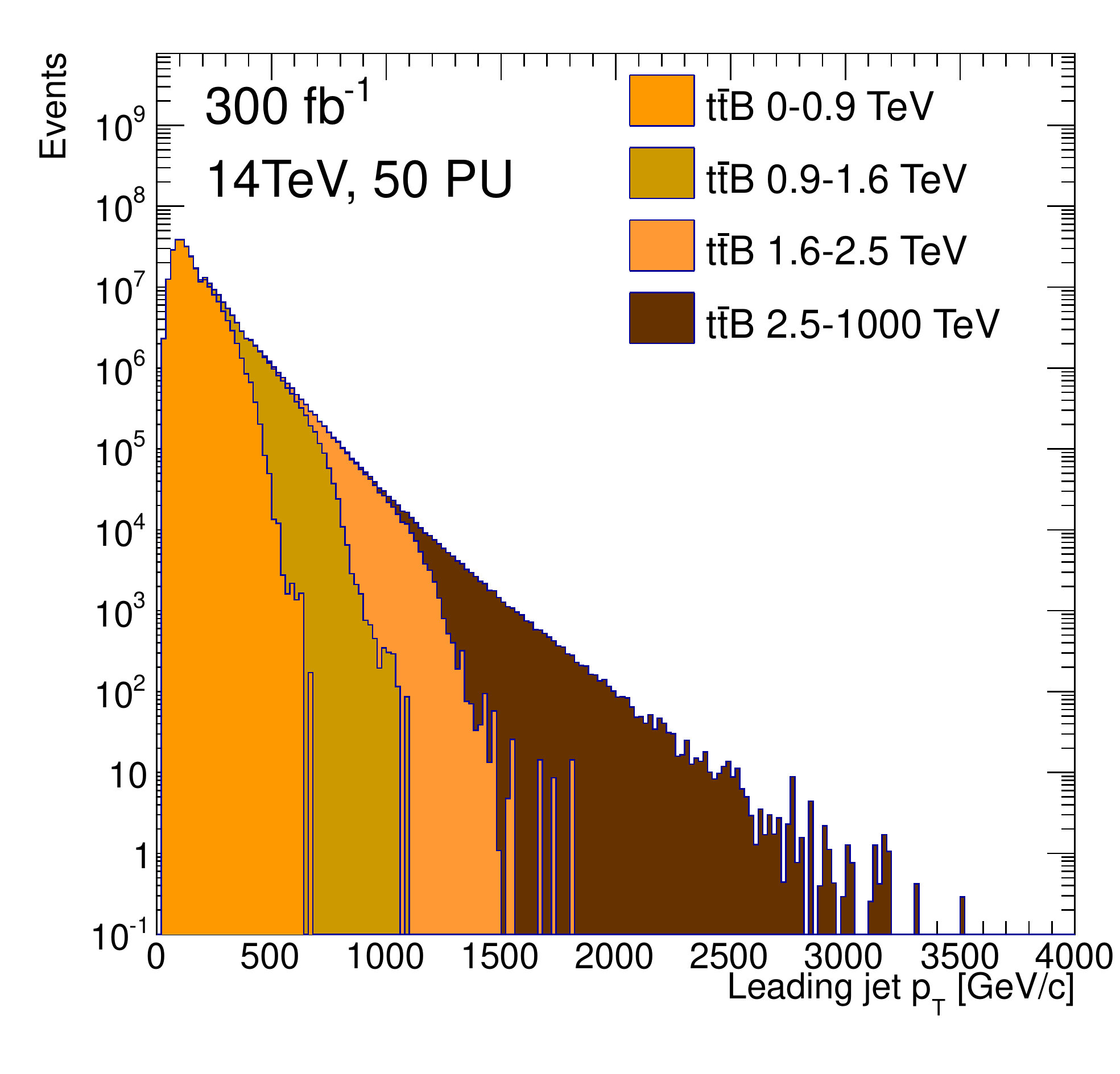}
\includegraphics[width=0.3\textwidth]{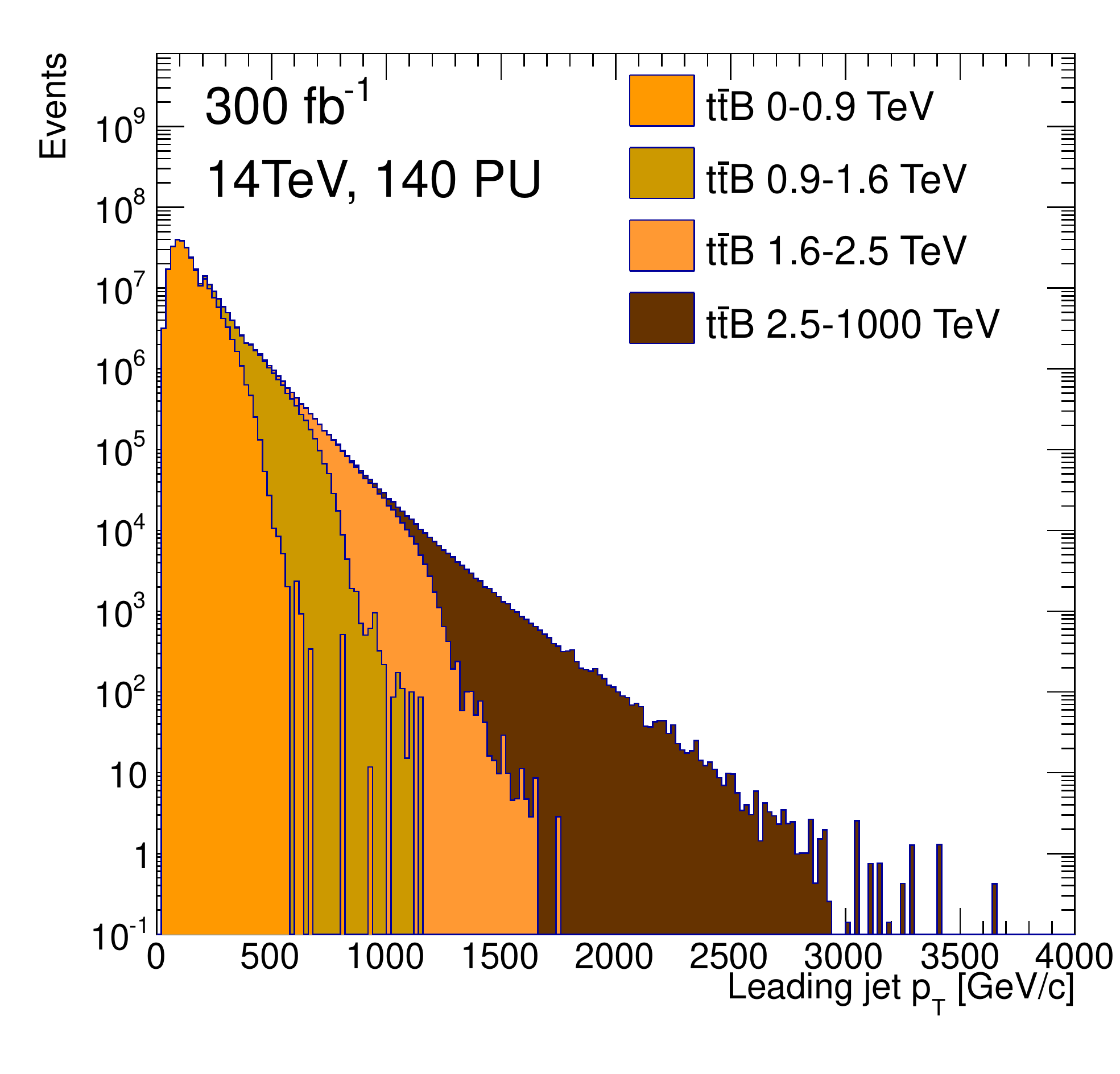}\\
\includegraphics[width=0.3\textwidth]{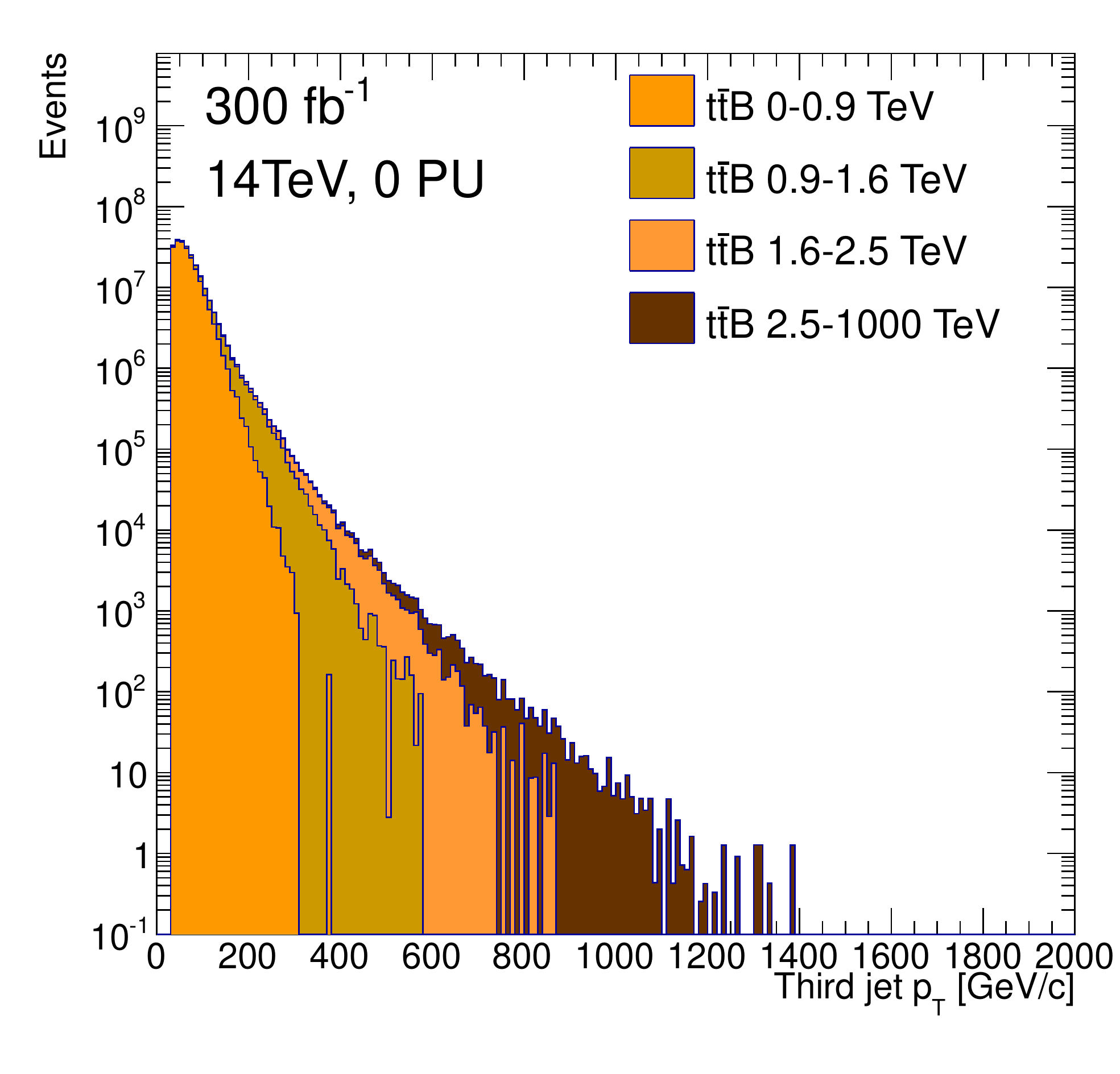}
\includegraphics[width=0.3\textwidth]{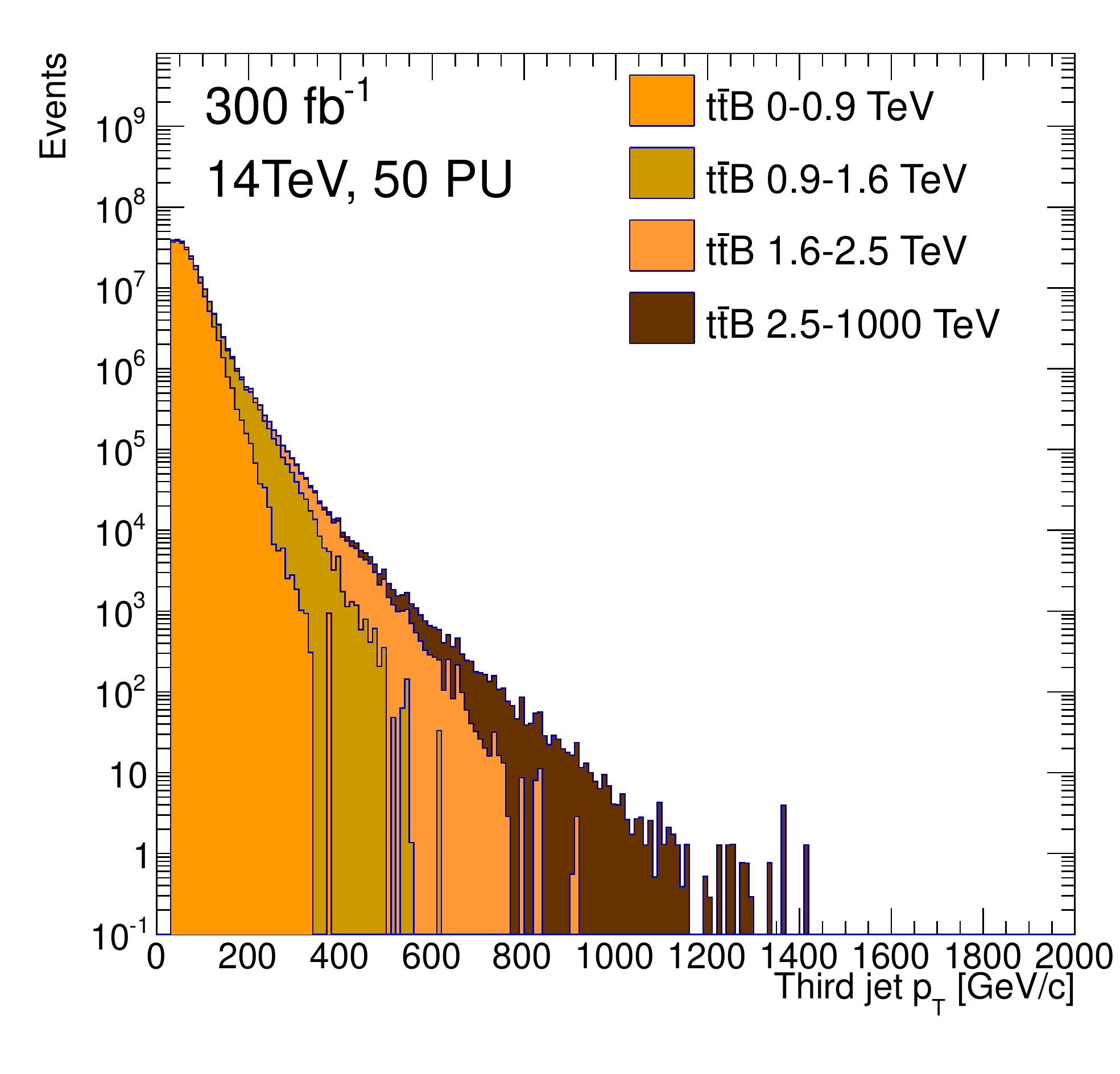}
\includegraphics[width=0.3\textwidth]{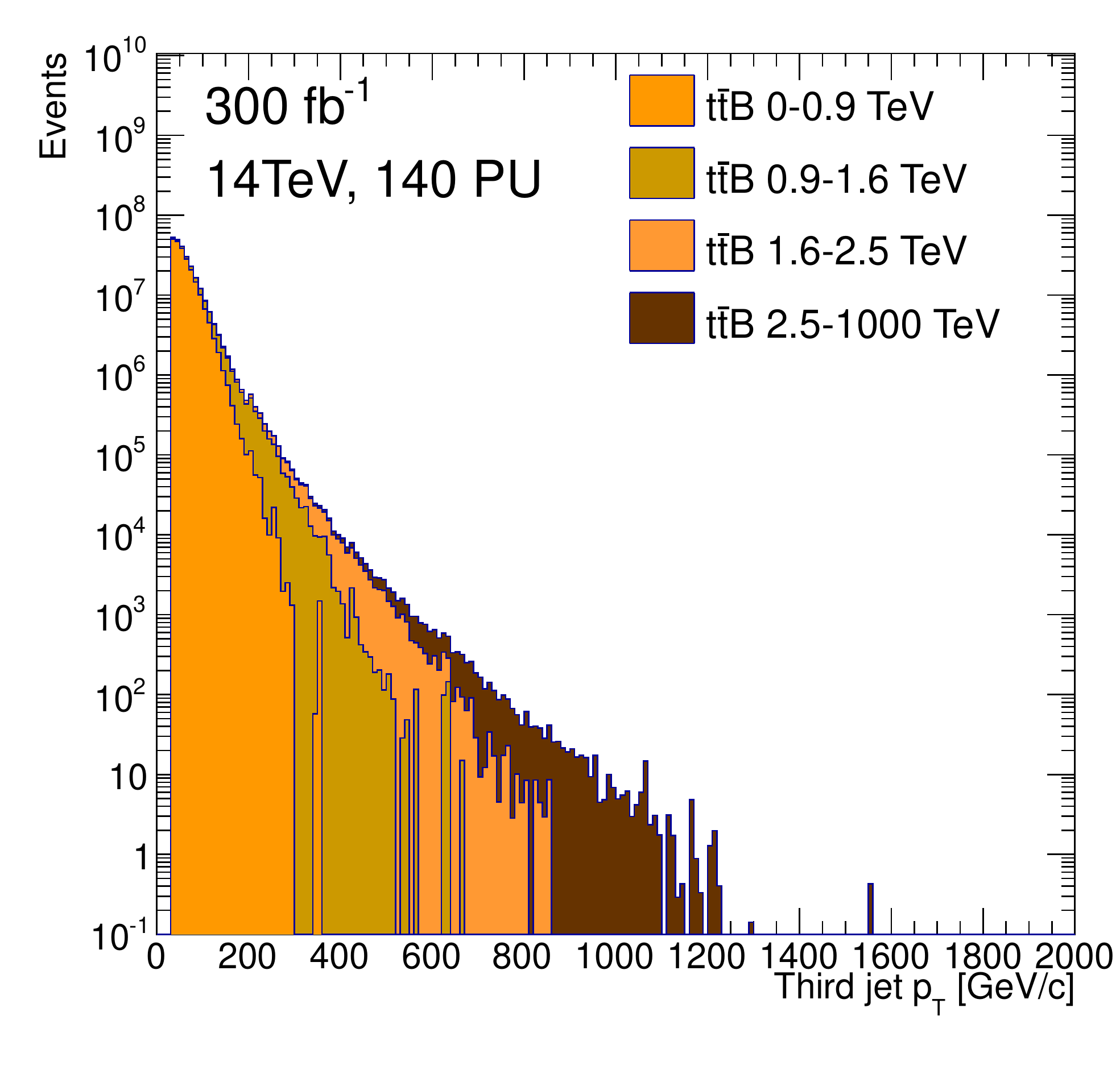}
\caption{Distribution of total number of selected jets (top row), $p_T$ of the 
highest $p_T$ (middle row) and the third highest $p_T$ jet in the event (bottom row). 
The three different columns represent conditions with 
an average of zero, 50 and 140 additional pile-up interactions.
\label{fig:binned03}}
\end{figure}

\begin{figure}[h!]
\centering
\includegraphics[width=0.3\textwidth]{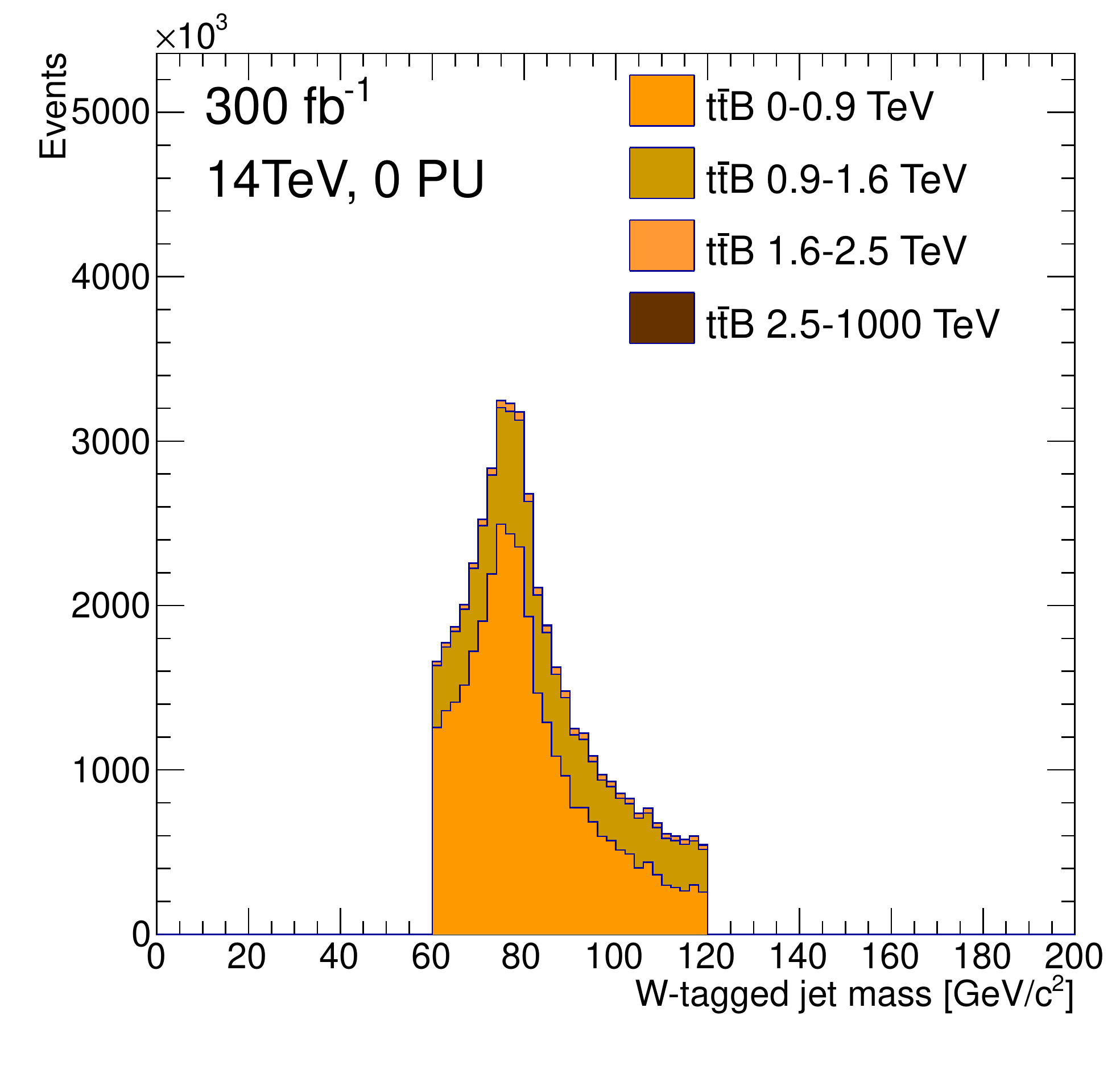}
\includegraphics[width=0.3\textwidth]{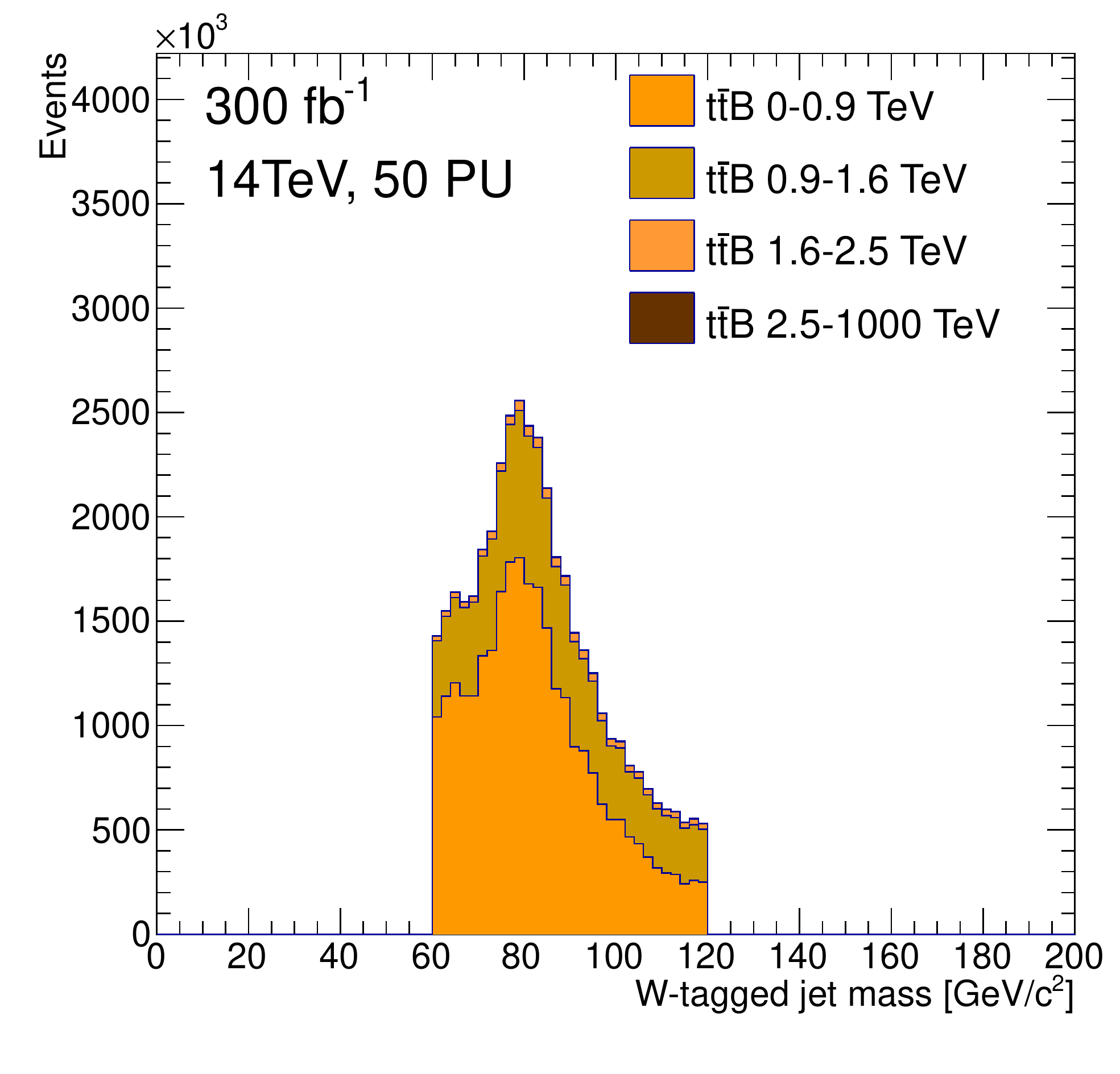}
\includegraphics[width=0.3\textwidth]{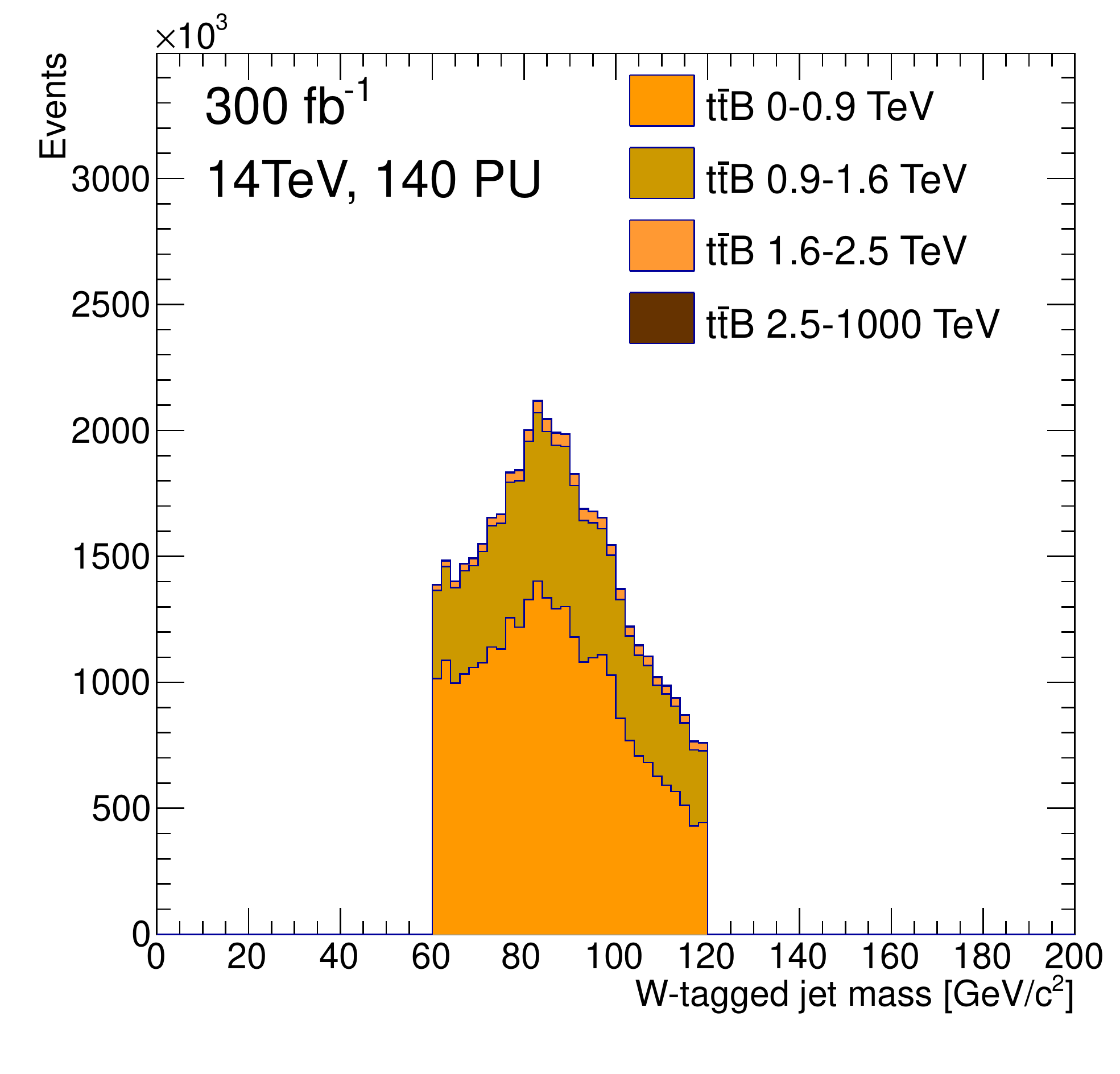}\\
\includegraphics[width=0.3\textwidth]{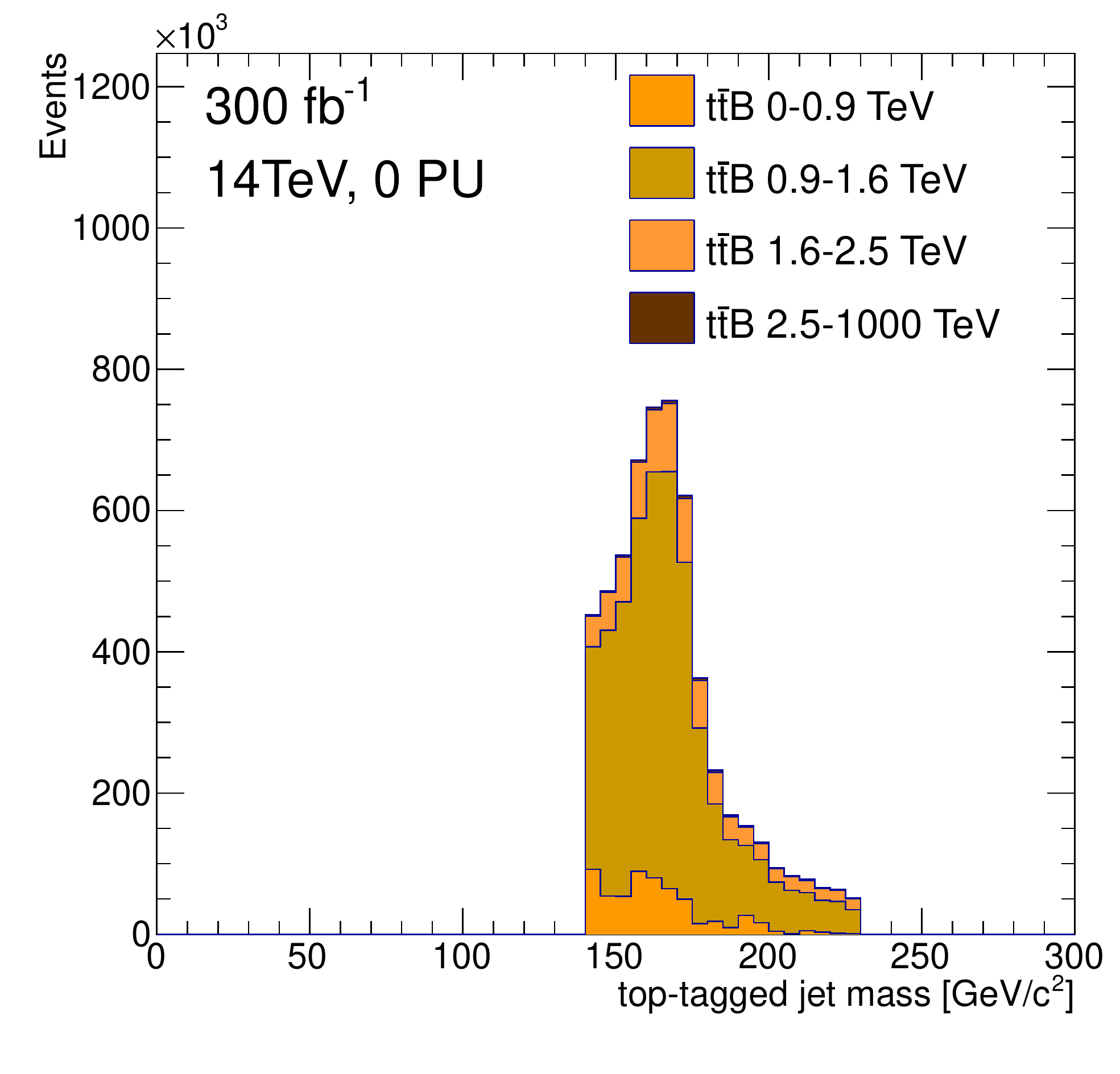}
\includegraphics[width=0.3\textwidth]{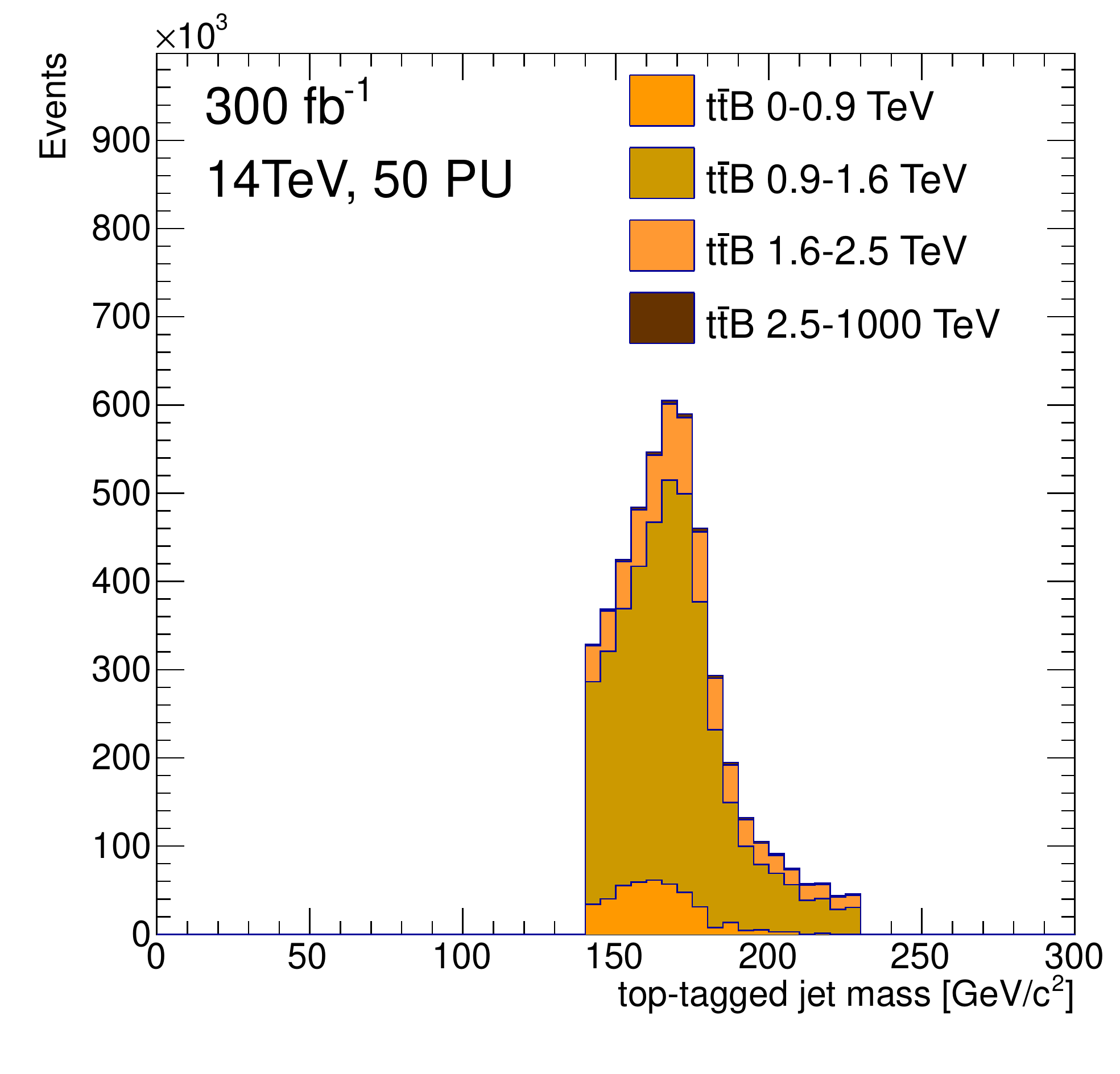}
\includegraphics[width=0.3\textwidth]{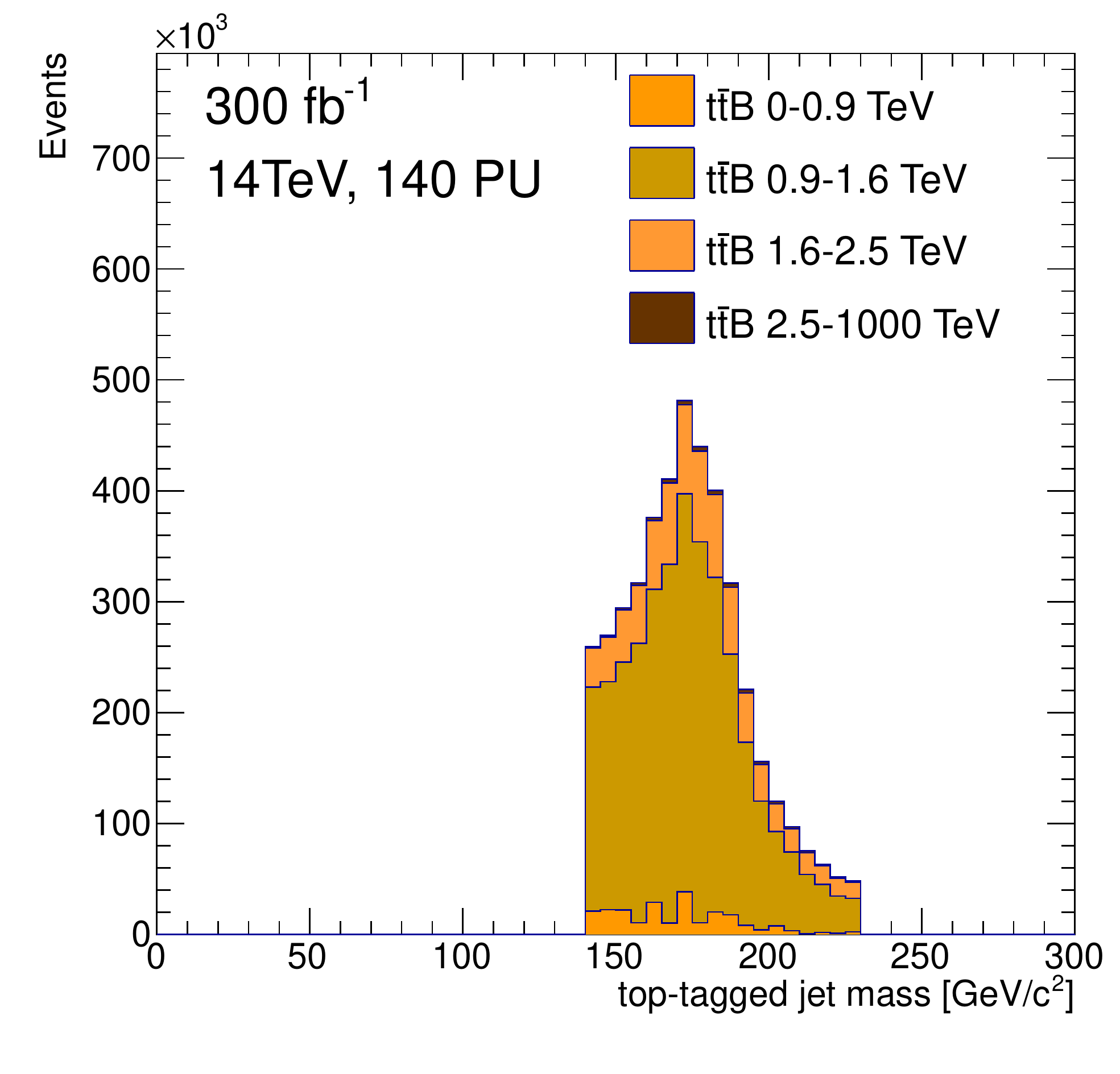}
\caption{Distribution of the ``trimmed jet mass'' computed by the fatjet algorithm for
merged $W$ and top jets. The three different columns represent conditions with 
an average of zero, 50 and 140 additional pile-up interactions.
\label{fig:binned04}}
\end{figure}

In Figures~\ref{fig:binned01}-\ref{fig:binned04}, the distributions of a few 
key kinematic variables are shown for a sample of $t\overline tV+nJ$ events.
In the figures, the effect of each $S^*_T$ bin is displayed. All 
$S^*_T$ bins for a particular sample have to be added together, using the appropriate weights.
The high end tails of the  distributions is reasonably well populated.

In these distributions, reconstructed jets and leptons are 
required to have  $p_T>$30 \GeV and $|\eta|<2.5$. $H_T$ is defined as scalar sum of $p_T$ of
all selected jets,   $S_T$ is defined as the scalar sum of $H_T$, $\ETmiss$ and $p_T$ of 
all selected leptons. The mass of the boosted jet tagged as a $W$ jet or a top-quark-jet is
denoted as the  ``trimmed mass''~\cite{Krohn}. The ``trimmed mass'' is calculated after the 
jet has been reclustered into subjets.

\subsection{Effect of Pile Up}
\label{sec:simSM-dist-pile-up}

\begin{figure}[h!]
\centering
\includegraphics[width=0.3\textwidth]{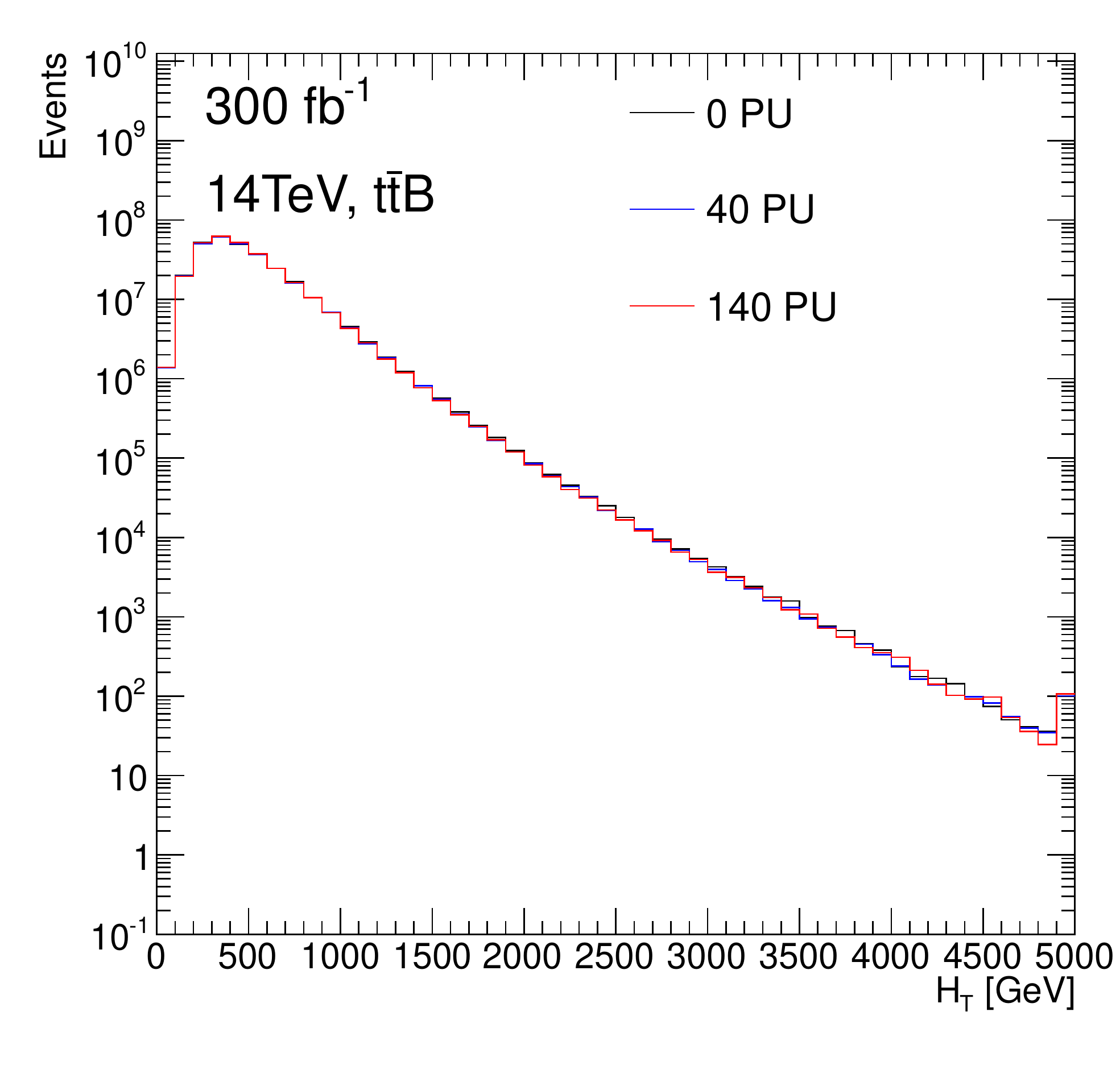}
\includegraphics[width=0.3\textwidth]{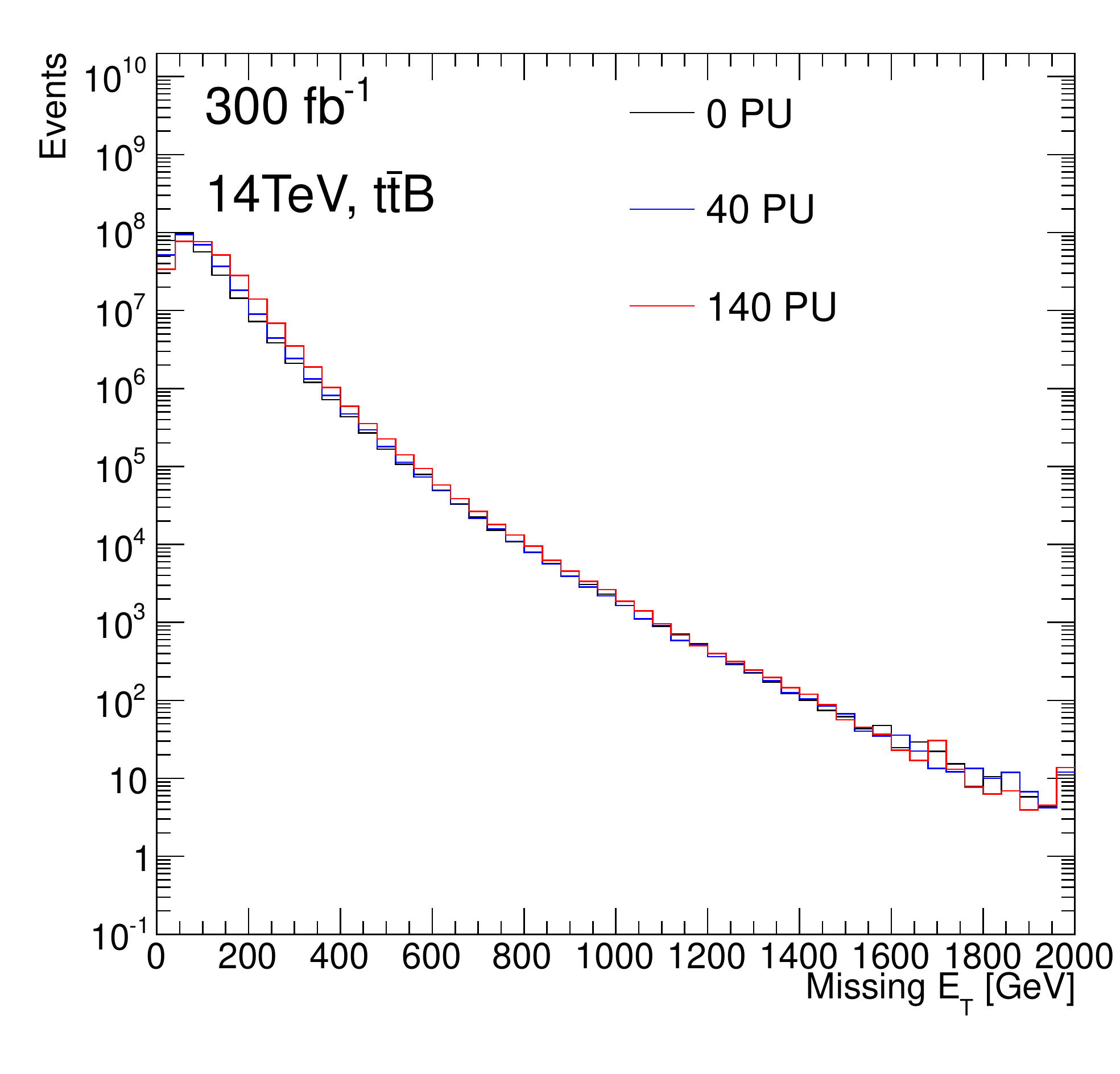}
\includegraphics[width=0.3\textwidth]{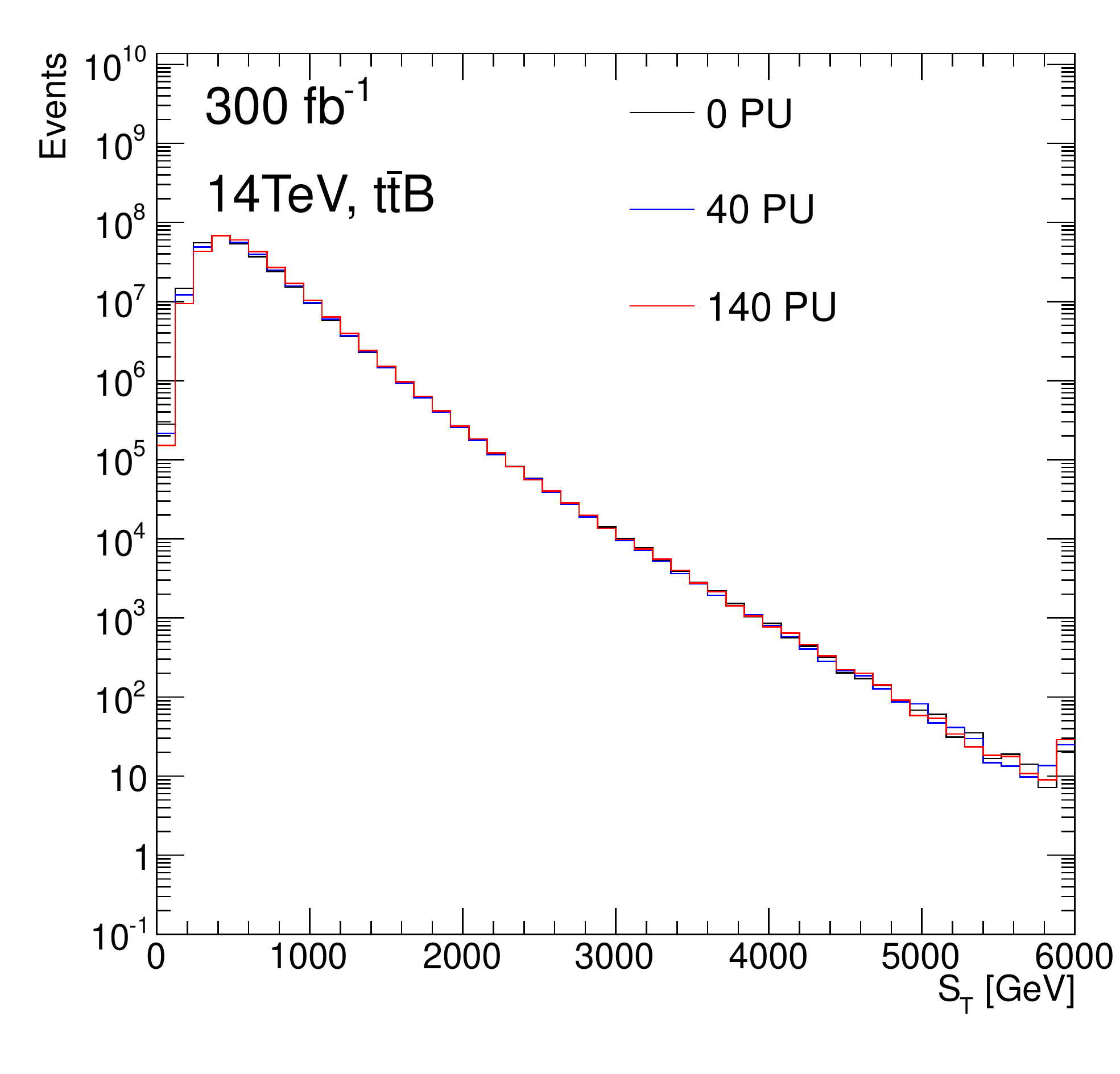}
\caption{Comparison of $H_T$ (left), $\ETmiss$ (middle) and $S_T$ (right)
in $t\overline tV+nJ$ events with different event pile-up conditions.
\label{fig:PU01}}
\end{figure}

\begin{figure}[h!]
\centering
\includegraphics[width=0.3\textwidth]{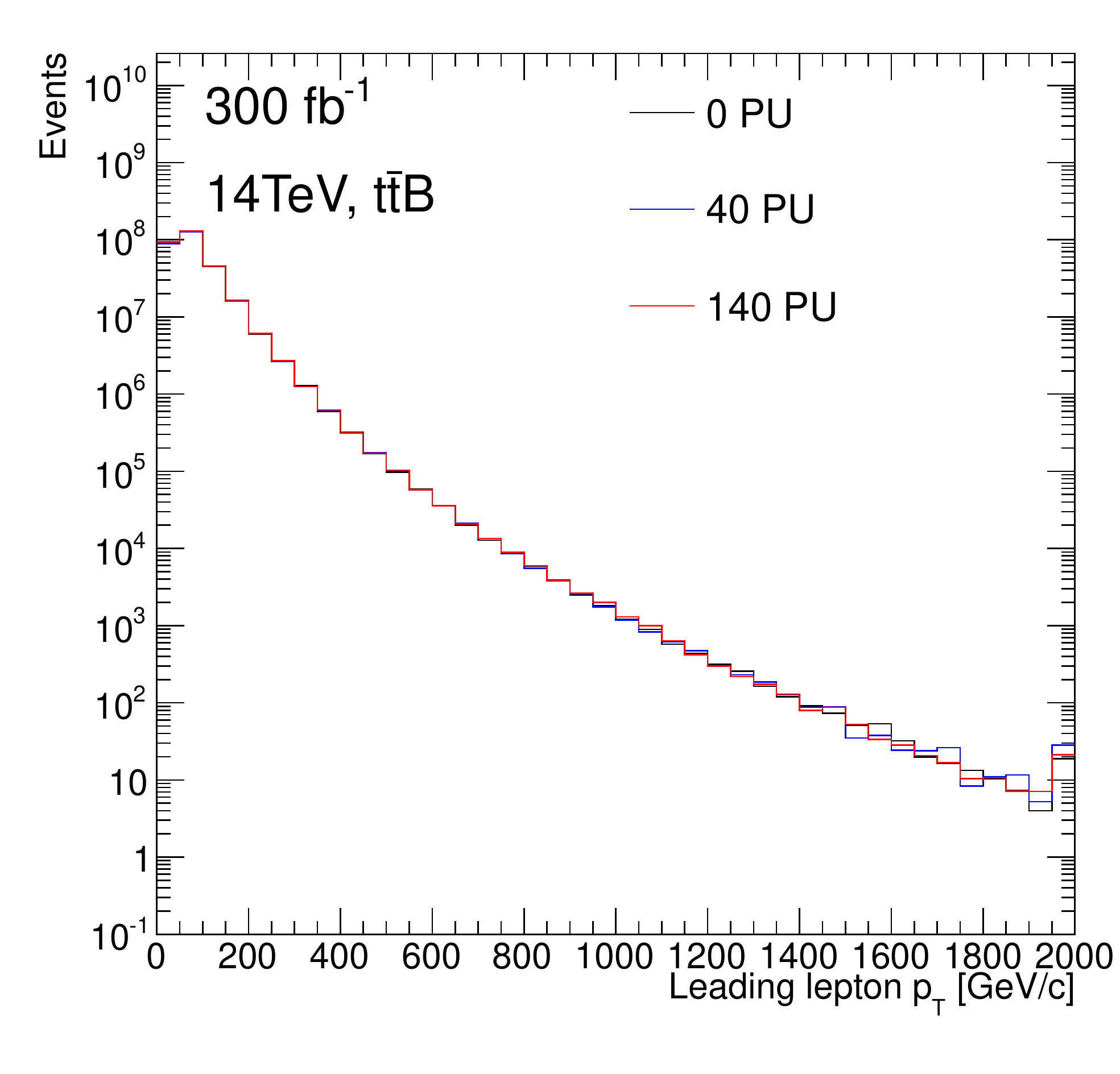}
\includegraphics[width=0.3\textwidth]{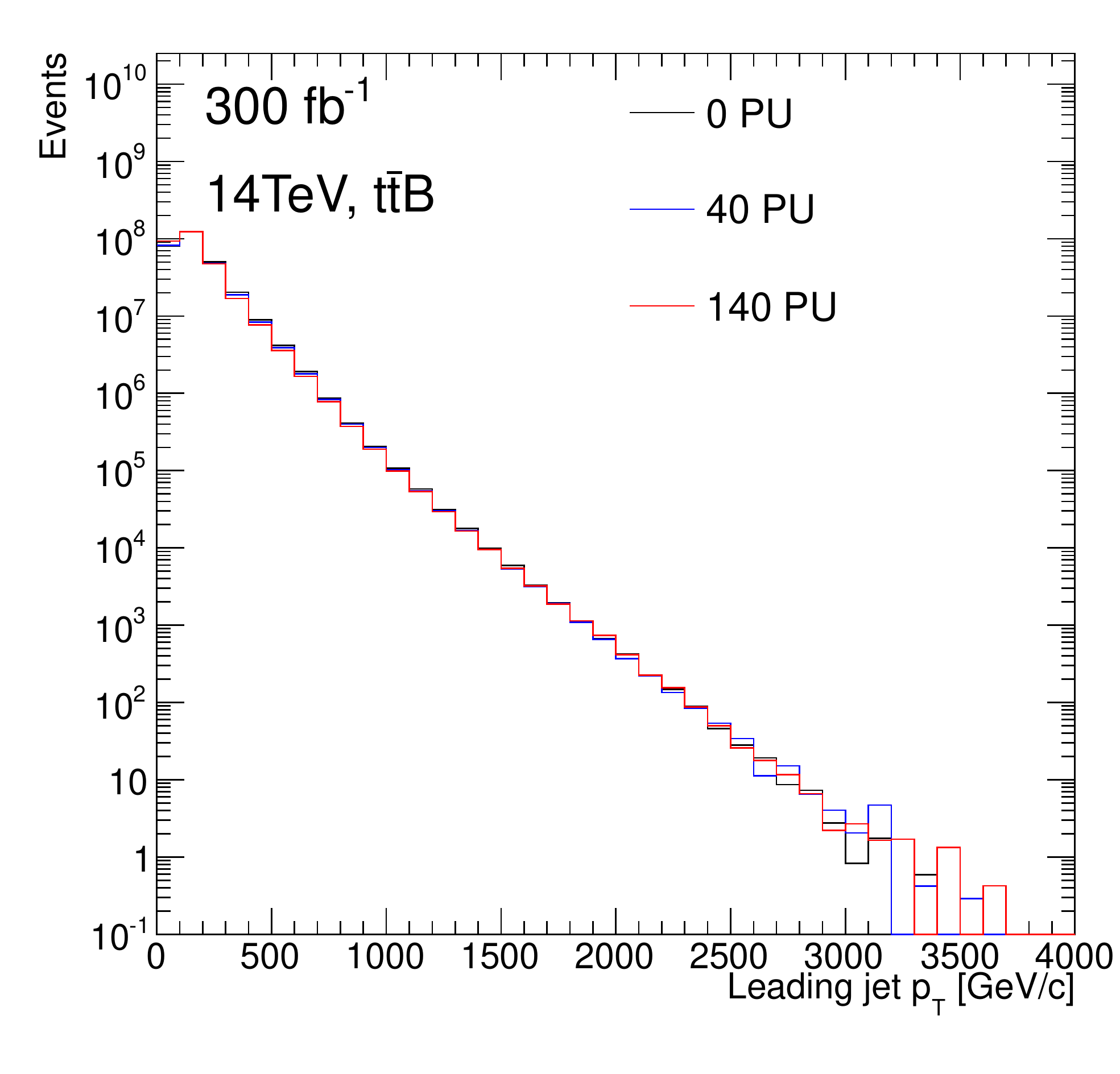}
\includegraphics[width=0.3\textwidth]{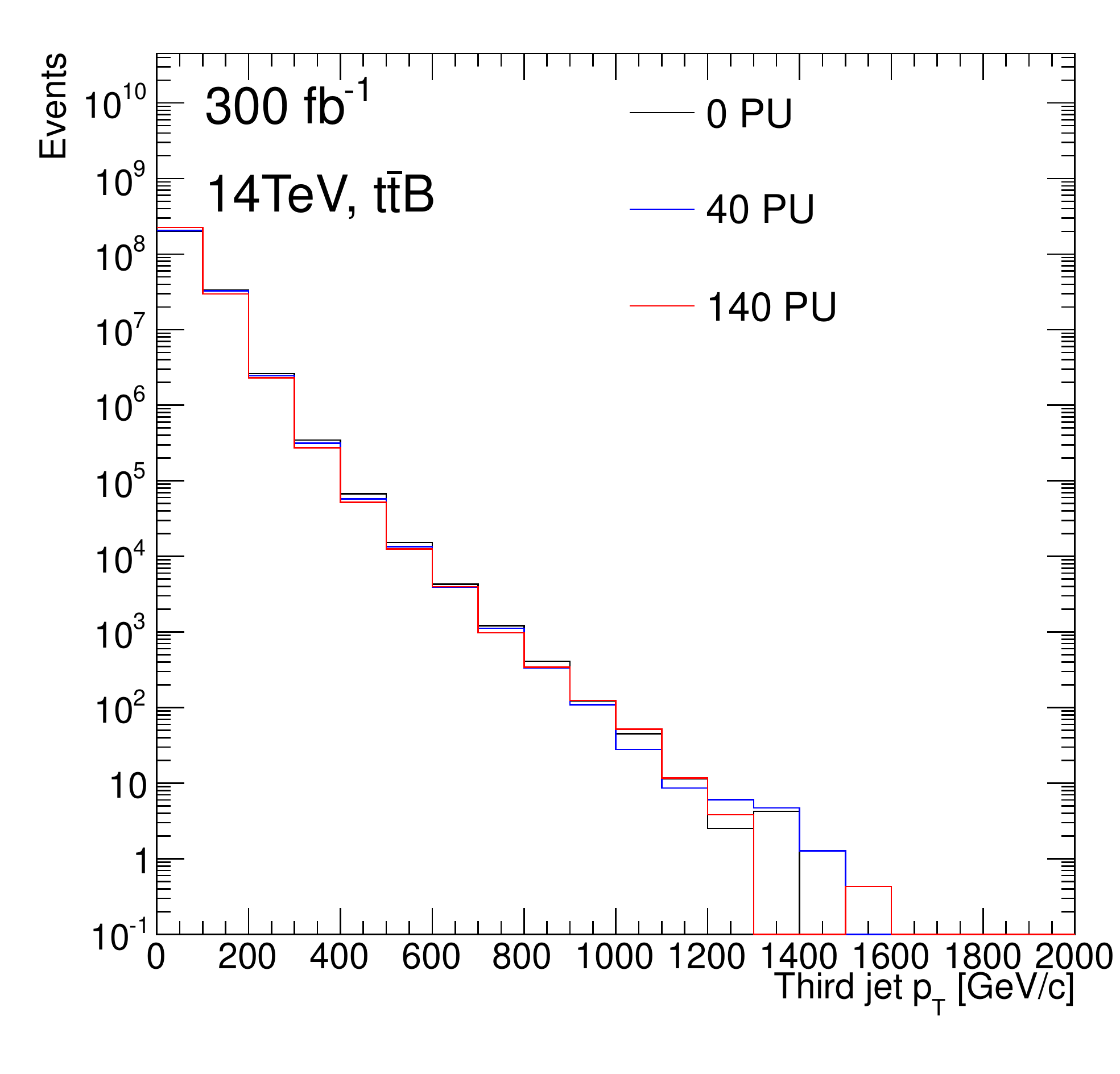}\\
\includegraphics[width=0.3\textwidth]{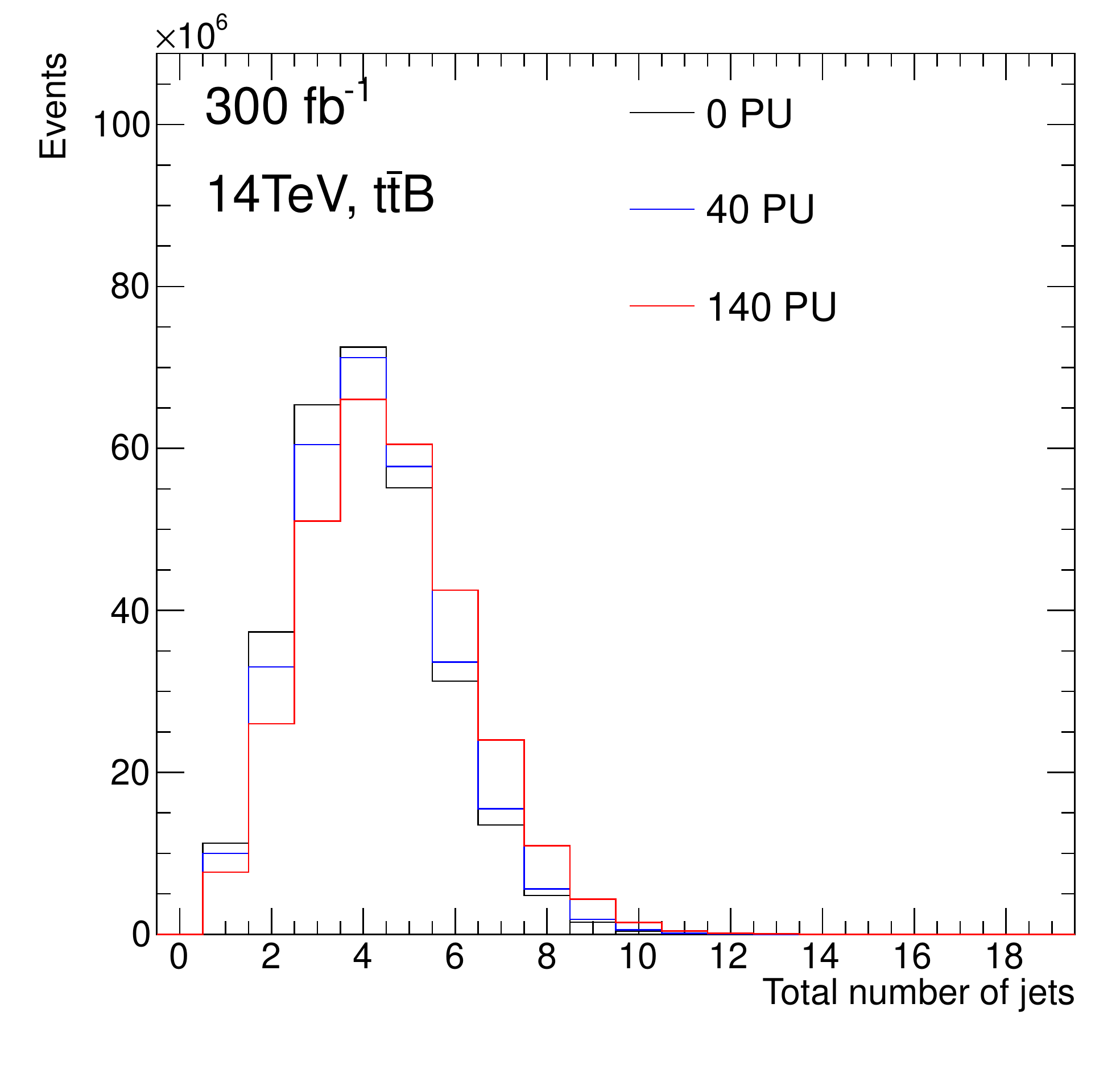}
\includegraphics[width=0.3\textwidth]{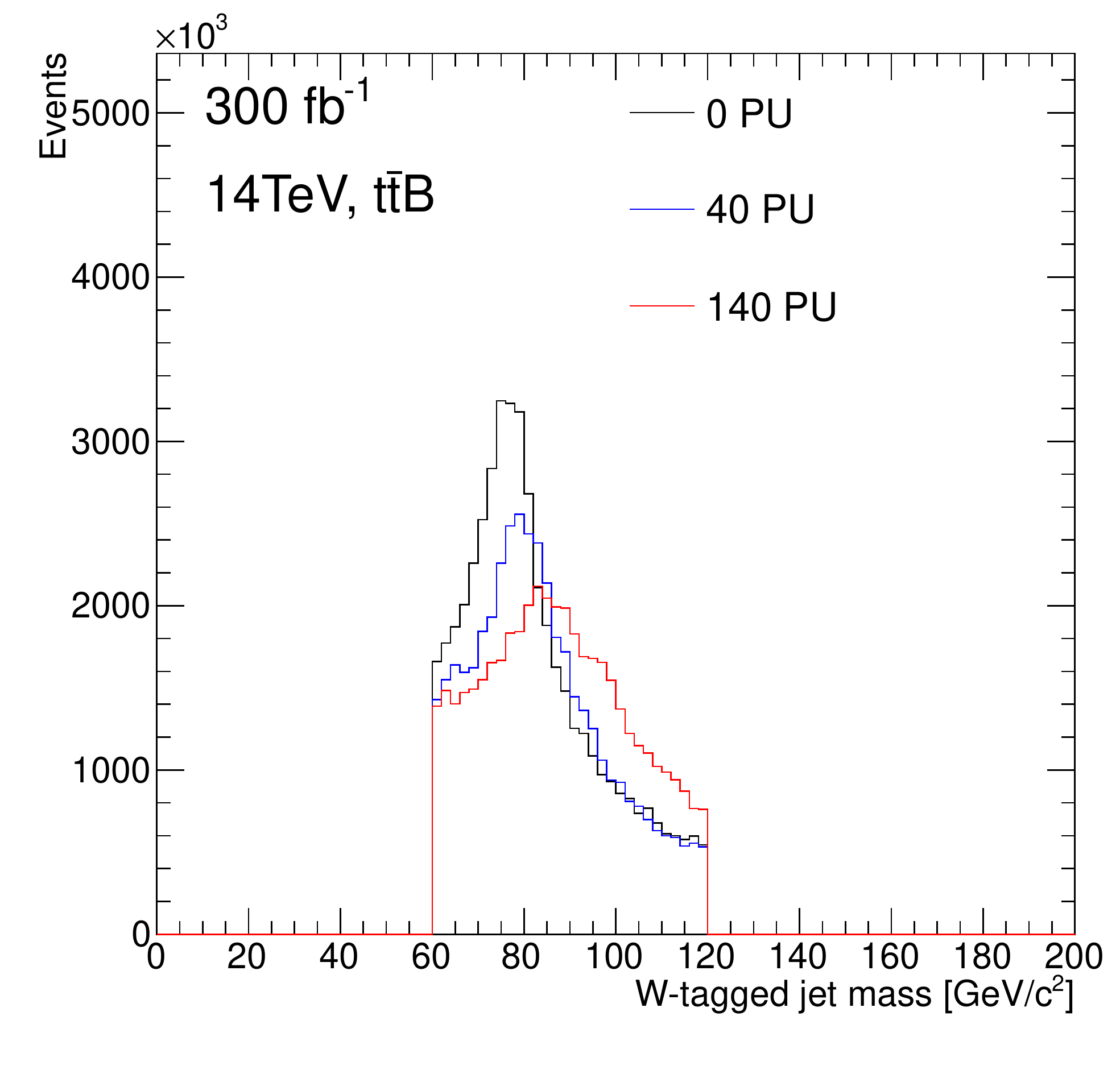}
\includegraphics[width=0.3\textwidth]{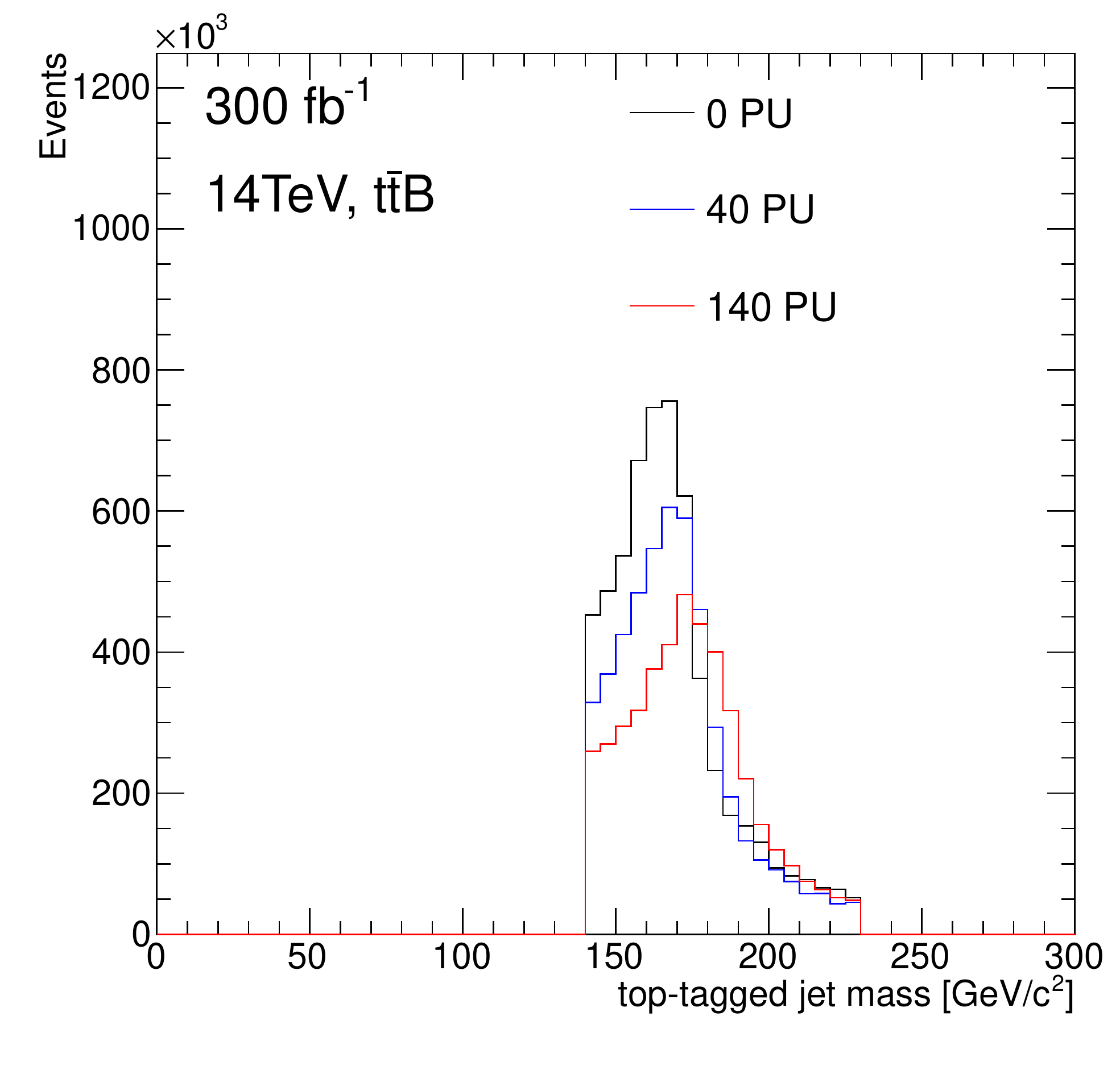}
\caption{Comparison of lepton $p_T$ (left), Leading jet $p_T$ (middle) and third leading
jet $p_T$ (right) in shown in the top row. In the bottom row the distributions for
total number of selected jets (left), the mass of the merged and $W$ jets (middle) 
and top jets (right) are shown. A sample of  $t\overline tV+nJ$ events 
with different event pile-up conditions is used. 
\label{fig:PU02}}
\end{figure}

For LHC Phase I operations, we expect an average of 50 pile-up interactions per bunch crossing,  
and around  140 pile-up interactions for the later high luminosity phase (HL-LHC). 
For the LHC energy upgrade scenario with  $\sqrt{s}$ = 33 TeV (HE-LHC), we also 
consider an average of 140 pile-up interactions per bunch crossing for the studies. 
In Figures~\ref{fig:PU01} and~\ref{fig:PU02}, using a sample of  $t\overline tV+nJ$ events,  
we show the dependence of the shape of 
distributions for the kinematic variables on additional pile-up interactions.
We note that after taking into account jet-by-jet corrections for pileup, based on jet areas, 
many of the distributions, for example $H_T$, $\ETmiss$, $S_T$, and jet $p_T$, show 
a mild variation with additional pile-up events. The ``trimmed mass'' variable computed using
subjets with $\Delta{R}=0.2$ in $W$ and top tagged jets shows a lesser dependence,
because the low $p_T$ subjets are discarded while computing the trimmed 
mass and hence less affected  by pile-up.

\subsection{Comparison between $S^*_T$ binned and inclusive Schemes}
\label{sec:simSMcomp}
\begin{figure}[h!]
\centering
\includegraphics[width=0.3\textwidth]{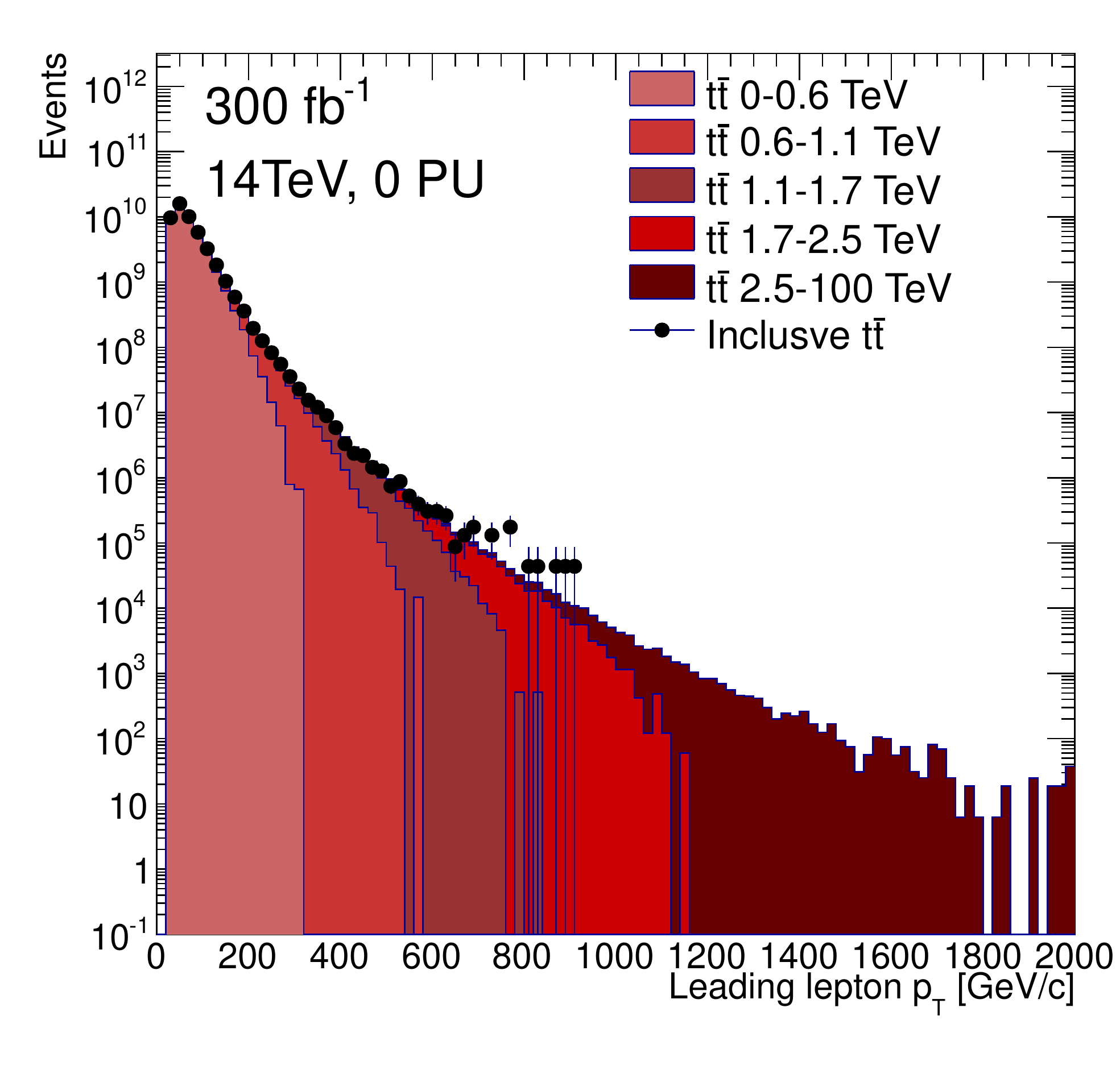}
\includegraphics[width=0.3\textwidth]{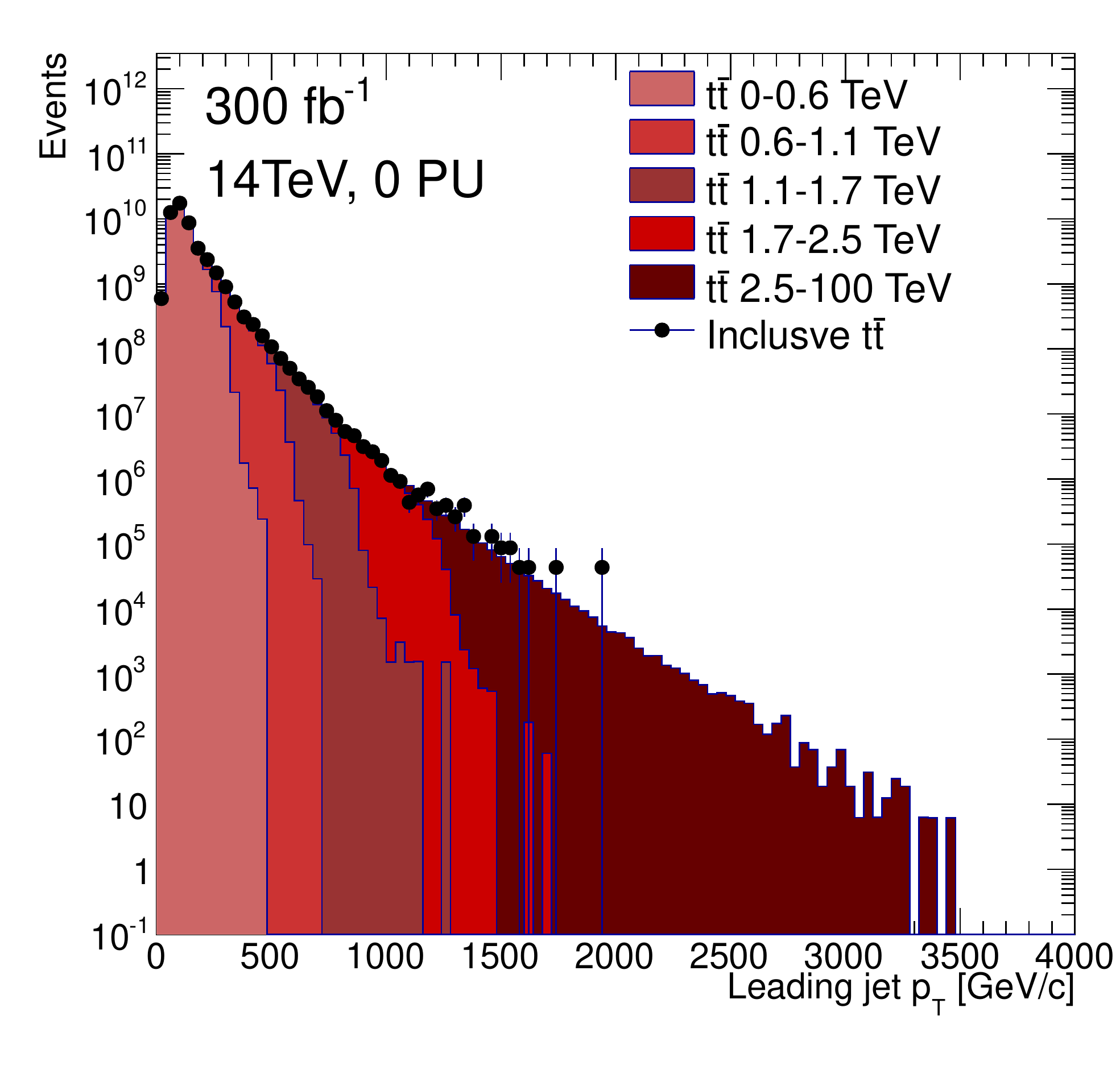}
\includegraphics[width=0.3\textwidth]{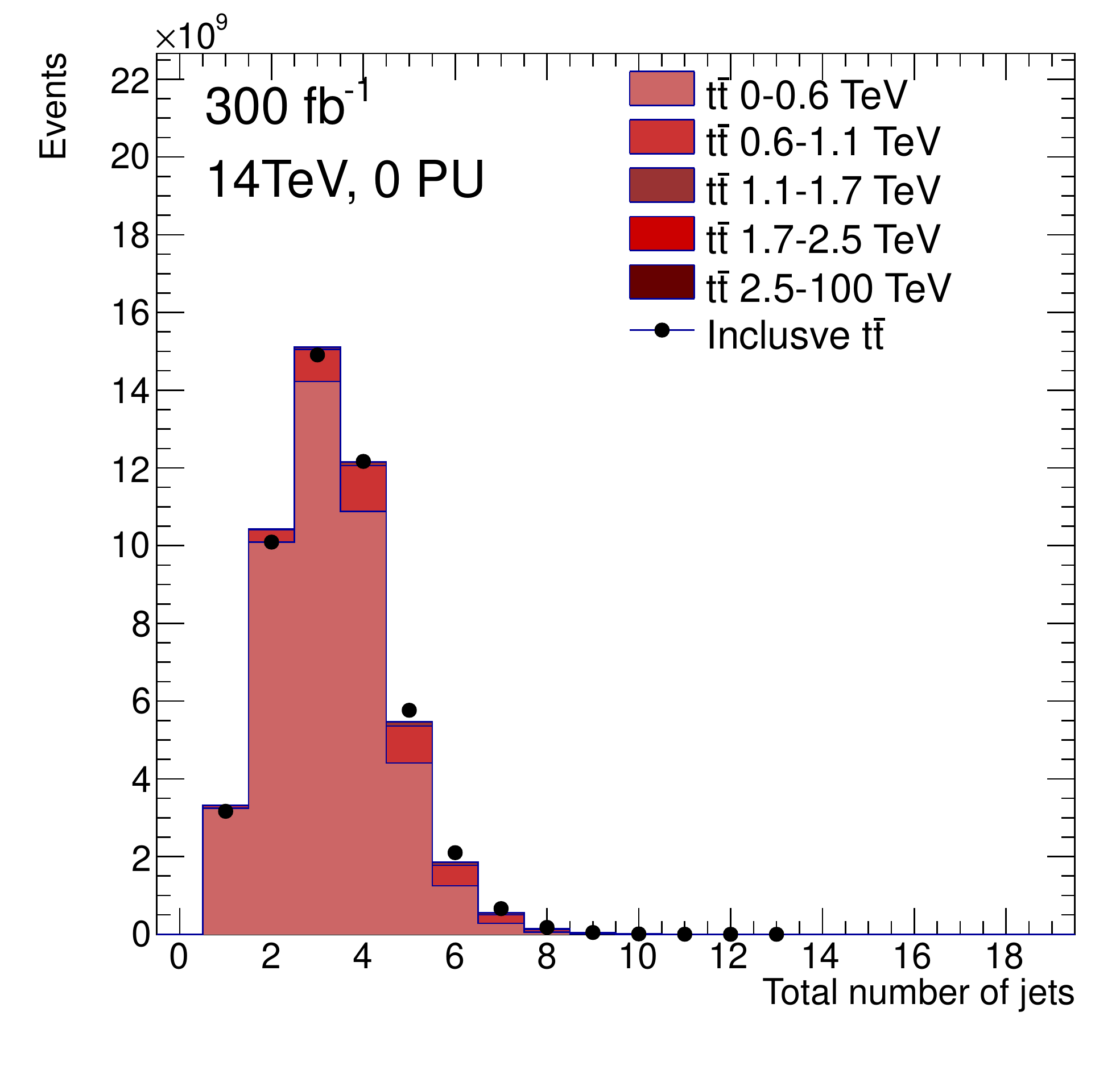}\\
\includegraphics[width=0.3\textwidth]{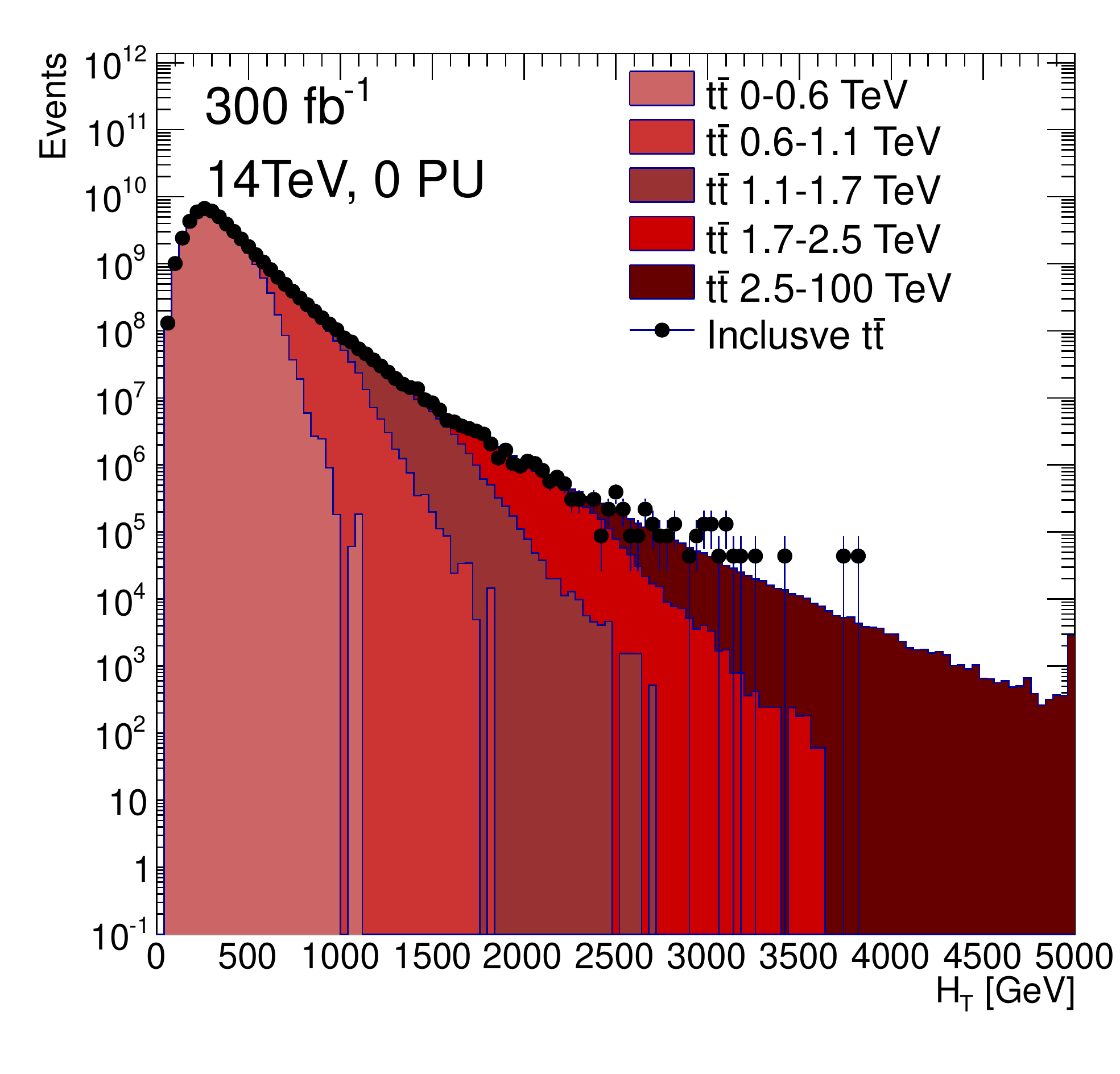}
\includegraphics[width=0.3\textwidth]{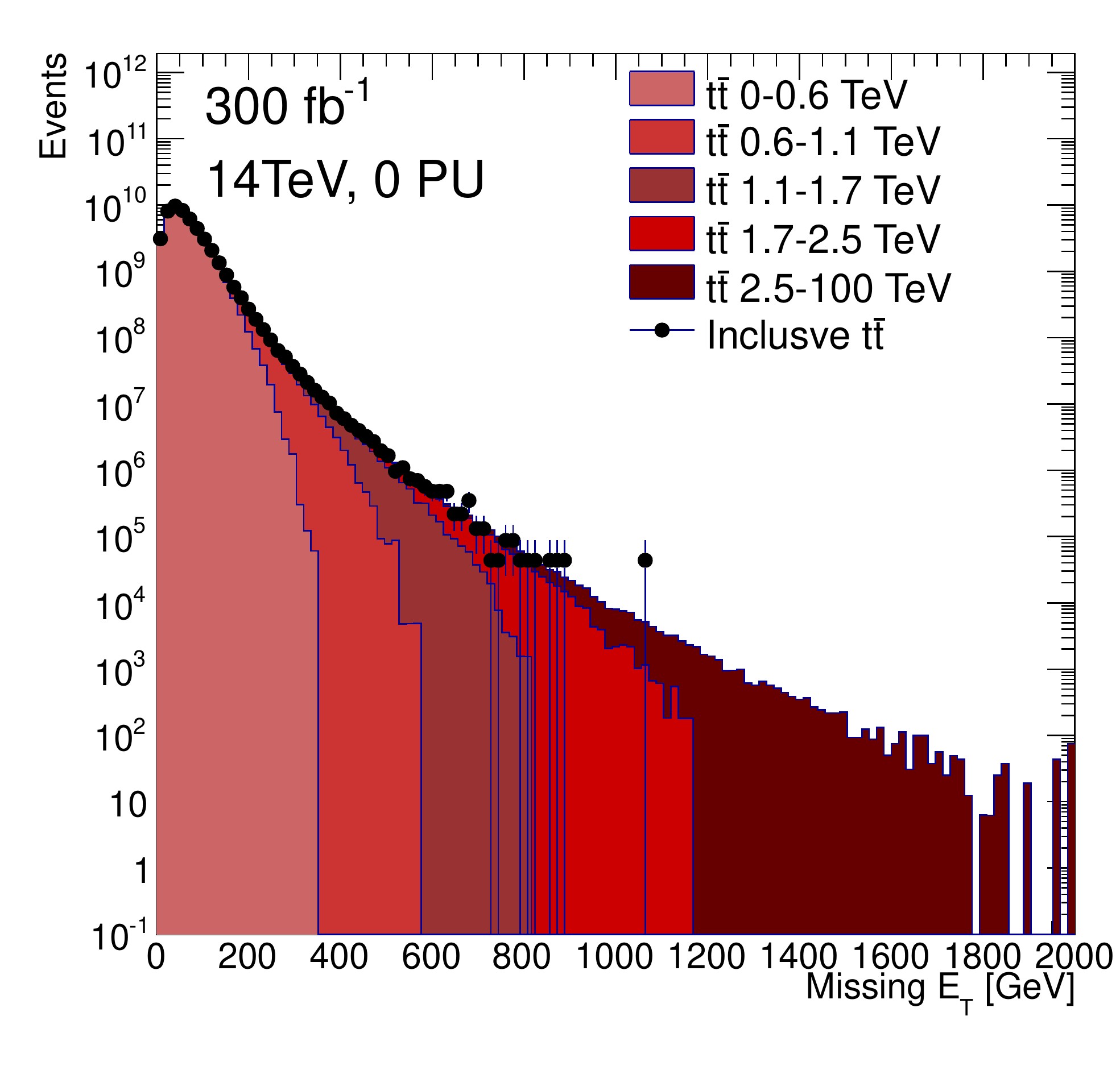}
\includegraphics[width=0.3\textwidth]{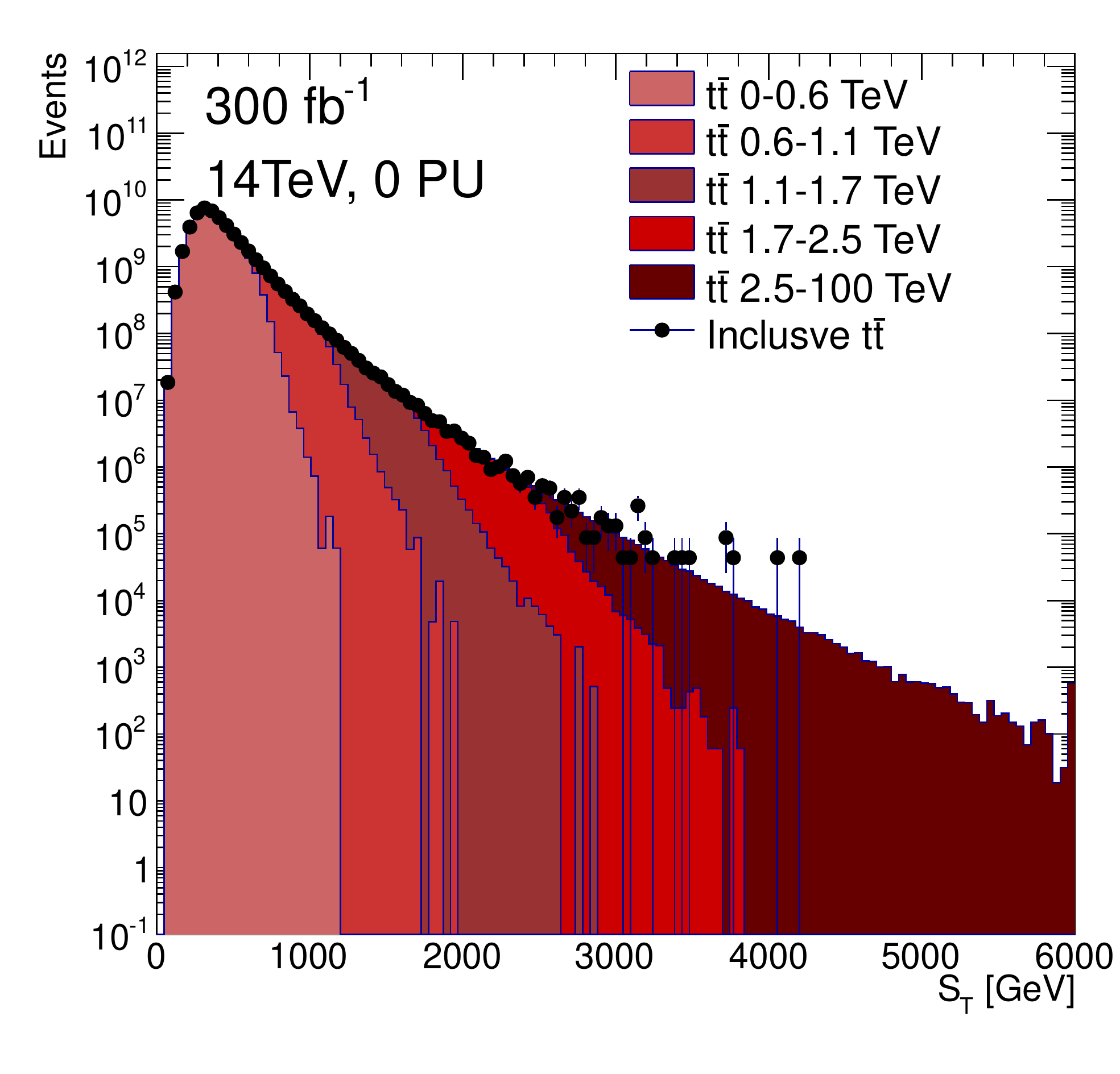}
\caption{Distributions of $p_T$ of leading lepton (top left), $p_T$ of leading jet (top middle), 
total number of selected jets (top right), 
$H_T$ (bottom left),  $\ETmiss$ (bottom middle), and $S_T$ (bottom right) 
in a sample of $t\overline t+nJ$ events. The filled histogram represents events
generated using the $S^*_T$ binned event generation scheme. The distributions
obtained using events generated by  inclusive generation scheme are shown as dots. 
\label{fig:validate}}
\end{figure}

In order to validate the weighted generation scheme based on binning events in $S^*_T$, 
we compare some distributions between  samples generated using the unweighted (inclusive) scheme  
and the $S^*_T$ binned weighted scheme. A few key variables used in many searches for new particles 
($H_T$, $\ETmiss$, $S_T$, lepton and jet $p_T$s) are shown in Fig.~\ref{fig:validate}, where 
a reasonable agreement between the distributions obtained from the two samples is demonstrated. 
Some disagreement at the low $H_T$ regions are due to the slight difference between the 
collision energies of the two  
samples; the inclusive events were generated at $\sqrt{s}=13$ TeV, compared to the
 $\sqrt{s}=14$ TeV used for the  $S^*_T$ binned samples. 

\section{Summary}
\label{sec:simSM-status}
We have developed methods for generating SM Monte Carlo samples with sufficient statistics to accurately model large integrated luminosity data sets.  The samples are relevant for studying the reach for new particle searches at proposed future accelerators.  The events were generated using \texttt{Madgraph}, utilizing a weighted scheme and binned in $S^*_T$, and were then processed via the \texttt{Delphes} framework, which simulates the ``Snowmass'' detector response and performs object reconstruction~\cite{SnowmassPerformanceWP}.  
The SM samples, generated 
using the the Open Science Grid infrastructure~\cite{Snowmass-OSG},
are available and have been used by several Snowmass Energy Frontier studies.

\clearpage

\Acknowledgements
The studies are done using resources provided by the Open Science Grid, 
which is supported by the National Science Foundation and 
the U.S. Department of Energy's Office of Science.  
We are grateful to Matt Reece for modifying BRIDGE for us to include Higgs
decays to photons. JMC is supported by the US DoE under contract number DE-AC02-07CH11359. 
TC and JGW are  supported by the US DoE under contract number DE-AC02-76SF00515.  
KH is supported by an NSF Graduate Research Fellowship under Grant number DGE-0645962. 
MN acknowledges the support from  US DoE via contract number DOE DE-FG02-13ER42023.


\begin{thebibliography}{99}


\bibitem{ATLAS-Higgs} 
  G.~Aad {\it et al.}  [ATLAS Collaboration],
  Phys.\ Lett.\ B {\bf 716}, 1 (2012)
  [arXiv:1207.7214 [hep-ex]].


\bibitem{CMS-Higgs} 
  S.~Chatrchyan {\it et al.}  [CMS Collaboration],
  Phys.\ Lett.\ B {\bf 716}, 30 (2012)
  [arXiv:1207.7235 [hep-ex]].


\bibitem{MCFM} 
  J.~M.~Campbell and R.~K.~Ellis,
  Nucl.\ Phys.\ Proc.\ Suppl.\  {\bf 205-206}, 10 (2010)
  [arXiv:1007.3492 [hep-ph]].

\bibitem{SnowmassPerformanceWP}
  J.~Anderson, A.~Avetisyan, R.~Brock, S.~Chekanov, T.~Cohen, N.~Dhingra, J.~Dolen and J.~Hirschauer {\it et al.},
  arXiv:1309.1057 [hep-ex].


\bibitem{Snowmass-OSG}
  A.~Avetisyan, S.~Bhattacharya, M.~Narain, S.~Padhi, J.~Hirschauer, T.~Levshina, P.~McBride and C.~Sehgal {\it et al.},
  arXiv:1308.0843 [hep-ex].


\bibitem{SnowmassSimulationsTwiki}
\url{http://www.snowmass2013.org/tiki-index.php?page=Energy_Frontier_FastSimulation}


\bibitem{Madgraph}
  J.~Alwall, M.~Herquet, F.~Maltoni, O.~Mattelaer and T.~Stelzer,
  JHEP {\bf 1106}, 128 (2011)
  [arXiv:1106.0522 [hep-ph]].


\bibitem{bridge}
  P.~Meade and M.~Reece,
  hep-ph/0703031.

\bibitem{Pythia6} 
  T.~Sjostrand, S.~Mrenna and P.~Z.~Skands,
  JHEP {\bf 0605}, 026 (2006)
  [hep-ph/0603175].

\bibitem{Delphes}
S.~Ovyn, X.~Rouby and V.~Lemaitre,
  arXiv:0903.2225 [hep-ph].
  J.~de Favereau, C.~Delaere, P.~Demin, A.~Giammanco, V.~Lemaitre, A.~Mertens and M.~Selvaggi,
  arXiv:1307.6346 [hep-ex].

\bibitem{rootA}
R.~Brun and F.~Rademakers,
``ROOT - An Object Oriented Data Analysis Framework,''
Proceedings AIHENP'96 Workshop, Lausanne, Sep. 1996, 
Nucl. Instrum. and Methods Phys. Res. {\bf A389} 81-86 (1997).
See also \url{http://root.cern.ch/}.


\bibitem{Krohn} 
  D.~Krohn, J.~Thaler and L.~-T.~Wang,
  JHEP {\bf 1002}, 084 (2010)
  [arXiv:0912.1342 [hep-ph]].

\end{thebibliography}
\end{document}